\documentclass[preprint,12pt]{elsarticle}

\usepackage{amssymb}
\usepackage{amsmath}
\usepackage{amsthm}

\usepackage{pdflscape}
\usepackage{geometry}
\usepackage{caption}

\usepackage{booktabs,threeparttable}

\journal{Journal of Computational Physics/Combustion and Flame}

\begin{document}

\begin{frontmatter}



\title{ A Split Random 
	Reaction Method for 
	Stiff and Nonstiff Chemically Reacting Flows } 

 \author[label1]{Jian-Hang Wang}
 \author[label1]{Shucheng Pan} 
 \author[label1]{Xiangyu Y. Hu\corref{xiangyu}}
 \author[label1]{Nikolaus A. Adams}
 \cortext[xiangyu]{Corresponding author}
 \address[label1]{Chair of Aerodynamics and Fluid Mechanics, Department of Mechanical Engineering, Technical University of Munich}

\begin{abstract} 
In this paper, a new fractional step method is proposed for 
simulating stiff and nonstiff chemically reacting flows. 
In stiff cases, a well-known spurious numerical phenomenon, i.e. the incorrect propagation speed of discontinuities, may be produced by general fractional step methods due to the under-resolved discretization in both space and time.
The previous random projection method has been successfully applied for stiff detonation capturing in under-resolved conditions.   
Not to randomly project the intermediate state into two presumed equilibrium states (completely burnt or unburnt) as in the random projection method, the present study is to randomly choose the time-dependent advance or stop of a reaction process. Each one-way reaction has been decoupled from the multi-reaction kinetics using operator splitting and the local smeared temperature due to numerical dissipation of shock-capturing schemes is compared with a random one within two limited temperatures corresponding to the advance and its inverse states, respectively, to control the random reaction. The random activation or deactivation in the reaction step is thus promising to correct the deterministic accumulative error of the propagation of discontinuities. Extensive numerical experiments, including model problems and realistic reacting flows in one and two dimensions, demonstrate this expectation as well as the effectiveness and robustness of the method. Meanwhile, for nonstiff problems when spatial and temporal resolutions are fine, the proposed random method recovers the results as general fractional step methods, owing to the increasing possibility of activation with diminishing randomness by adding a shift term. 
\end{abstract}


\end{frontmatter}




\section{Introduction}

One of the main numerical challenges for chemically reacting flows is that the chemical kinetics often includes reactions with widely varying time scales, which may be orders of magnitude faster than the fluid dynamical time scale \cite{bao2000random,bao2001random,bao2002random}. Consider a combustion problem, where the chemical reaction, i.e. the burning process, may be much faster than the gas flow for example. This leads to severe problems of numerical stiffness due to the source terms representing reactions \cite{yee2013spurious}. When the chemical scales are not resolved numerically in time and space (using a grid size larger than the width of the reaction zone), it is not only impossible to capture the detailed structure of the reaction zone (such as the von Neumann spike), but also might calculate a spurious solution with the incorrect propagation of discontinuities and nonphysical states, even though standard dissipative numerical methods that were developed for non-reacting flows with good performance are employed.

The latter numerical phenomenon is well-known and has been an active area of research in the past three decades. It was first observed by Colella \textit{et al.} \cite{colella1986theoretical} in 1986 who considered both the reactive Euler equations and a simplified system obtained by coupling the inviscid Bergers equation with a single reaction equation. LeVeque \& Yee \cite{leveque1990study} showed that a similar spurious propagation phenomenon can happen even with scalar equations, by properly defining a model problem with a stiff source term. By analysis of such a simple scalar problem, they found that the propagation error is mainly due to numerical dissipation contained in the scheme, which smears the discontinuity front and activates the source term in a nonphysical manner. To overcome this difficulty, a natural strategy is to avoid any numerical dissipation in the scheme \cite{bao2002random,zhang2014equilibrium} or to use sufficiently fine mesh. By using a front-tracking approach such as the ghost fluid/level-set method \cite{nguyen2002fully,bourlioux1991theoretical,bourlioux1992theoretical} or the local grid/timestep refinement \cite{jeltsch1999error,bihari1999multiresolution}, the correct propagation speed of the reactive front may be obtained. The random choice method proposed by Chorin in \cite{chorin1976random,chorin1977random} had been successfully used in \cite{colella1986theoretical,majda1990numerical} for the solution of under-resolved detonation waves, which is based on the exact solution of the Riemann problem at randomly chosen locations within the computational cells and does not introduce any viscosity. In \cite{deng2017new}, Deng \textit{et al.} introduced a hybrid reconstruction scheme named MUSCL-THINC-BVD to reduce numerical dissipation around discontinuities significantly to a tolerable level for the examined model experiments. 

However, in wider areas resolution of fine scale is not always realistic due to expensive computational costs, unless one is interested in the detailed structure of a detonation wave. The best one can hope is to capture the speed of the discontinuity as well as other global features of the fluid dynamics \cite{bao2002random}. Also, since numerical dissipation/viscosity is an essential feature of modern shock-capturing schemes with considerable popularity, there is another category of works focused on accepting the diffused profiles by shock-capturing schemes and then make careful use of the averaged information for the correct ignition of source terms in the following reaction step. Engquist \cite{engquist1991robust} presented a simple temperature extrapolation method, which uses an extrapolated temperature from outside the shock profile to activate the chemical source term. This approach is easily extended to multi-dimensions, but it does not work well in insufficient spatial resolutions. In \cite{berkenbosch1998detonation}, Berkernbosch suggested introducing a suitable ignition temperature which is considerably lower than any temperature actually found in the reaction zone of a resolved detonation. Helzel \cite{helzel2000modified} proposed a modified fractional step method for under-resolved detonation waves, in which the exact Riemann solution is required to determine where burning should occur. Tosatto \& Vigevano \cite{tosatto2008numerical} proposed a MinMax method, based on a two-value variable reconstruction within each cell, where the appropriate maximum and minimum values of the unknown are considered within the local neighbouring cells. In \cite{wang2012high} Wang \textit{et al.} proposed a new high-order finite-difference method utilizing the idea of Harten ENO subcell resolution method for stiff source terms with a single reaction and in \cite{yee2013spurious} well-balanced high-order nonlinear filter schemes were added to the subcell resolution method for reacting flows, effectively delaying the onset of wrong speed of propagation in coarse grids and moderate stiff source terms. When the grid is refined, a counter-intuitive spurious behavior (see \cite{yee2013spurious,zhang2015short}) with incorrect shock location was observed.
All these methods are confronted with difficulties in the extension to either high-dimensional or multi-species/multi-reaction kinetics based reacting flows.      
Zhang \textit{et al.} \cite{zhang2014equilibrium} reported their equilibrium state method with the idea of replacing the cell average representation with a two-equilibrium-state reconstruction. The two equilibrium states are locally defined in each transition cell, making its extension to high dimensions straightforward. They also extended the method to a simple multi-reaction system by treating the two one-way reactions totally independent. Unfortunately, realistic nonequilibrium chemical kinetics with multiple finite-rate reversible reactions has not been discussed in any literature so far.
       
In \cite{bao2000random,bao2001random,bao2002random}, Bao \& Jin introduced a random projection method for the reaction step by replacing the ignition temperature with a uniformly distributed random variable. Although the random projection method cannot avoid the introduction of numerical dissipation by shock-capturing schemes, it can eliminate the effect of any numerical dissipation, even with a 1st-order shock-capturing scheme, owing to its random nature. The method was strictly proved using a scalar problem and successfully applied to various model problems of 1D or 2D reactive Euler equations. With the presumption of two time-independent equilibrium states of totally burnt and unburnt gases (regardless of the detailed reaction process), the method is only suitable for under-resolved stiff cases.     
%
%
%
%
 
Here we further discuss the fractional step method using an arbitrary shock-capturing scheme to capture stiff detonation waves in under-resolved conditions. More generally, the main goal of this study is to simulate chemically reacting flows with real-world multi-species multi-reaction nonequilibrium chemistry in a unified manner, regardless of the stiff/nonstiff source terms or the under-/well-resolved conditions in grid and timestep. For the convection step, any modern shock-capturing scheme can be used to solve the homogeneous conservation laws apart from the ordinary differential equations (ODEs) for reaction source terms. Following the convection step, the zero-dimensional ODEs based on the present local smeared state in each cell/point is to be solved in the reaction step of fractional step methods. 
The idea of random projection method implies the reaction being activated or deactivated in one reaction step has no direct correlation with the final correct shock location, unless the activation (in both scalar and Euler problems) or deactivation (only in scalar problems) constantly occurs without restriction. The accumulative error will grow with time and leads to spurious propagation of discontinuities in a long run. On the contrary, if the activation or deactivation can occur alternatively and randomly according to a certain possibility, the correct shock location can be obtained with temporal convergence. Unlike Bao \& Jin's random projection method, the activation and deactivation of chemical reactions in our proposed method will not be projected into two prescribed equilibrium states, as a \textit{priori}, but two time-dependent states corresponding to advancing the reaction in one timestep forward and making the reaction stand still, respectively. The criterion to the progress of a reaction is by comparing the local smeared temperature with a randomized temperature depending on the advance state and the state of the inverse of advance. In this way, every reaction step contains an effect of the predictor-corrector algorithm (predictor is the advance state and corrector draws the predictor back to the current state) for the correct and controllable propagation of the reacting front. 
Besides, by adding a shift term into the random temperature sampling when the resolution is improving, the chosen random temperature tends to be below the mean value of the two limited temperatures and thus activation of the reaction is increasingly possible to happen as the deterministic methods always do.    
That is, the proposed method recovers the solution of a general fractional step method in nonstiff cases when the spatial and temporal resolutions are fine to resolve the reaction scales. Consequently, the method is promising for both stiff and nonstiff problems in under-resolved and resolved conditions.

On the other hand, different from the famous ODEs solvers such as the implicit solver VODE \cite{brown1989vode}, explicit CHEMEQ2 \cite{mott2001chemeq2}, scale-separated MTS/HMTS \cite{gou2010dynamic} and the recent quasi-steady-state approximation based ERENA \cite{morii2016erena}, the present random ODEs solver basically takes the advantages of the Split Single Reaction Integrator (SSRI) \cite{nguyen2009mass} for chemical kinetics in both mass conservation and preserving the positivity of mass fractions. Using analytical solutions in SSRI or the approximate exact solution in our development, almost unconditional stability can be a promise for the present ODEs solver. Therefore, even when the timestep is large and under-resolved for small chemical time scales, the ODEs solver is still able to work effectively and also paves the way for subsequent randomization of each reaction. Not limited to model problems with simplified kinetics reported in previous literature, operator splitting upon the reaction system makes the proposed random method applicable for real-world reacting flows with complicated nonequilibrium chemistry involving multiple species and reactions, e.g. the hydrogen-air combustion kinetics.           

The paper is organized as follows. In Section \ref{Formulation}, we introduce the concerned reactive Euler equations with chemical reaction source terms. A standard fractional step method to solve the Euler system is outlined by operator splitting into the convection step and reaction step. In the reaction step, a new ODEs solver, as the generalization of SSRI, is developed to approximate the exact solution with advantages of exact mass conservation and strict definite positivity as well as almost unconditional stability. Based on the split reaction-by-reaction ODEs solver for general chemical kinetics, individual random reaction between advancing and stopping its process can be realized to correct the deterministic spurious propagation of discontinuities in stiff and under-resolved conditions. Next in Section \ref{section2}, by comparing with other standard methods, we examine the pure ODEs solver and the split random reaction method as a new fractional step method for capturing stiff detonations, respectively, by extensive classical model examples and realistic reacting flows in both 1D and 2D numerically. Conclusions will be drawn in the last section. More information about the ODEs solver and reaction mechanism used in numerical tests are provided in the final appendices.

\section{Formulation}
\label{Formulation}

We have a first glance at the mathematical model of the time-dependent reacting flows involving nonequilibrium chemical kinetics, i.e. reactive Euler equations with chemical source terms. 
Assuming the flow is compressible, inviscid and in two dimensions for simplicity, the multi-species Euler equations coupled with reaction source terms take the form
\begin{equation}\label{Euler_eq}
\begin{aligned}
U_t + F(U)_x + G(U)_y = S(U),
\end{aligned}
\end{equation} 
where 
\begin{equation}\label{Euler_eq1}
\begin{aligned}
U=
\begin{pmatrix}
\rho\\
\rho u\\
\rho v\\
\rho e_t\\
\rho y_1\\
\rho y_2\\
\cdots\\
\rho y_{N_s-1}\\
\end{pmatrix},
F(U)=
\begin{pmatrix}
\rho u\\
\rho u^2 + p\\
\rho u v\\
(\rho e_t + p) u\\
\rho u y_1\\
\rho u y_2\\
\cdots\\
\rho u y_{N_s-1}\\
\end{pmatrix},
G(U)=
\begin{pmatrix}
\rho v\\
\rho u v\\
\rho v^2 + p\\
(\rho e_t + p) v\\
\rho v y_1\\
\rho v y_2\\
\cdots\\
\rho v y_{N_s-1}\\
\end{pmatrix},
S(U)=
\begin{pmatrix}
0\\
0\\
0\\
0\\
\dot{\omega_1}\\
\dot{\omega_2}\\
\cdots\\
\dot{\omega_{N_s-1}}\\
\end{pmatrix}
\end{aligned}
\end{equation} 
are vectors of the conserved variables, advection flux in the x- or y-direction and source terms, respectively, with $\dot{\omega_i}$ representing the rate of change of species $i$ in the reactive gas mixture due to the chemical kinetics consisting of $N_r$ reactions and $N_s$ species. Furthermore, $e_t = e + \frac{1}{2}(u^2+v^2)$ is the specific total energy including the specific internal energy $e$. To the closure of the system, the equation of state (EoS) for the chemically reactive mixture should be added. Thus the density $\rho$, pressure $p$ and temperature $T$ of the gas mixture can be explicitly connected by
\begin{equation}\label{EoS}
\begin{aligned}
 p = \rho \sum^{N_s}_{i=1} y_i \frac{R_u}{W_i} T,
\end{aligned}
\end{equation}   
with $y_i$ and $W_i$ denoting the mass fraction and molecular weight of the $i$-th species, respectively, and $R_u$ being the universal gas constant.

The above conservation laws of mass, momentums and energy with source terms are usually solved numerically in a fractional step manner, i.e. based on operator splitting, we have a set of partial differential equations (PDEs) for the homogeneous fluid transport dynamics 
\begin{equation}\label{S_c}
\begin{aligned}
S_c:\quad U_t + F(U)_x + G(U)_y = 0
\end{aligned}
\end{equation} 
assuming the chemical reactions are frozen and mass fractions of all species are transported during the pure convection process,
apart from the system of ODEs in the chemical kinetics
\begin{equation}\label{S_r}
\begin{aligned}
S_r:\quad \frac{dy_i}{dt} = \frac{\dot{\omega_i}}{\rho},\quad i=1,\dots,N_s,
\end{aligned}
\end{equation} 
under adiabatic and constant-volume conditions with fixed total density and constant specific internal energy. The first-order accurate Lie splitting scheme \cite{mclachlan2002splitting} (also known as Godunov splitting \cite{toro2013riemann}) or the second-order Strang splitting \cite{strang1968construction} can be employed to approximate the solution from the discrete time level $n$ to $n+1$ with a timestep of $\Delta t$, in the following forms 
\begin{equation}\label{Lie}
\begin{aligned}
U^{n+1} = S_r^{(\Delta t)} \circ S_c^{(\Delta t)} U^{n},
\end{aligned}
\end{equation}
or
\begin{equation}\label{Strang}
\begin{aligned}
U^{n+1} = S_c^{(\frac{\Delta t}{2})} \circ S_r^{(\Delta t)} \circ S_c^{(\frac{\Delta t}{2})} U^{n}.
\end{aligned}
\end{equation}
In many practical cases nearly identical results are obtained with both splitting schemes \cite{deiterding2003parallel}. 
Regardless of the selection of operator splitting schemes, the method of fractional steps decouples the physical processes of hydrodynamic transport and chemical reaction, i.e. a convection step and a reaction step from a computational viewpoint. Accordingly, for the convection operator $S_c$, any modern shock-capturing methods especially some high-order low-dissipation schemes such as WENO-JS5 \cite{jiang1996efficient}, WENO-CU6 \cite{hu2010adaptive} and the recently proposed TENO6 \cite{fu2016family} with local/global Lax-Friedrich flux splitting can be adopted. Also, in the reaction step, for $S_r$, any ODEs solver such as VODE, CHEMEQ2 and MTS, etc., can be conveniently implemented as a "black-box", intaking $\{y_1,\dots,y_{N_s}\}^{n}$ and outputting $\{y_1,\dots,y_{N_s}\}^{n+1}$ with several case-dependent constant inputs such as $e$, $\rho$, $T$, etc. Besides, local sub-stepping/cycling can be presumed or executed adaptively in the ODEs solver.  

Despite using high-order shock-capturing schemes in $S_c$, numerical dissipation or viscosity is inherently existing. The captured discontinuities in the discrete space will therefore be smeared instead of sharp jumps, which indicates the predicted properties in such smeared locations/areas of the flowfield are numerically averaged properties rather than physically realistic ones. It is the nonphysical properties in the smeared discontinuities that further induce the incorrect (too early) ignition of chemical reactions by pointwisely evaluating the source terms and finally lead to a spurious solution with a bifurcating wave pattern and wrong propagation speed of the reacting front. On the other hand, with more or less numerical viscosity, modern high-resolution shock-capturing schemes benefit in good robustness and accuracy, being widely accepted for solving homogeneous conservation laws in practice. From this aspect, it is highly desirable to develop methods for reacting flows that, instead of avoiding the numerical viscosity, make correct use of it. 

Therefore, attention has to be paid from $S_c$ to $S_r$: we introduce and improve SSRI for solving ODEs of the nonequilibrium chemical kinetics so that the multi-reaction system can be decoupled into a series of single reaction steps. Then we introduce the idea of random projection into the ODEs solver in order to realize the random ignition of reactions. In our development, each reaction process will be randomly advanced one timestep forward (activation) or be ceased (deactivation) instead of being projected into two prescribed equilibrium states (completely burnt and unburnt). In this way, the randomization of reactions can be achieved for the general real-world nonequilibrium kinetics of multiple finite-rate reactions, no matter the source terms are stiff or nonstiff and the numerical discretization in space and time is under-resolved or resolved, in a unified manner. Hereafter, we term the randomized and reaction-by-reaction ODEs solver for the nonequilibrium chemistry, to be Split Random Reaction Method (SRR) in the reaction step $S_r$, independent of the convection operator $S_c$.         
 
\subsection{Split reaction-by-reaction ODEs solver for chemical kinetics}

In a common nonequilibrium chemical kinetics accounting for the ODEs in Eq. \eqref{S_r}, chemical production rates are derived from a reaction mechanism that consists of $N_s$ species and $N_r$ reactions
\begin{equation}
\sum_{i=1}^{N_s} \nu_{ji}^f X_i  \Longleftrightarrow \sum_{i=1}^{N_s} \nu_{ji}^b X_i,  \quad j=1,\dots,N_r,
\end{equation} 
where $\nu_{ji}^f$ and $\nu_{ji}^b$ are the stoichiometric coefficients of species $i$ appearing as a reactant and as a product in reaction $j$. The net production rate of species $i$ in Eqs. \eqref{Euler_eq1} and \eqref{S_r} is usually the summation of the production rate from each single elementary reaction as
\begin{equation}
\dot{\omega_i} = W_i \sum_{j=1}^{N_r} (\nu_{ji}^b-\nu_{ji}^f) \left[ k_j^f \prod_{l=1}^{N_s} \left[\frac{\rho_l}{W_l}\right]^{\nu_{jl}^f} - k_j^b \prod_{l=1}^{N_s} \left[\frac{\rho_l}{W_l}\right]^{\nu_{jl}^b}  \right]
\end{equation} 
with $k_j^f$ and $k_j^b$ denoting the forward and backward reaction rate of each chemical reaction. Note that reactions are reversible here for the sake of generality.

In SSRI, Nguyen \textit{et al.} successfully utilize operator splitting in a reaction-by-reaction manner to decouple the above multi-reaction system in order to achieve definite positivity and mass conservation during the temporal integration. However, only simple one-way reactions with constant rates and two or three reactants at most are considered, using an analytical exact solution. For reactions with more than three reactants or the stoichiometric coefficients of reactants are larger than one, which indicates the overall order of the reaction is usually higher than two, analytical solutions are explicitly unavailable or difficult to derive. Alternatively, numerical solutions which require root-finding algorithms result in additional computational costs.     

Following the idea of SSRI, we also decouple the multi-reaction system by operator splitting at first, taking the Lie splitting for example, which means during a given timestep, we traverse all the reactions by visiting each reaction separately and successively. That is, one simply needs to consider the effect of one reaction on the mass production or consumption of species involved in this reaction and then move on to the next one till the completeness of traversal 
\begin{equation}\label{R1st}
\begin{aligned}
S_r: \quad R_{1st}^{(\Delta t)} = R_{N_r}^{(\Delta t)} \circ R_{N_r-1}^{(\Delta t)} \circ \cdots \circ R_2^{(\Delta t)} \circ R_1^{(\Delta t)},
\end{aligned}
\end{equation}
where each $R_j$ corresponds to a single reaction channel, independent of all other reactions. The reaction-by-reaction idea agrees with the physical reality that in a microscopic scale, one molecule/atom can only experience one reaction or event with others or by itself solely at one time instance, which is the case in the stochastic simulation of chemical kinetics \cite{gibson1999efficient}. Unsurprisingly in a macroscopic scale, the reactions involving large numbers of species molecules/atoms can be treated as simultaneously occurring processes. In the original SSRI, the second-order accurate Strang splitting is adopted and the traversal goes forward first from the fastest reaction to the lowest one for half a timestep and goes backward in a reverse direction afterwards for the rest half timestep. Here we take the traversal order not according to reaction rates but to the number of index in the reaction mechanism that we adopt, which is more general but simpler without loss of the convergence rate, i.e.
\begin{equation}\label{R2nd}
\begin{aligned}
S_r: \quad R_{2nd}^{(\Delta t)} & = R_1^{(\frac{\Delta t}{2})} \circ R_2^{(\frac{\Delta t}{2})} \circ \cdots \circ R_{N_r-1}^{(\frac{\Delta t}{2})} \circ R_{N_r}^{(\frac{\Delta t}{2})} \circ
R_{N_r}^{(\frac{\Delta t}{2})} \circ R_{N_r-1}^{(\frac{\Delta t}{2})} \circ \cdots \circ R_2^{(\frac{\Delta t}{2})} \circ R_1^{(\frac{\Delta t}{2})} \\
& = \overline{ R_{1st}^{(\frac{\Delta t}{2})} } \circ R_{1st}^{(\frac{\Delta t}{2})},
\end{aligned}
\end{equation}
where $\overline{ R_{1st} } $ is the inverse operator of $R_{1st}$. 
Accordingly for each $R_j$, we have 
\begin{equation}\label{Rj}
\begin{aligned}
R_j:\quad & \sum_{i=1}^{N_s} \nu_{ji}^f S_i  \Longleftrightarrow \sum_{i=1}^{N_s} \nu_{ji}^b S_i, \\
&\frac{dy_i}{dt} = \frac{\dot{\omega_i}^j}{\rho}, \quad i=1,\dots,N_s, \\
&\dot{\omega_i}^j = W_i (\nu_{ji}^b-\nu_{ji}^f) \left[ k_j^f \prod_{l=1}^{N_s} \left[\frac{\rho_l}{W_l}\right]^{\nu_{jl}^f} - k_j^b \prod_{l=1}^{N_s} \left[\frac{\rho_l}{W_l}\right]^{\nu_{jl}^b}  \right].
\end{aligned}
\end{equation}
We now rewrite the ODEs in Eq. \eqref{Rj} in the following form \cite{mott2001chemeq2}
\begin{equation}\label{qss}
\frac{dy_i}{dt} = q_i^j - p_i^j y_i, \quad i=1,\dots,N_s,
\end{equation}
where $q_i^j \geq 0$ is the production rate and $p_i^j y_i^j \geq 0$ is the loss rate for the $i^{th}$ species through reaction $j$.

Following the operator splitting of reactions, we continue to split the reversible reaction, e.g. reaction $j$ if applicable, apart into the forward reaction and backward reaction as 
\begin{equation}\label{Rfb}
R_j^{(\Delta t)} = R_{j,b}^{(\Delta t)} \circ R_{j,f}^{(\Delta t)}
\end{equation}
such that the species involved will either gain mass or lose mass through the one-way forward/backward reaction from Eq. \eqref{qss}, i.e.
\begin{equation}\label{pq}
\begin{aligned}
\text{if gain mass}: q_i^j \geq 0, \, p_i^j y_i = 0, \\
\text{else lose mass}: q_i^j = 0, \, p_i^j y_i \geq 0,
\end{aligned}
\end{equation}
with the simplified
\begin{equation}\label{pq1}
\begin{aligned}
q_i^j &= \frac{W_i}{\rho} \nu_{ji}^b \left[ k_j^f \prod_{l=1}^{N_s} \left[\frac{\rho_l}{W_l}\right]^{\nu_{jl}^f} \right], \, p_i^j y_i = 0 \quad \text{for product species}, \\
q_i^j &= 0, \, p_i^j y_i = \frac{W_i}{\rho} \nu_{ji}^f \left[ k_j^f \prod_{l=1}^{N_s} \left[\frac{\rho_l}{W_l}\right]^{\nu_{jl}^f} \right] \quad \text{for reactant species} \\
\end{aligned}
\end{equation} 
in a forward reaction for example. It is clear that a backward reaction can be thought of as a forward one inversely if we exchange the reactants and products. Also, an irreversible reaction can be treated as a reversible one with a backward reaction rate being equal to zero such that the idea of splitting is still applicable. 

Since each elementary reaction has been numerically decoupled from the rest and each reversible reaction again has been split into two oppositely unidirectional reactions, one finally merely ought to solve a single reaction equation as
\begin{equation}\label{one-way}
a A + b B + \cdots \longrightarrow x X + y Y + \cdots
\end{equation}
in every operation. Mass conservation and positivity of mass fractions, the two highly significant requirements for either accuracy or stability of the numerical integration, can be carefully and properly treated.

In some simple cases for the reaction Eq. \eqref{one-way} from wide applications, with the following forms  
\begin{equation}\label{one-way-simple}
\begin{aligned}
A \longrightarrow \text{products}, \\
\text{or} \quad A + B \longrightarrow \text{products}, \\
\text{or} \quad 2A \longrightarrow \text{products}, \\ 
\text{or} \quad A + B + C \longrightarrow \text{products}, \\
\text{or} \quad 3A \longrightarrow \text{products}, \\
\end{aligned}
\end{equation}
one may easily find their analytical solutions, see \ref{appendix1}. It is thus natural to employ the analytical solutions rather than numerical solutions, with the advantages of avoiding introducing any numerical scheme error and being unconditionally stable \cite{nguyen2009mass}. However, as previously stated, for the general form of Eq. \eqref{one-way} (usually with a higher overall order than two) whose analytical solution is explicitly unavailable or difficult to derive, a more convenient alternative is to perform quasi-steady-state (QSS) methods to obtain the approximate exact solution. 

The QSS methods are based on the exact solution of Eq. \eqref{qss} if $p_i^j$ and $q_i^j$ are constant \cite{jay1997improved,verwer1995explicit}, i.e.
\begin{equation}\label{exact_solution}
y_i^{n+1} = y_i^n e^{-p_i^j \Delta t} + \frac{q_i^j}{p_i^j}(1-e^{-p_i^j \Delta t}), \quad i=1,\dots,N_s.
\end{equation}
However, in practice $p_i^j$ and $q_i^j$ inherently depend on $\{y_1,\dots,y_{N_s}\}$ from Eq. \eqref{pq} or \eqref{pq1} and Eq. \eqref{exact_solution} provides an approximate solution if one assumes $p_i^j$ and $q_i^j$ are fixed during the timestep. The present SRR method is based on this plain approximate exact solution without invoking traditional time-integration schemes such as the Euler scheme with a poor stability \cite{nguyen2009mass}. Consequently, the QSS-based SRR method is almost unconditionally stable, which means the timestep size is not limited to the characteristic time sizes of chemical species and thus a larger timestep rendering less computational efforts is possible.

\newtheorem{mydef}{Remark}
\theoremstyle{remark}
\begin{mydef}
The plain QSS approximation adopted here in the SRR method is first-order accurate. But given that fluid dynamic calculations are seldom accurate to better than a few percent, any requirement of the chemical integrator to calculate the species concentrations more accurately than a few tenths of a percent is usually extensive. And the chemical integrator may be relatively low-order \cite{mott2001chemeq2}.  
\end{mydef}

\subsubsection{treatment for mass conservation}

If we straightforwardly employ the approximate solution of QSS in Eq. \eqref{exact_solution} for all the species through a reaction, we will have
\begin{equation}\label{sum_exact_solution}
\begin{aligned}
\sum_{i=1}^{N_s} y_i^{n+1} & = \sum_{i=1}^{N_s} \left( y_i^n e^{-p_i^j \Delta t} + \frac{q_i^j}{p_i^j}(1-e^{-p_i^j \Delta t}) \right) \\
& \neq 1. 
\end{aligned}
\end{equation}
It is obvious to see that mass conservation is not preserved. To cure this problem and utilize the excellent stability of the QSS approximation, instead of advancing $y^n$ to $y^{n+1}$ for all the species involved, one can choose to only advance $y_k^n$ to $y_k^{n+1}$ of a reactant species $k$ by Eq. \eqref{exact_solution} and update other $\{y_{i,i\neq k}\}^{n+1}$ by the law of mass conservation of a single reaction equation in Eq. \eqref{Rj}. This merit of knowing the exact net gain or loss of mass of other species originates from the operation upon only one reaction decoupled from others in both the present method and the original SSRI. Therefore, for the reactant $k$, combining Eqs. \eqref{exact_solution} and \eqref{pq1} we have
\begin{equation}\label{reactant_k}
y_k^{n+1} = y_k^n e^{-p_k^j \Delta t}
\end{equation}
and for the rest species including other reactants and all the products in the reaction $j$, taking species $i$ for example, its change of mass fraction $\Delta y_i=y_i^{n+1} - y_i^{n}$ should obey
\begin{equation}\label{delta}  
\frac{\Delta y_i/W_i}{ \nu_{ji}^b-\nu_{ji}^f } = \frac{\Delta y_k/W_k}{ \nu_{jk}^b-\nu_{jk}^f }, 
\end{equation}
(which is essentially the conservation of the number of particles involved in a reaction system,) giving the below update 
\begin{equation}\label{mass_delta}
\begin{aligned}
 y_i^{n+1} &= y_i^{n} + \Delta y_i \\
 		   &= y_i^{n} + \frac{\nu_{ji}^b-\nu_{ji}^f}{\nu_{jk}^b-\nu_{jk}^f} \frac{W_i}{W_k} \Delta y_k.
\end{aligned}
\end{equation}
It is easy to prove that $\sum_{i=1}^{N_s} \Delta y_i = 0 $ which is equivalent to $\sum_{i=1}^{N_s} y_i =1$ for mass conservation. 
 
\subsubsection{Positivity-preserving treatment}

Since we only need to consider a single one-way reaction (forward or backward reaction) after two splitting procedures, the mass loss of reactants are exactly known and the non-negative mass fraction should be promised for the reactant species which are suffering mass loss. Without loss of generality, considering the forward reaction of the $j^{th}$ reaction and assuming that reactant $k$ with $\nu_{jk}^b=0$ is imposed by the QSS approximation in Eq. \eqref{reactant_k}, we further look into another reactant species, e.g. $i$ with $\nu_{ji}^b=0$, and we combine Eqs. \eqref{reactant_k} and \eqref{mass_delta} to obtain
\begin{equation}\label{mass_delta1}
\begin{aligned}
 y_i^{n+1} 
 		   &= y_i^{n} - \frac{\nu_{ji}^f}{\nu_{jk}^f} \frac{W_i}{W_k} y_k^n + \frac{\nu_{ji}^f}{\nu_{jk}^f} \frac{W_i}{W_k} y_k^n e^{-p_k^j\Delta t}.
\end{aligned}
\end{equation}
Recalling Eq. \eqref{pq1} for reactants $i$ and $k$, we have 
\begin{equation}\label{pq2}
	\frac{p_i^j y_i}{p_k^j y_k} = \frac{\nu_{ji}^f}{\nu_{jk}^f} \frac{W_i}{W_k}.
\end{equation}
Then rearrange Eq. \eqref{pq2} and substitute it into Eq. \eqref{mass_delta1} we can obtain
\begin{equation}\label{ynew}
\begin{aligned}
 y_i^{n+1} 
		   &= y_i^{n}\frac{p_k^j-p_i^j}{p_k^j} + \frac{\nu_{ji}^f}{\nu_{jk}^f} \frac{W_i}{W_k} y_k^n e^{-p_k^j\Delta t}.
\end{aligned}
\end{equation}
With the aid of Eq. \eqref{ynew}, it is readily to see that we can guarantee the positivity of $y_i^{n+1}$, i.e. $y_i^{n+1} \geq 0$, by choosing $p_k^j \geq p_i^j$ since the second right-hand term is always non-negative. Therefore, for this reaction, in order to preserve the positivity of species mass fractions, especially for the reactants involved, the reactant $k$ using the QSS approximation should satisfy
\begin{equation}
	p_k^j = max\{p_i^j\} \quad \text{among all the reactant species in reaction \,} j. 
\end{equation}
Regarding the positivity preserving for the choosen reactant $k$, according to Eq. \eqref{reactant_k}, it is naturally satisfied owing to the positivity of the exponential function. 

\theoremstyle{remark}
\begin{mydef}
The original SSRI and its improved counterpart in this study both can perform sufficiently well for the pure system of ODEs in chemical kinetics as a stand-alone solver. Randomization of this ODEs solver in the next subsection is not designed for integrating the ODEs accurately and individually, but mainly aimed at cancelling the effect of the harmful but unavoidable introduction of numerical dissipation resulting from the hydrodynamic solver $S_c$ using shock-capturing schemes into the reaction step $S_r$, i.e. the early ignition.
\end{mydef}

\subsection{Finite randomization of chemical reactions} 

Bao \& Jin \cite{bao2000random,bao2001random,bao2002random} first proposed the idea of random projection into the ODEs solver instead of the deterministic projection which strictly obeys the time-dependent integration based on the local smeared information around the discontinuities. They also theoretically proved the random projection method gives basically first-order convergence for the scalar problem. For both scalar problems and Euler equations with stiff source terms, their random projection method is numerically demonstrated to be of excellent performance in obtaining the correct propagation of shocks and reacting fronts in under-resolved spatial and temporal discretizations.     

After two steps of operation splitting upon the ODEs system in $S_r$, one only needs to consider the randomization of a single one-way reaction from time point $t_n$ to $t_{n+1}$ for an interval $\Delta t$. In Bao \& Jin's formulation, temperature will be a randomized variable instead of its local value to determine the progress (completely burnt or not) of the entire reaction system, by comparing with a pre-known ignition temperature, $T_{ign}$. A upper and lower limit of temperature are needed, i.e. $T_u$ and $T_b$ (corresponding to the two equilibrium states of the initial combustible gas mixture being completely burnt and unburnt) as a \textit{priori}. Therefore, in such cases the equilibrium states are presumed and distributed before and behind the discontinuity as initial conditions, having not taken into account the far more complicated time-dependent finite-rate nonequilibrium chemistry 
without
defined equilibrium states.

By the above split reaction method, we advance the current state vector $\{y_1,\dots,y_{N_s}\}$ through a single one-way reaction indexed by the subscript $j$ for generality, as in Eq. \eqref{one-way}, as
\begin{equation}\label{y+}
\begin{aligned}
	 \{y_1,\dots,y_{N_s}\}^{+} = R_j^{(\Delta t)} \{y_1,\dots,y_{N_s}\},
\end{aligned}
\end{equation}
where $\{y_1,\dots,y_{N_s}\}^{+}$ represents the advance in time by one operation $R_j$ (i.e. $R_j^f$ or $R_j^b$ after splitting the reversible reaction in Eq. \eqref{Rfb}). Thus, we can obtain the change of mass fractions for the species involved in this reaction, i.e.
\begin{equation}\label{dy}
\begin{aligned}
	   \{\Delta y_1,\dots,\Delta y_{N_s}\}_j = \{y_1,\dots,y_{N_s}\}^{+} - \{y_1,\dots,y_{N_s}\}.
\end{aligned}
\end{equation}
An inverse operation from time level $n$ back for a timestep $\Delta t$ is therefore upon the current state vector, giving
\begin{equation}\label{y-}
\begin{aligned}
	   \{y_1,\dots,y_{N_s}\}^{-} = \{y_1,\dots,y_{N_s}\} - \{\Delta y_1,\dots,\Delta y_{N_s}\}_j.
\end{aligned}
\end{equation}
It is to be noted that during either advance or its inverse operation, any mass fraction of species involved should be inside [0,1] and once a species' mass fraction exceeds the range (usually larger than one because the positivity-preserving QSS approximation prevents negative mass fractions), all the mass fractions should be rescaled properly according to Eq. \eqref{delta}. For the two limited states with superscripts $+$ and $-$, two limited temperature $T^{+}$ and $T^{-}$ can be derived according to the EoS in Eq. \eqref{EoS} with the help of the basic thermodynamic relation which is implicit about temperature,
\begin{equation}\label{thermodynamic}
\begin{aligned}
	& h-e = \frac{p}{\rho}, \\
	& p = p(y_1,\dots,y_{N_s}, T), \\
	& h = h(y_1,\dots,y_{N_s}, T), \\
\end{aligned}
\end{equation}
where $\rho$ and $e$ are fixed during the constant-volume adiabatic reaction and $h$ represents the specific enthalpy. If we assume the present reaction is exothermic, $T^{+}$ will be a high temperature and $T^{-}$ will be a low temperature, with the local temperature $T$ falling between the two limits, i.e. $T^{-}<T<T^{+}$, and \textit{vice versa}. $T^{+}$ will thus be naturally imagined as the $T_b$ in the original random projection method while $T^{-}$ corresponds to $T_u$. Given the two limited values of temperature, we can assemble the local random temperature by
\begin{equation}\label{random_T}
\begin{aligned}
	T^{*} = T^{-} + \theta_n ( T^{+} - T^{-} ),
\end{aligned}
\end{equation}    
where $\theta_n$ is a random real number between 0 and 1 and $T^{*}$ is the randomized local temperature with $min\{T^{-},T^{+}\}<T^{*}<max\{T^{-},T^{+}\}$ and $T^{*}\neq T$ in general. Regarding the generation of random number $\theta_n$, Bao \& Jin suggested the van der Corput's sampling scheme since it produces an equidistributed sequence on the interval [0,1], and among all known uniformly distributed sequences the deviation of van der Corput's sequence is minimal \cite{hammersley2013monte}. Besides, we have also tested the in-built random number generator in Fortran 95, trivial distinctions were detected except for the different degree of statistical noise/fluctuation. 

Provided the random temperature $T^{*}$, the single unidirectional reaction $j$ can be controlled by
\begin{equation}\label{random_projection}
\begin{aligned}
P_j^{(\Delta t)}: \quad 
\{y_1,\dots,y_{N_s}\}_j= 
\begin{cases}
\{y_1,\dots,y_{N_s}\}^{+}, & if \: T>T^{*}, \\
\{y_1,\dots,y_{N_s}\}, & otherwise,
\end{cases}
\end{aligned}
\end{equation}
which indicates the reaction can be activated only if the local temperature is sufficiently high; otherwise, the reaction is to be ceased and the reacting front stops developing for this moment. This is the mechanism of preventing a too fast detonation wave. Having considered reaction $j$ by random projection to either the advance state or the current state, the updated state vector $\{y_1,\dots,y_{N_s}\}_j$ will be taken in as the initial state, as $\{y_1,\dots,y_{N_s}\}$ in Eq. \eqref{y+}, for the next reaction $j+1$ in a new operation till the end of the multi-reaction system.

\theoremstyle{remark}
\begin{mydef}
The random process from the current state to a new state in the forward direction of time or not plays a similar role as the predictor-corrector algorithm. Since the random temperature $T^{*}$ and the beforehand predicted local temperature $T$ both lie between the two temperature limits, activation and deactivation both can happen for enough times in a long-term period of time. Thus the accumulative propagation of the discontinuity over many time steps converges to the correct position, taking into account the possibilities of both moving forward and standing still, as proved in \cite{bao2000random}. On the contrary, with traditional deterministic ODEs solvers, once the early triggering of the chemical reaction occurs 
, the reacting front will be forced to move one grid point forward. But no mechanism in such solvers is invented to halt this moving forward, thus a faster and faster shock will develop unrestrictedly to a spurious one.            
\end{mydef}

Inserting Eq. \eqref{random_projection} into the split reaction method in Eqs. \eqref{R1st} and \eqref{R2nd}, the present SRR method, denoted by $P$, is more than an ODEs solver, having the following form
\begin{equation}\label{SRR1}
\begin{aligned} 
P_{1st}^{(\Delta t)} = P_{N_r}^{(\Delta t)} \circ P_{N_r-1}^{(\Delta t)}  \circ \cdots \circ P_2^{(\Delta t)} \circ P_1^{(\Delta t)}
\end{aligned}
\end{equation} 
corresponding to the Lie's reaction-by-reaction splitting
or
\begin{equation}\label{SRR2}
\begin{aligned}
P_{2nd}^{(\Delta t)} = 
P_1^{(\frac{\Delta t}{2})} \circ P_2^{(\frac{\Delta t}{2})}  \circ \cdots \circ P_{N_r-1}^{(\frac{\Delta t}{2})} \circ P_{N_r}^{(\frac{\Delta t}{2})} \circ
P_{N_r}^{(\frac{\Delta t}{2})} \circ P_{N_r-1}^{(\frac{\Delta t}{2})}  \circ \cdots \circ P_2^{(\frac{\Delta t}{2})} \circ P_1^{(\frac{\Delta t}{2})},
\end{aligned}
\end{equation} 
corresponding to the Strang splitting. It thus transforms the state vector of species mass fractions by
\begin{equation}\label{SRR2}
\begin{aligned}
\{y_1,\dots,y_{N_s}\}^{n+1} = P^{(\Delta t)} \{y_1,\dots,y_{N_s}\}^{n}
\end{aligned}
\end{equation} 
through the entire multi-reaction system of chemical kinetics.

\theoremstyle{remark}
\begin{mydef}
Due to the randomization of integrating the reaction system in $P$, the present SRR method can overcome the disadvantage of numerical dissipation introduced by the convection term, $S_c$. So when reacting flows are of interest to solve in many applications, SRR is very likely to be suitable, especially for stiff cases in under-resolved conditions. If only an ODEs system, such as a zero-dimensional ignition problem, is under consideration, the above reaction-by-reaction ODEs solver or the original SSRI is sufficient to provide deterministic solutions with good accuracy and robustness.       
\end{mydef}

Last but not the least, in nonstiff cases when the spatial and temporal resolutions are fine to resolve the reaction area (usually at least tens of points are required in the reacting front \cite{jones2016passive} and the time interval $\Delta t$ is also very small according to the CFL condition), the present SRR method will gradually reduce to a deterministic ODEs solver if we shift the sampling interval of random temperature in Eq. \eqref{random_T} by 
\begin{equation}
\begin{aligned}
T^{**} =
\begin{cases}
T^{*} - \frac{1}{2}(T^{+} - T^{-}) 
(1-f), & if \: f<1, \\  
T^{*}, & otherwise,
\end{cases}
\end{aligned}
\end{equation} 
where
\begin{equation}
\begin{aligned}
f =  N \left| \frac{T^{+}-T^{-} }{T^{++}-T^{--}+\epsilon} \right|
\end{aligned}
\end{equation} 
with $T^{++}$ representing the temperature corresponding to a state in $N$ timesteps forward (e.g. $N=5$) and $T^{--}$ corresponding to its inverse state according to Eqs. \eqref{dy} and \eqref{y-} and $\epsilon$ is a small positive number. Thus $f$ is a dynamic measure for the resolution of the concerned reaction. When $f$ is large, e.g. $f > 1$, random projection plays an important role for the under-resolved stiff case.    
When the resolution is fine enough, $f$ is small and $T^{*}$ tends to shift downwards for up to a half bandwidth of $\left[ T^{-},T^{+}\right]$ to be lower than $T$ (linearly approximated to be $\left( T^{+}+T^{-} \right)/2$) such that activation will happen for an increasing possibility according to Eq. \eqref{random_projection}. The random reaction reduces to a deterministic process with consistency in non-stiff cases.      
However, for the original random projection method, its relying on two presumed equilibrium states (including $T_b$ and $T_u$) essentially conflicts with the finite-rate nonequilibrium kinetics when the time scale is resolved and stiffness tends to diminish.

\theoremstyle{remark}
\begin{mydef}
Due to the reduced randomness between activation and deactivation, the proposed SRR method can also cope with nonstiff problems while the original random projection method is merely suitable for under-resolved stiff cases.        
\end{mydef}
           
\section{Numerical results and discussion}
\label{section2}

In this section, we have three parts of numerical experiments: the first subsection validates the split reaction-by-reaction ODEs solver based on either analytical solutions if available or the plain QSS approximation for the zero-dimensional reaction operator, ignoring the fluid transport. The following two parts consider the coupled fluid dynamics with chemical kinetics by using simplified model kinetics and real-world finite-rate kinetics, respectively. Both 1D and 2D problems are taken into account, showing the dimensional independence of the present method.      

\subsection{Reaction-split ODEs solver for chemical kinetics}

\subsubsection{Michaelis-Menten test}
The first case concerns the Michaelis-Menten system \cite{higham2008modeling} with four species through three reactions as
\begin{equation*}
\begin{aligned}
S_1 + S_2 \xrightarrow{k_1} S_3,\\
S_3 \xrightarrow{k_2} S_1 + S_2,\\
S_3 \xrightarrow{k_3} S_2 + S_4,
\end{aligned}
\end{equation*}
where the rate constants $k_1$, $k_2$ and $k_3$ are $10^6$, $10^{−4}$ and $10^{−1}$, respectively. We can see the second reaction is the reverse counterpart of the first. The initial concentration data from \cite{wilkinson2011stochastic,higham2008modeling} are $5 \times 10^{-7}$ for $S_1$ and $2 \times 10^{-7}$ for $S_2$ with void $S_3$ and $S_4$. For this case, analytical solutions are provided for each reaction, see \ref{appendix1}, and we easily compare the convergence rates of the reaction splitting schemes of Lie and Strang, respectively. Reactions are simulated until $t=50$. In Table \ref{test1_convergence}, the $L_1$ and $L_{\infty}$ error norms of species $S_1$ and $S_4$ are detailed, showing the expected convergence rate, i.e. 1st order for Lie splitting and 2nd order for Strang splitting.           

\begin{table}
\setlength{\belowcaptionskip}{10pt}
\caption{Convergence rates for $S_1$ and $S_4$ using Lie and Strang splittings}
\centering
\label{test1_convergence}
\resizebox{\textwidth}{!}
{
\begin{tabular}{lcccccccccc}
\hline
				&				& \multicolumn{4}{c}{$S_1$}		&& \multicolumn{4}{c}{$S_4$}	\\
				\cline{3-6}\cline{8-11}
				& $\Delta t$	& $L_1$ error	&	rate	& $L_{\infty}$ error	&	rate && $L_1$ error	&	rate	& $L_{\infty}$ error	&	rate \\
\hline	
Lie	&	6.25E-03	&	3.47E-15	&	---	&	5.01E-15	&	---	&&	1.47E-12	&	---	&	2.27E-12	&	---	\\
	&	1.25E-02	&	7.30E-15	&	1.0709	&	1.05E-14	&	1.07166	&&	2.94E-12	&	0.999772	&	4.53E-12	&	0.999815	\\
	&	2.50E-02	&	1.60E-14	&	1.13228	&	2.32E-14	&	1.13699	&&	5.89E-12	&	0.999544	&	9.07E-12	&	0.999631	\\
	&	5.00E-02	&	3.76E-14	&	1.23291	&	5.51E-14	&	1.24985	&&	1.18E-11	&	0.999088	&	1.81E-11	&	0.999261	\\
	&	1.00E-01	&	9.76E-14	&	1.37647	&	1.47E-13	&	1.41746	&&	2.35E-11	&	0.998174	&	3.62E-11	&	0.99852	\\
Strang	&	6.25E-03	&	3.14E-17	&	---	&	1.00E-16	&	---	&&	5.25E-17	&	---	&	8.32E-17	&	---	\\
	&	1.25E-02	&	1.24E-16	&	1.97793	&	4.00E-16	&	1.99959	&&	2.10E-16	&	1.99745	&	3.34E-16	&	2.00663	\\
	&	2.50E-02	&	4.94E-16	&	1.99949	&	1.60E-15	&	2.0001	&&	8.39E-16	&	1.99996	&	1.34E-15	&	2.00002	\\
	&	5.00E-02	&	1.98E-15	&	2.00021	&	6.40E-15	&	1.99999	&&	3.36E-15	&	1.99999	&	5.35E-15	&	1.9999	\\
	&	1.00E-01	&	7.91E-15	&	1.99997	&	2.56E-14	&	2	&&	1.34E-14	&	1.99999	&	2.14E-14	&	2	\\
\hline
\end{tabular}
}
\end{table}

\subsubsection{Hydrogen-air ignition delay test}
\label{h2combustion}

For this case, we apply the reaction-split solver for more complicated chemical kinetics. The hydrogen ignition in air considers not only temperature-dependent reversible reactions but also third-body reactions, making the approximate solution to each reaction is practically preferred. Herein the mechanism of H$_2$-air combustion is from O'Conaire \textit{et al.} \cite{o2004comprehensive}, consisting of nine species (including the inert $\text{N}_2$) with twenty-three reversible reactions (equivalently forty-six one-way reactions), as listed in \ref{appendix2}. This mechanism has exhibited good prediction for the ignition delay time in \cite{zhukov2012verification}. All the temperature-dependent reaction rates are calculated using the Arrhenius law
\begin{equation}\label{Arrhenius}
k_r = A T^B \text{exp}(-T_{ign}/T),
\end{equation} 
where the subscript $r$ denotes $f$ for forward reactions or $b$ for backward reactions and $T$ is the temperature. The parameters $A$, $B$ and $T_{ign}$ for the forward rate of each reaction are often given in the mechanism. When parameters are not provided associatedly, the backward rate needs to be calculated from the equilibrium constant $K_{eq}$ and $k_f$ by assuming the corresponding reaction to be in chemical equilibrium, i.e. $K_{eq} = k_f/k_b$, where 
\begin{equation*}
K_{eq} = \left( \frac{1\text{atm}}{R_uT}\right)^{\sum_{i=1}^{N_s}(\nu_i^b-\nu_i^f)} \text{exp}\left( {-\sum_{i=1}^{N_s} (\frac{h_i}{R_i T}-\frac{s_i}{R_i})(\nu_i^b-\nu_i^f)} \right)
\end{equation*} 
including the species gas constant $R_i$, specific enthalpy $h_i$ and specific entropy $s_i$ (to be approximated by thermodynamical polynomials as in \cite{mcbride2002nasa}). The third-body effect is accounted for by the summation of the third-body collision efficiencies times the corresponding molar densities of species.

The ignition delay problem is a zero-dimensional homogeneous case in space since we assume a constant-volume and adiabatic environment. Initially the reactive H$_2$-air mixture is at a pressure of 1 atm, and in the molar ratio $2:1:3.76$ for $\text{H}_2:\text{O}_2:\text{N}_2$. Nitrogen is inert for the mechanism, and thus acts as a diluent. The initial temperature of the mixture is highly important for hydrogen ignition induction. All simulations end at $t=1\times 10^{-3}$s. 

\begin{figure}
  \centering
  \includegraphics[scale=0.4]{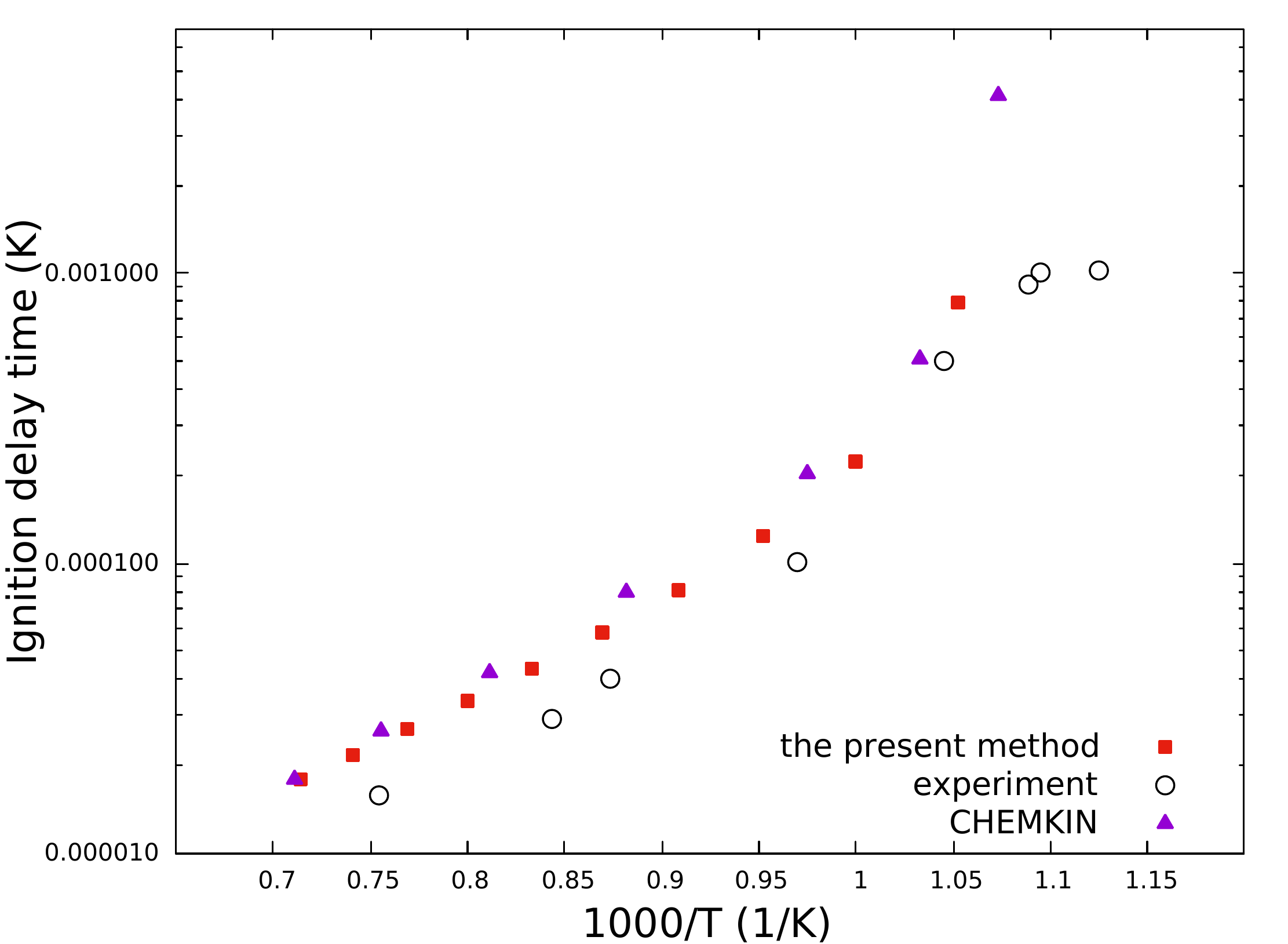}
  \caption{Ignition delay times with different initial temperatures}
  \label{test2_delay_time}
\end{figure}

\begin{figure}
  \centering
  \includegraphics[scale=0.4]{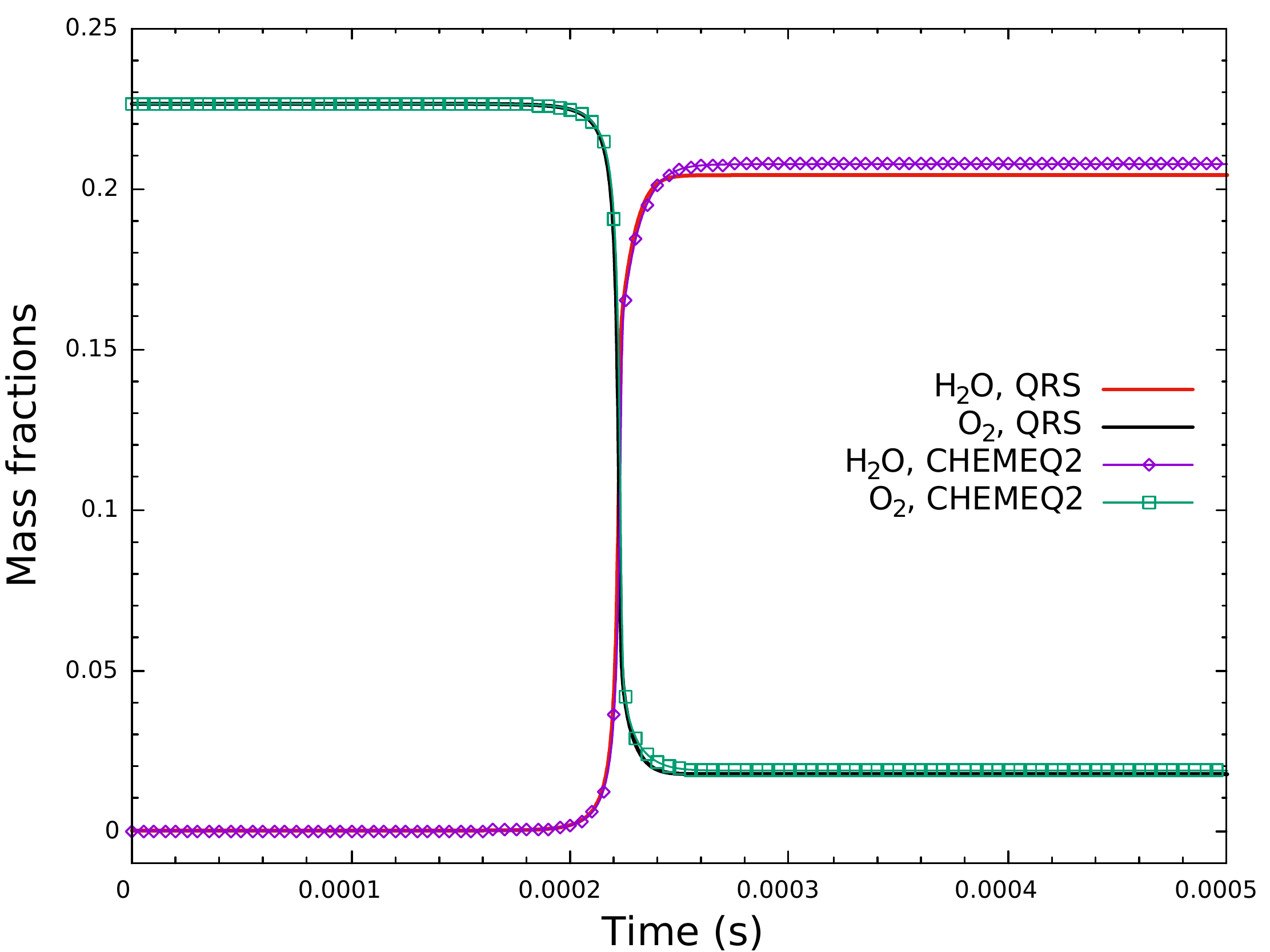}
  \caption{Time histories of mass fractions of H and $\text{H}_2\text{O}$}
  \label{mass_fraction}
\end{figure}

\begin{figure}
  \centering
  \includegraphics[scale=0.4]{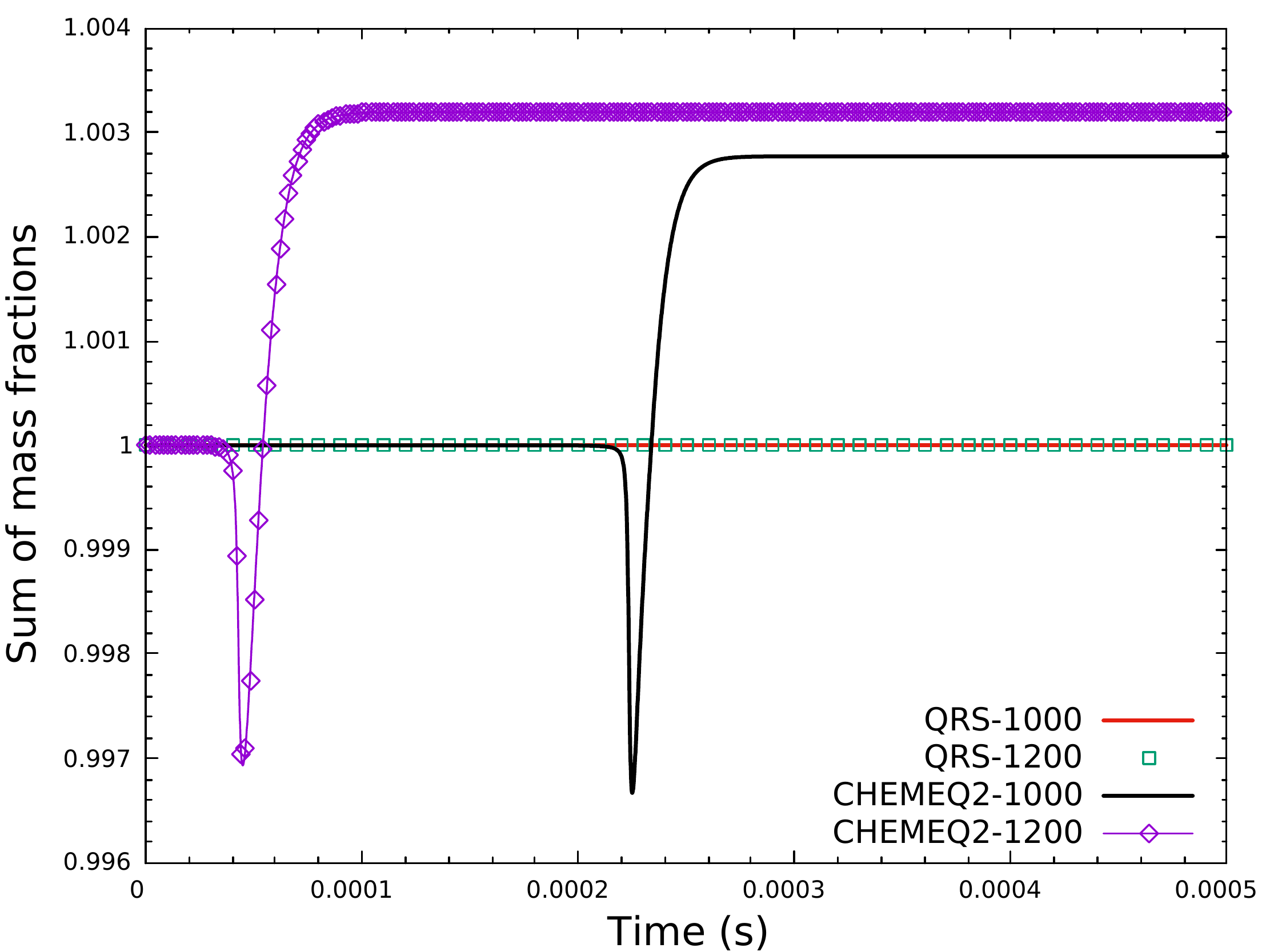}
  \caption{Time histories of the sum of mass fractions; '1000' $\sim$ $T_0=1000$ K, '1200' $\sim$ $T_0=1200$ K}
  \label{sum_mass_fraction}
\end{figure}

We firstly vary the initial temperature $T_0$ from $950$ K to $1400$ K with an equal interval of $50$ K. A fixed timestep of $1 \times 10^{-8}$s is applied, in which condition Lie splitting is sufficiently accurate. With an increasing initial temperature, the reaction rates are usually accelerated; thus the ignition delay time, corresponding to the time instance when the mixture temperature ascends most rapidly with time, generally decreases. We compare the ignition delay times predicted by the present solver with the experimental data and the CHEMKIN \cite{kee1989chemkin} results from Ref. \cite{zhukov2012verification} (see its Fig. 3) in Fig. \ref{test2_delay_time}. We can see that, in spite of varying setups, the QSS-based reaction-split method exhibits good predictions for the ignition induction of hydrogen using the present mechanism, especially in the high initial temperature range. In Fig. \ref{mass_fraction}, we compare the computed mass fractions with CHEMEQ2 at an initial temperature of $1000$ K, good agreement being reached especially at the ignition time. By setting the initial temperature at $1000$ K and $1200$ K, respectively, we consider the mass conservation resulted from the reaction-split method (abbreviated as QRS) and CHEMEQ2 in Fig. \ref{sum_mass_fraction}. It is readily to see that QRS can always preserve the mass conservation, whereas the CHEMEQ2 results show that total mass loss or gain occurs obviously around the ignition time when species concentrations vary most dramatically.

\subsection{Reactive Euler equations with simplified model kinetics}

In this part, we consider reactive Euler equations coupled with simplified model kinetics in several stiff detonation problems. In severe stiff cases, the Arrhenius form of reaction rates in Eq. \eqref{Arrhenius} also can be expressed in the Heaviside form as
\begin{equation*}\label{Heaviside}
k_r = 
\begin{cases}
A T^B, & T \geq T_{ign}, \\
0, & T < T_{ign}. \\
\end{cases}
\end{equation*} 
The EoS in Eq. \eqref{EoS} for the model problems is also simplified by
\begin{equation*}
p = (\gamma-1) \left( \rho e - q_1 \rho y_1 - q_2 \rho y_2 - \cdots - q_{N_s} \rho y_{N_s} \right)
\end{equation*} 
and $T=p/\rho$. Numerical experiments cover single reaction to multi-reaction system in 1D and 2D detonation problems. In our computation, the AUSM+ scheme \cite{liou1996sequel} is employed together with MUSCL reconstruction using a TVD Minmod limiter \cite{leveque1992numerical} in the convection step; the reaction step adopts the SRR method or merely the reaction-split solver as a deterministic method.

EXAMPLE 1 (A Chapman-Jouguet (CJ) Detonation). The first case considers the simplest reacting model, which has been studied in \cite{zhang2014equilibrium}, with only one reaction and two mutually dependent species
\begin{equation*}\label{case1_model}
\begin{aligned}
A \longrightarrow B, 
\end{aligned}
\end{equation*}
where $A$ represents the fuel being burnt by the one-way reaction and the mass fraction of the product can be directly given by $y_B=1-y_A$. 

The parameters for the reaction model and species properties are
\begin{equation*}\label{case1_para}
\begin{aligned}
\left( \gamma ,q_A, q_B \right) &= \left( 1.4, 25, 0     \right), \\
\left( A, B, T_{ign}    \right) &= \left( 16418, 0.1, 15 \right). 
\end{aligned}
\end{equation*}
The initial condition to generate the detonation wave consists of two parts in only one spatial dimension, with piecewise constants given by
\begin{equation*}\label{case1_initial}
\begin{aligned}
\left( p,T,u,y_A,y_B \right) = 
\begin{cases}
\left( 21.435,12.75134,2.899,0, 1 \right), & x<10, \\
\left( 1,1,0,1,0 \right), & x \geq 10. \\
\end{cases}
\end{aligned}
\end{equation*}
The left part gas is at the burnt equilibrium state and it is moving at a speed $u_{CJ}$ relative to the stationary unburnt gas of the right part. In fact, for any given initial state on the right, the initial CJ state on the left can be obtained in theory \cite{bao2000random,yee2013spurious,zhang2014equilibrium}. This problem is solved on the interval $\left[0,30\right]$. The left-end boundary condition is the inflow condition with fixed identical constants as the initial data on the left; the boundary condition for the right end is extrapolation from the mirror image points inside the domain. 
 
\begin{figure}
  \centering
  \includegraphics[scale=0.3]{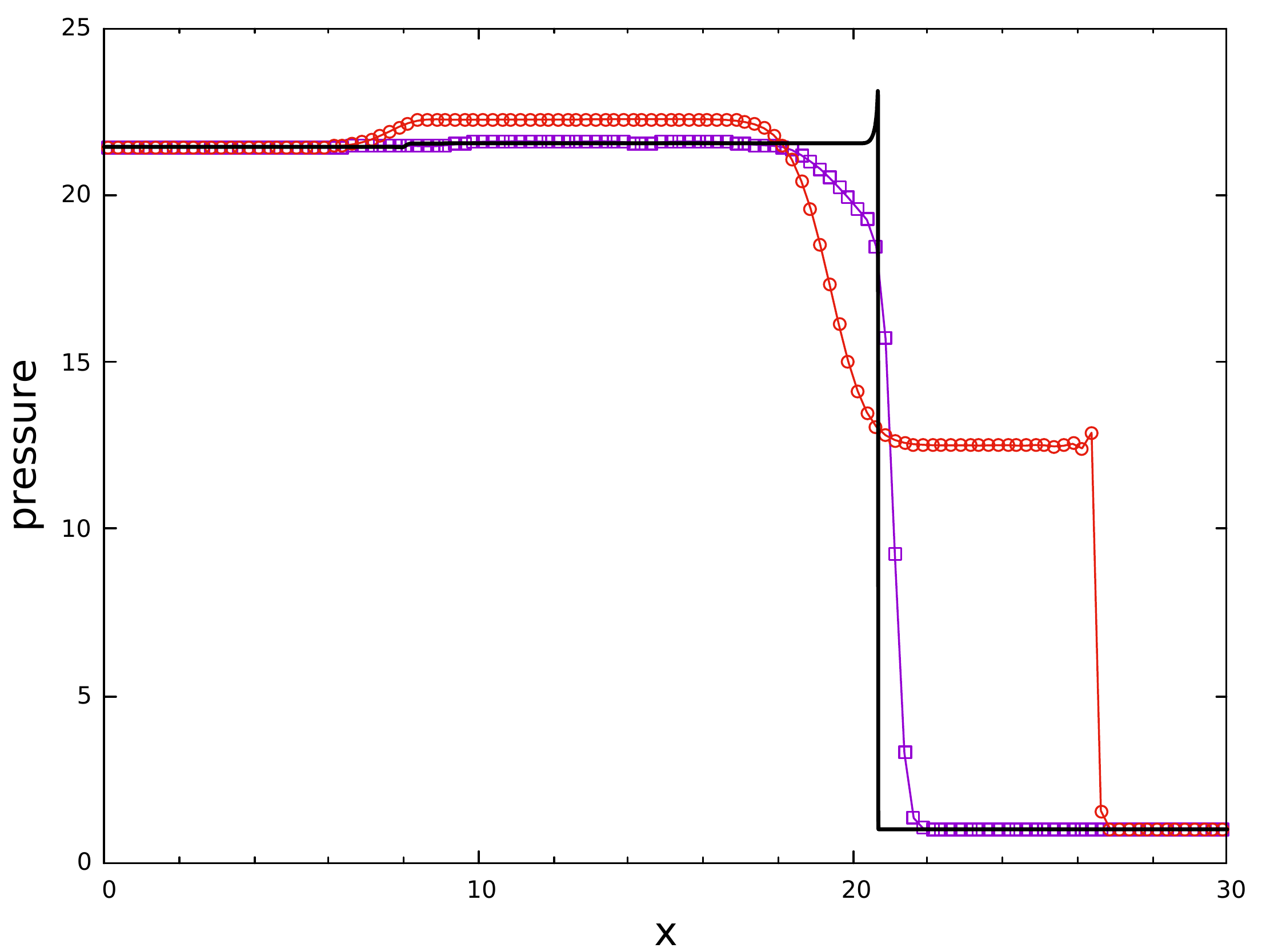} 
  \includegraphics[scale=0.3]{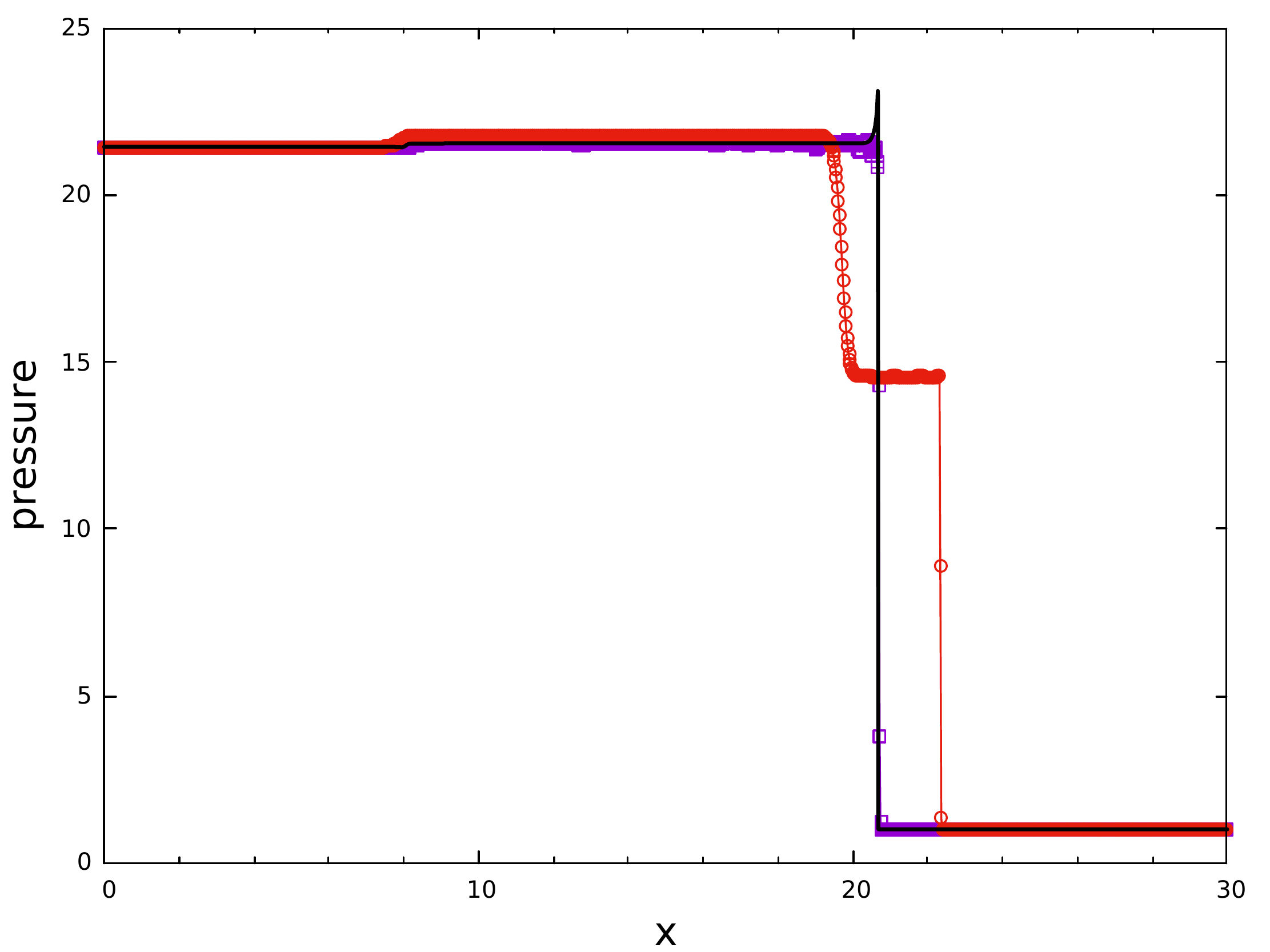} \\
  \includegraphics[scale=0.3]{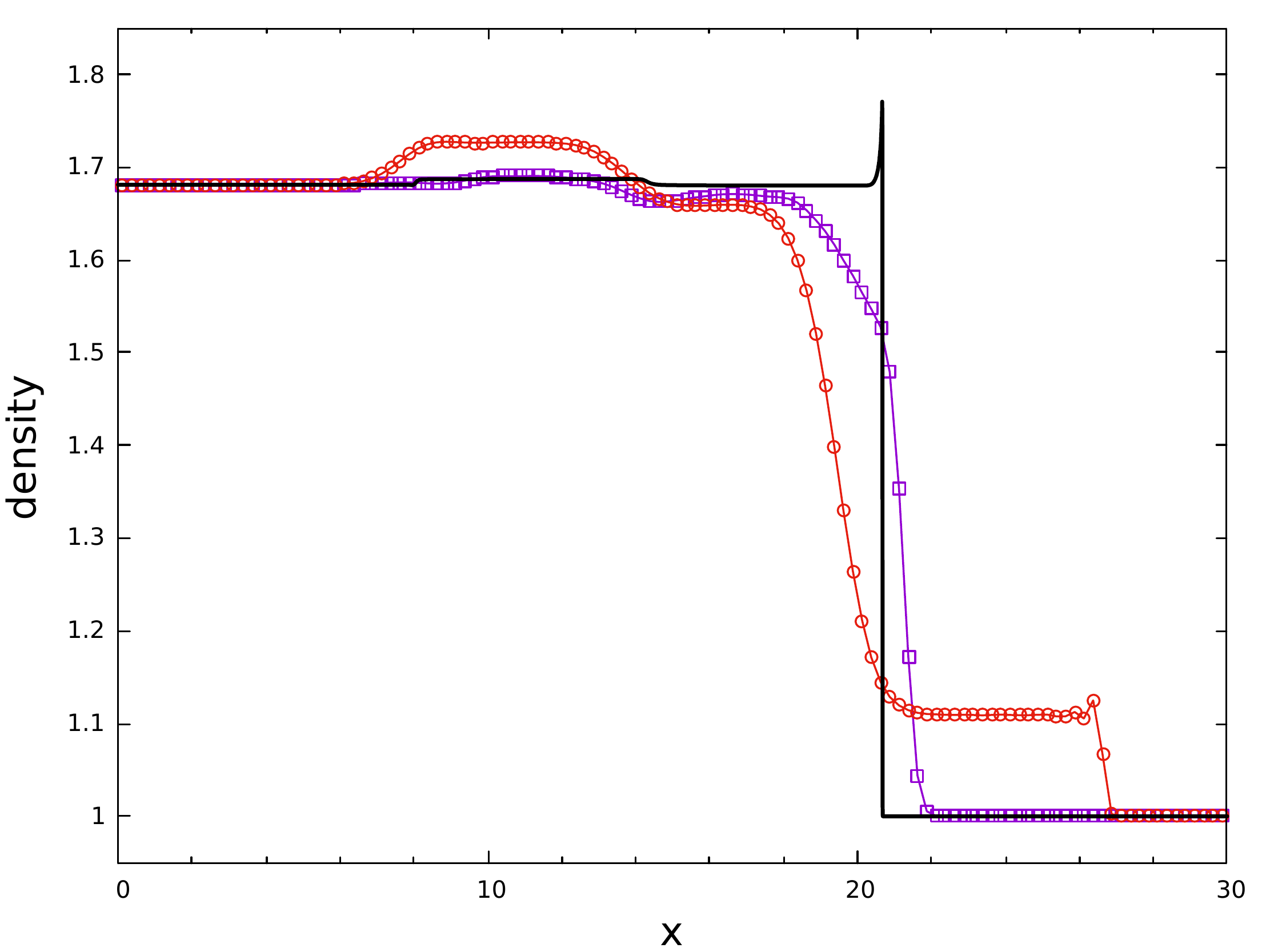} 
  \includegraphics[scale=0.3]{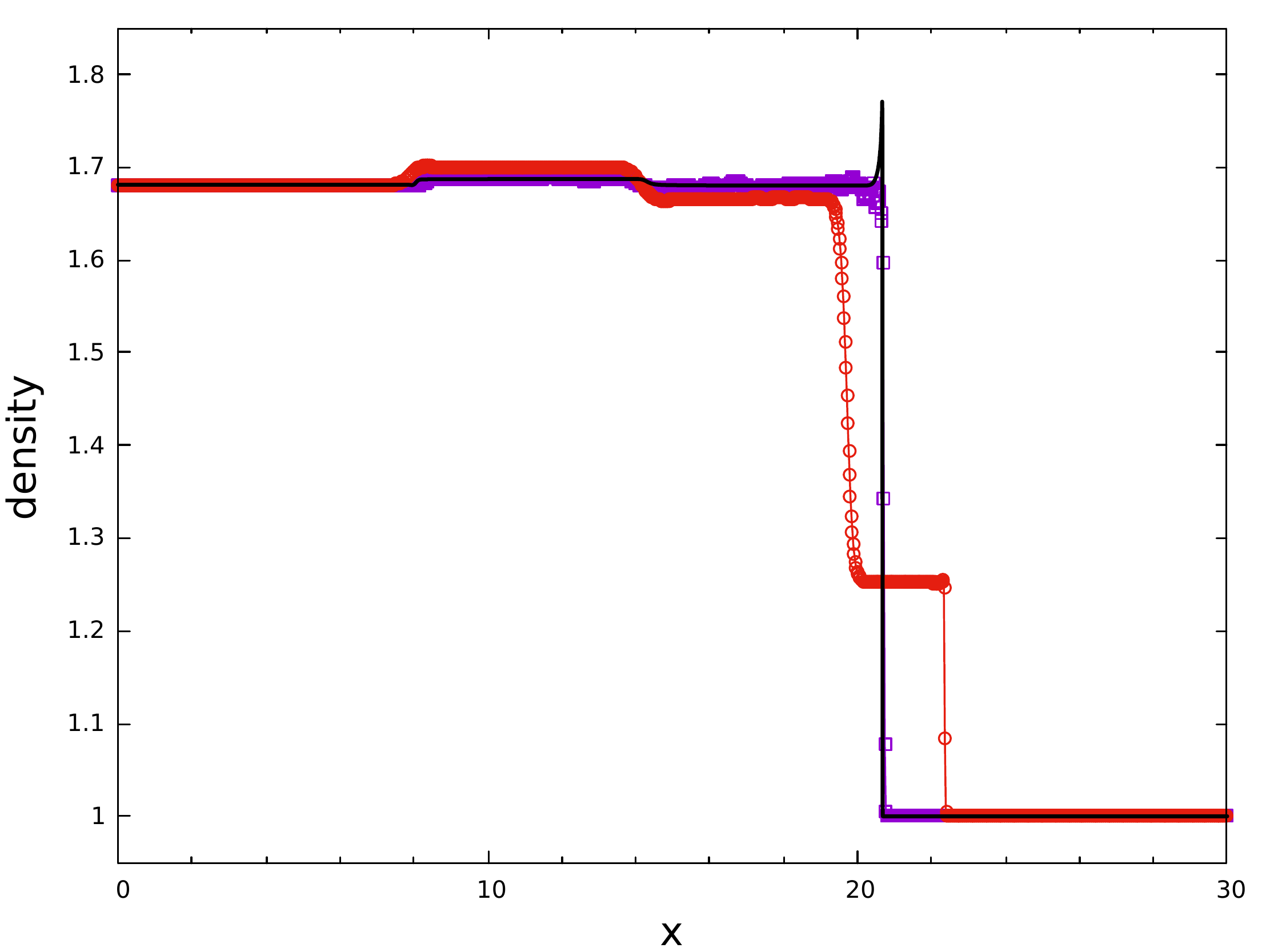} \\
  \includegraphics[scale=0.3]{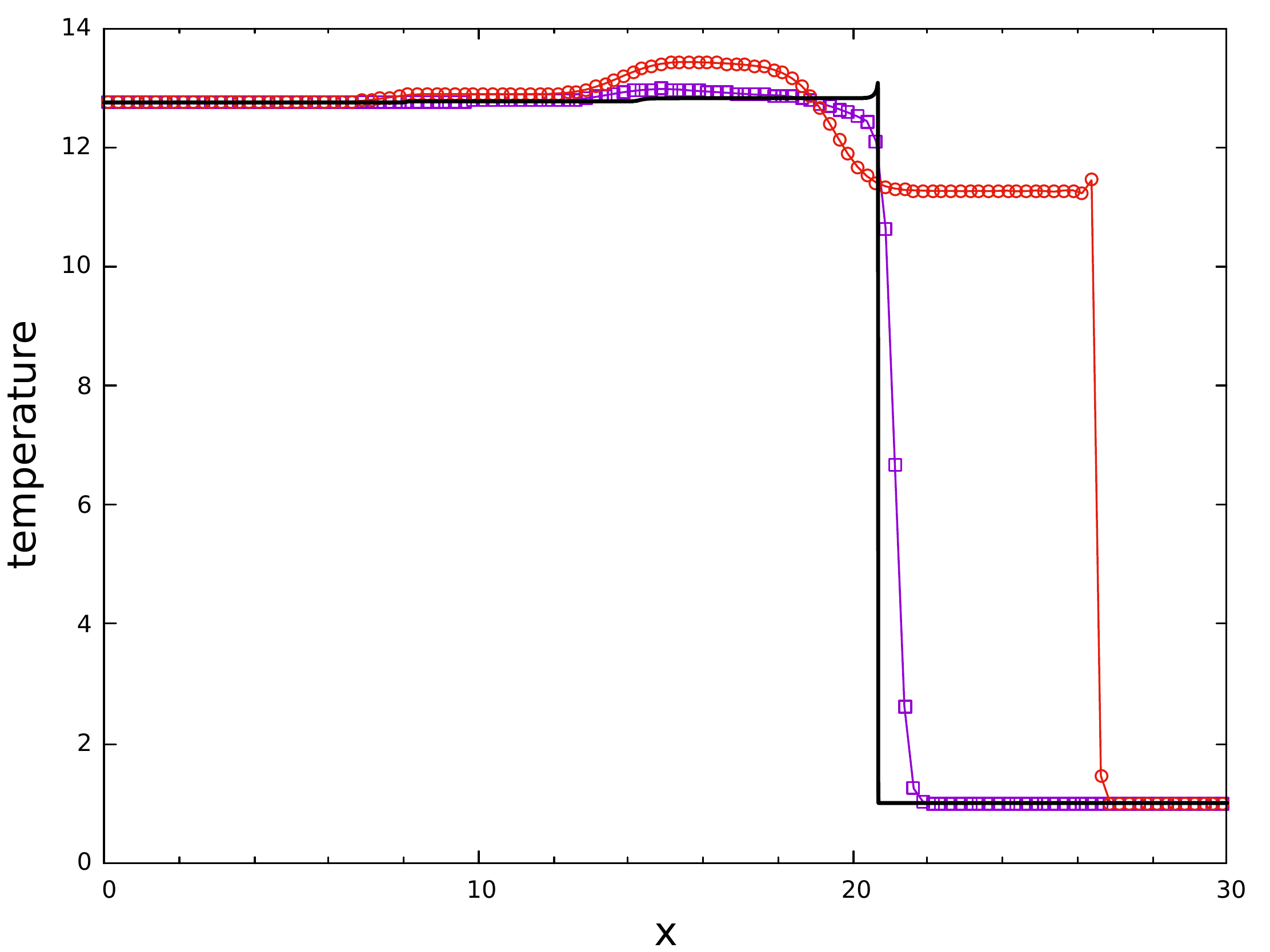} 
  \includegraphics[scale=0.3]{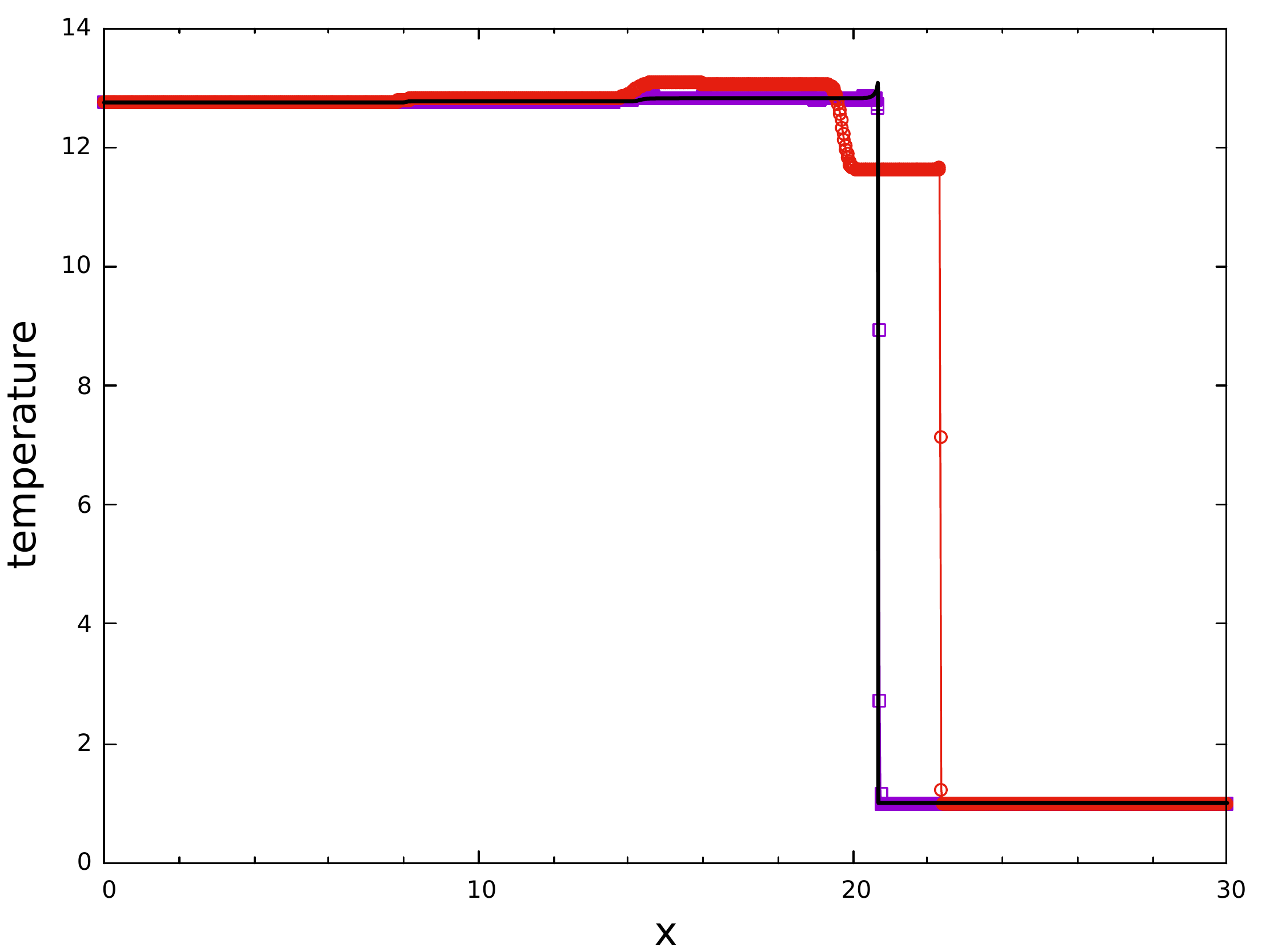} \\
  \includegraphics[scale=0.3]{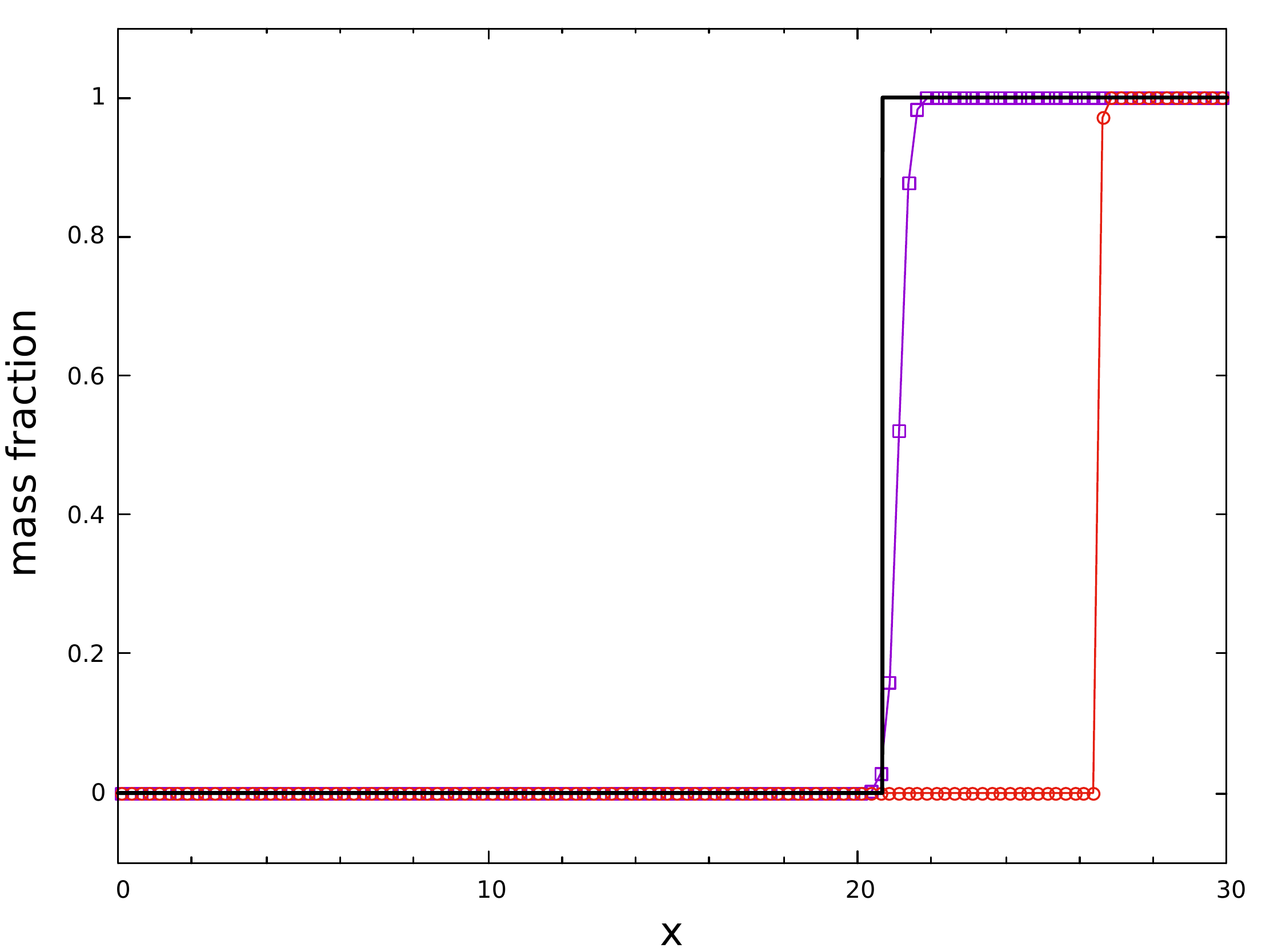} 
  \includegraphics[scale=0.3]{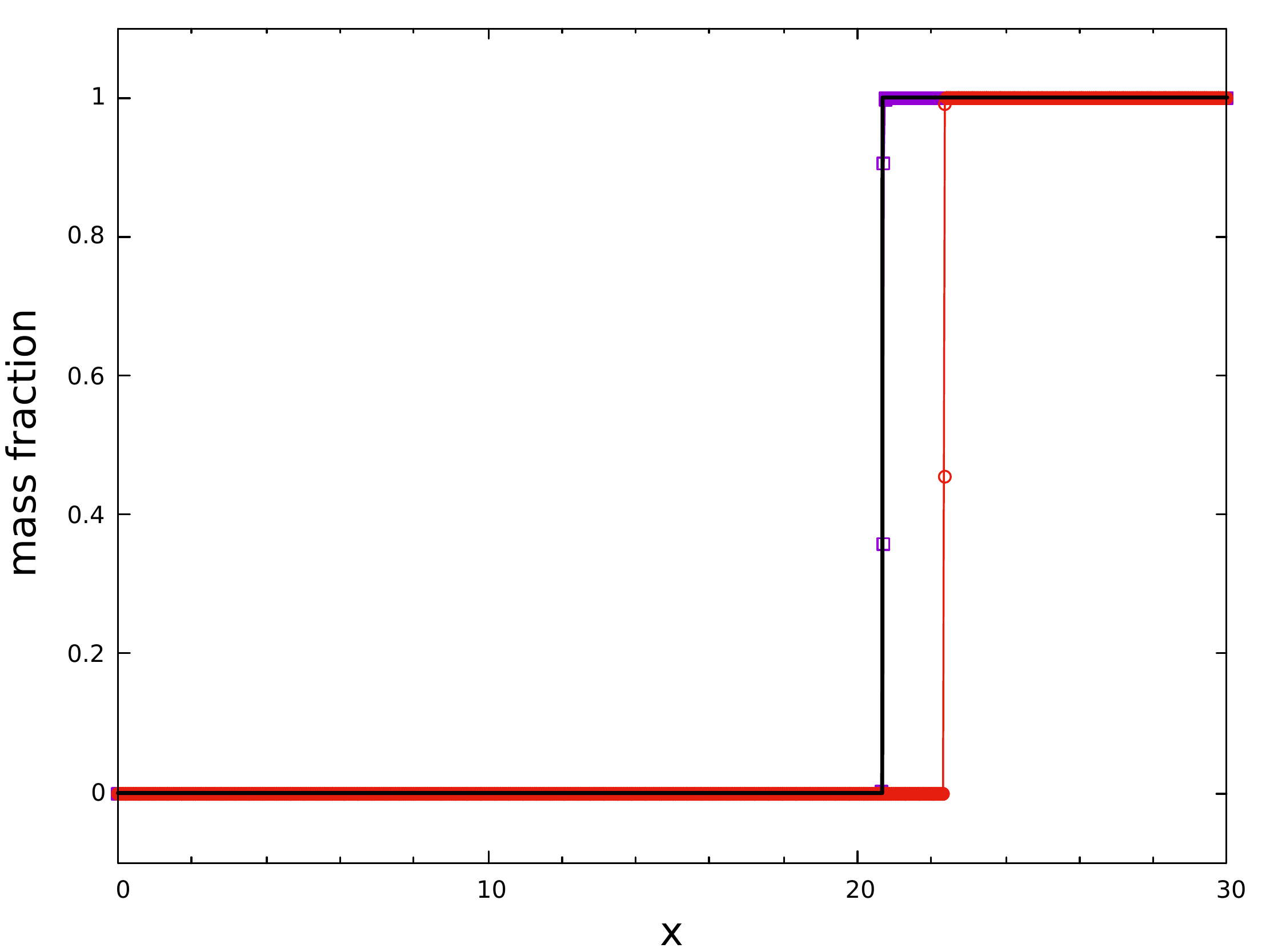} \\
  \caption{Example 1 one reaction, CJ detonation at $t=1.5$: purple square line $\sim$ SRR solution; red circle line $\sim$ deterministic solution with Arrhenius kinetics; black solid line $\sim$ reference solution; left column $\sim$ $\Delta x=0.25$, $\Delta t=0.01$; right column $\sim$ $\Delta x=0.025$, $\Delta t=0.001$.}
  \label{example1}
\end{figure}

The exact solution is simply a CJ detonation wave moving to the right and we obtain the reference 'exact' solution by the deterministic method using a resolved grid ($\Delta x = 0.0025$) and a tiny timestep of $\Delta t=0.0001$. We compare the results given by SRR and the deterministic method, respectively, using two sets of grid ($\Delta x=0.25, 0.025$) and timestep ($\Delta t = 0.01, 0.001$). Figure \ref{example1} shows the computed pressure, density, temperature and mass fraction. Clearly, the proposed random method can capture the correct propagation of the detonation wave with both coarse and fine grids, while the deterministic method produces the spurious solutions in the same under-resolved conditions, i.e. a weak detonation wave propagates faster than the theoretical detonation speed of $D_{CJ}=7.124$ in this case \cite{zhang2014equilibrium}. Besides, since a coarser grid with a larger timestep indicates the stiffness is more severe, the deterministic method produces far more nonphysical weak detonation wave compared to our SRR or the reference solution.         
Also to be noted, the location of mass fraction on the coarse grid may be few grid points away from the exact location due to random effect, but such a deviation does not grow in time \cite{bao2000random}, essentially unlike the error accumulation of the deterministic method. 

EXAMPLE 2 (A Strong Detonation). This example considers a reacting model, which has been studied in \cite{zhang2014equilibrium}, with one reaction and three species
\begin{equation*}\label{case2_model}
\begin{aligned}
2 \text{H}_2 + \text{O}_2 \longrightarrow 2 \text{H}_2\text{O}. 
\end{aligned}
\end{equation*}

The parameters for the reaction kinetics and species properties are
\begin{equation*}\label{case2_para}
\begin{aligned}
\left( \gamma ,q_{\text{H}_2}, q_{\text{O}_2}, q_{\text{H}_2\text{O}},W_{\text{H}_2}, W_{\text{O}_2}, W_{\text{H}_2\text{O}} \right) &= \left( 1.4, 300, 0, 0, 2, 32, 18     \right), \\
\left( A, B, T_{ign}    \right) &= \left( 10^6, 0, 2 \right). 
\end{aligned}
\end{equation*}
The initial condition of piecewise constants is given by
\begin{equation*}\label{case2_initial}
\begin{aligned}
\left( p,T,u,y_{\text{H}_2},y_{\text{O}_2},y_{\text{H}_2\text{O}} \right) = 
\begin{cases}
\left( 20,10,8, 0,0,1 \right), & x<2.5, \\
\left( 1,1,0,\frac{1}{9},\frac{8}{9},0 \right), & x \geq 2.5. \\
\end{cases}
\end{aligned}
\end{equation*}
The left part gas is at the burnt equilibrium state and it is moving at a speed larger than $u_{CJ}$ relative to the stationary unburnt gas of the right part so that a strong detonation wave is to occur. This problem is solved on the interval $\left[0,50\right]$.
 
\begin{figure}
  \centering
  \includegraphics[scale=0.3]{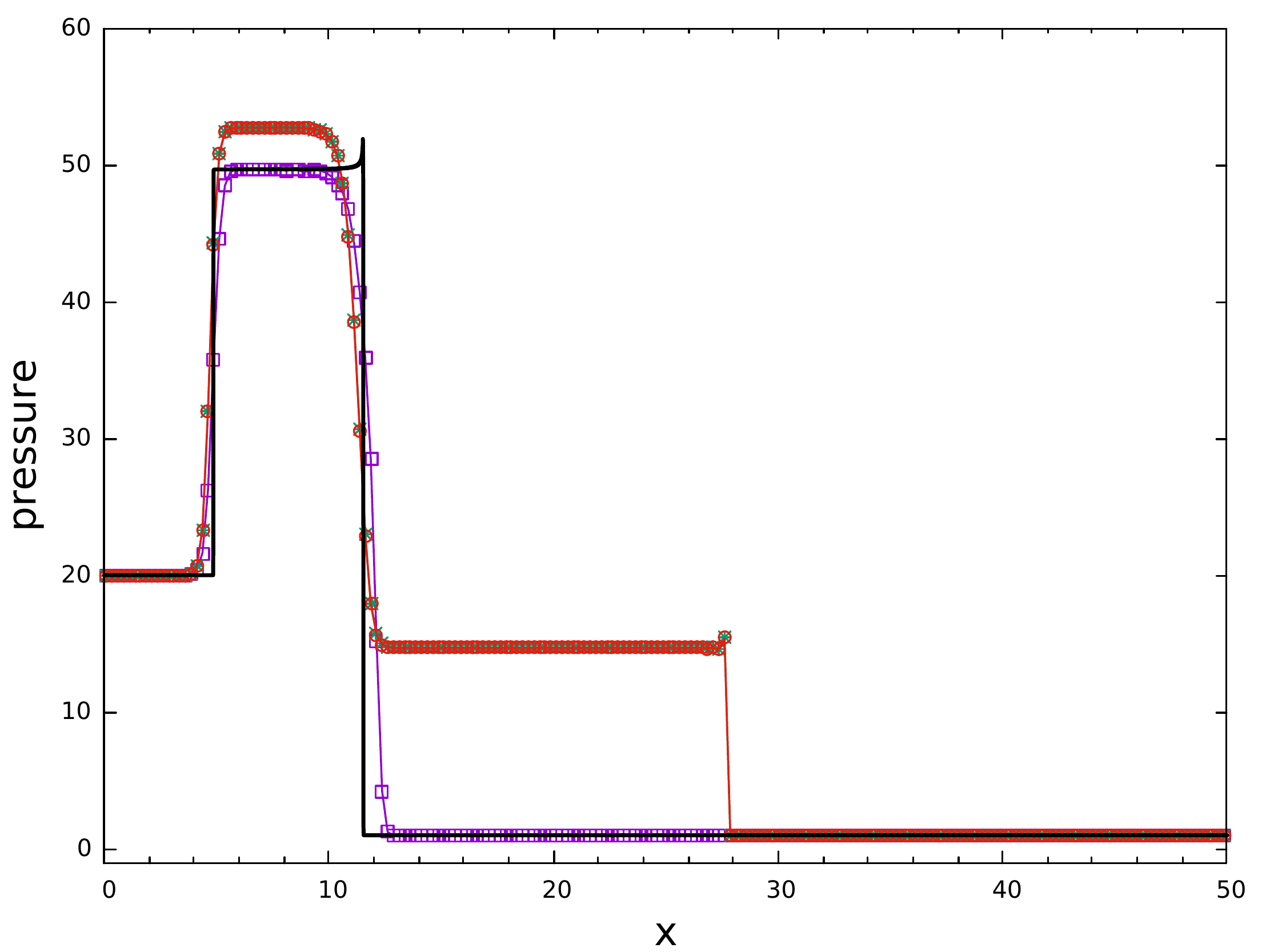} 
  \includegraphics[scale=0.3]{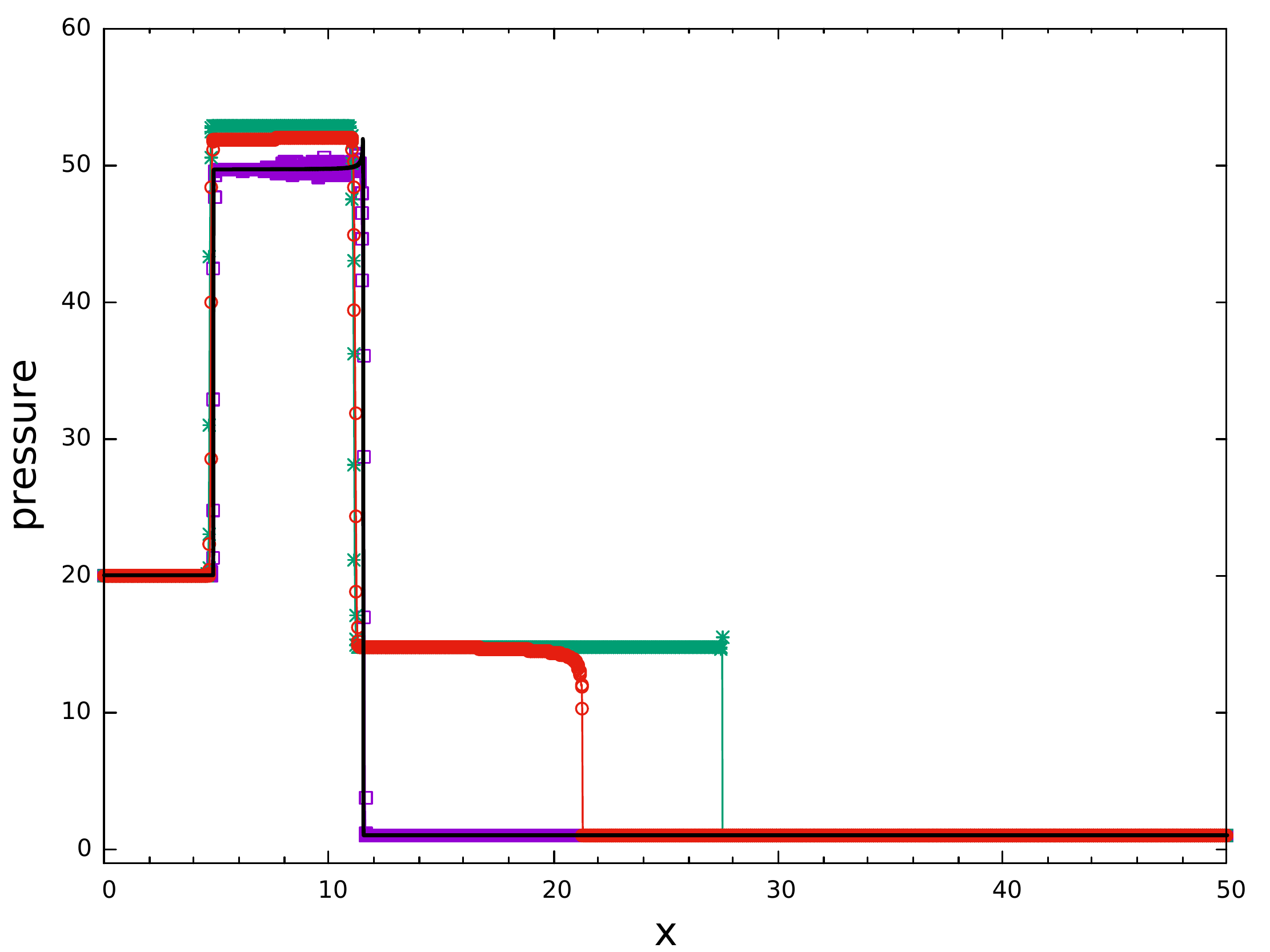} \\
  \includegraphics[scale=0.3]{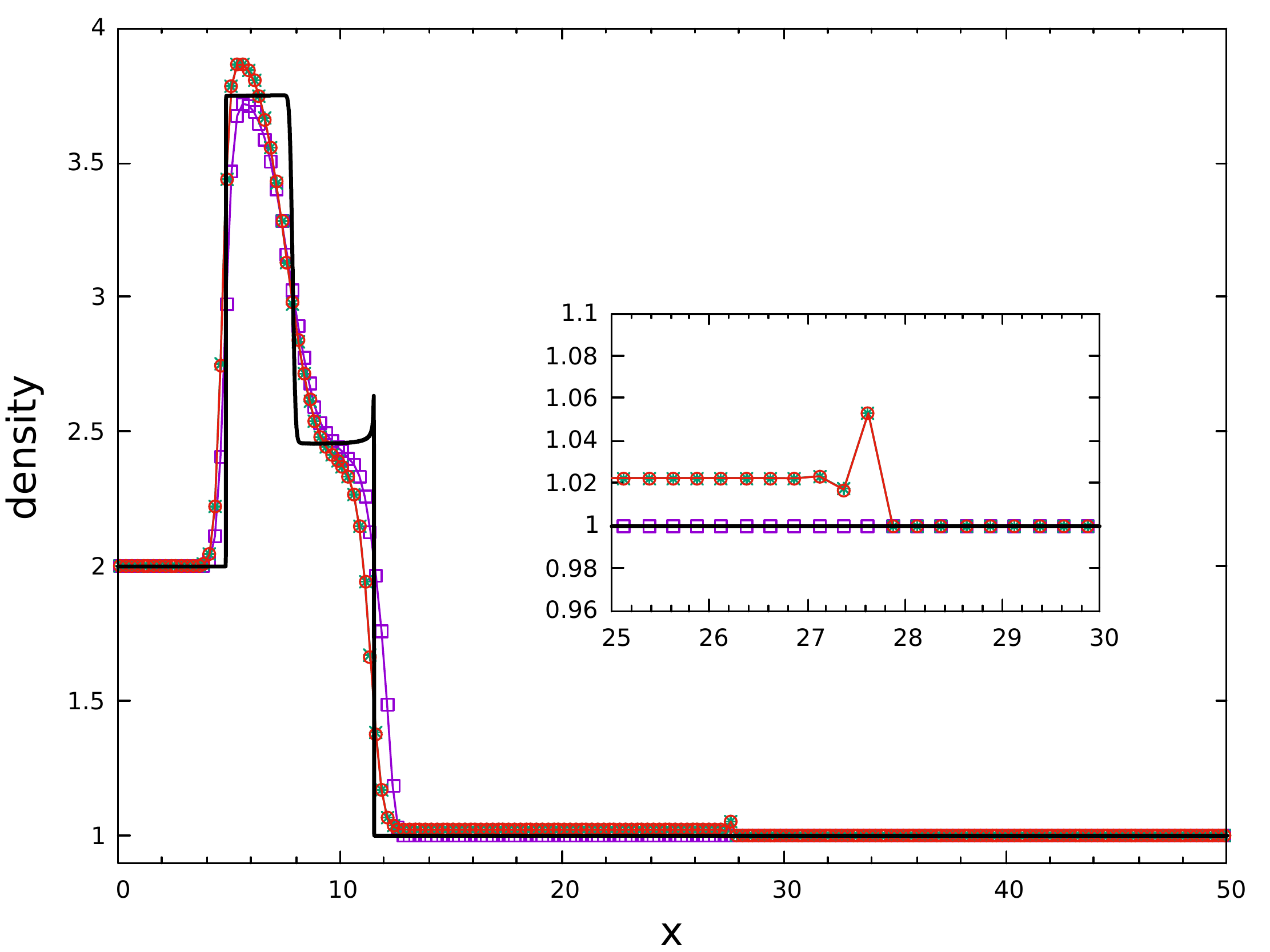} 
  \includegraphics[scale=0.3]{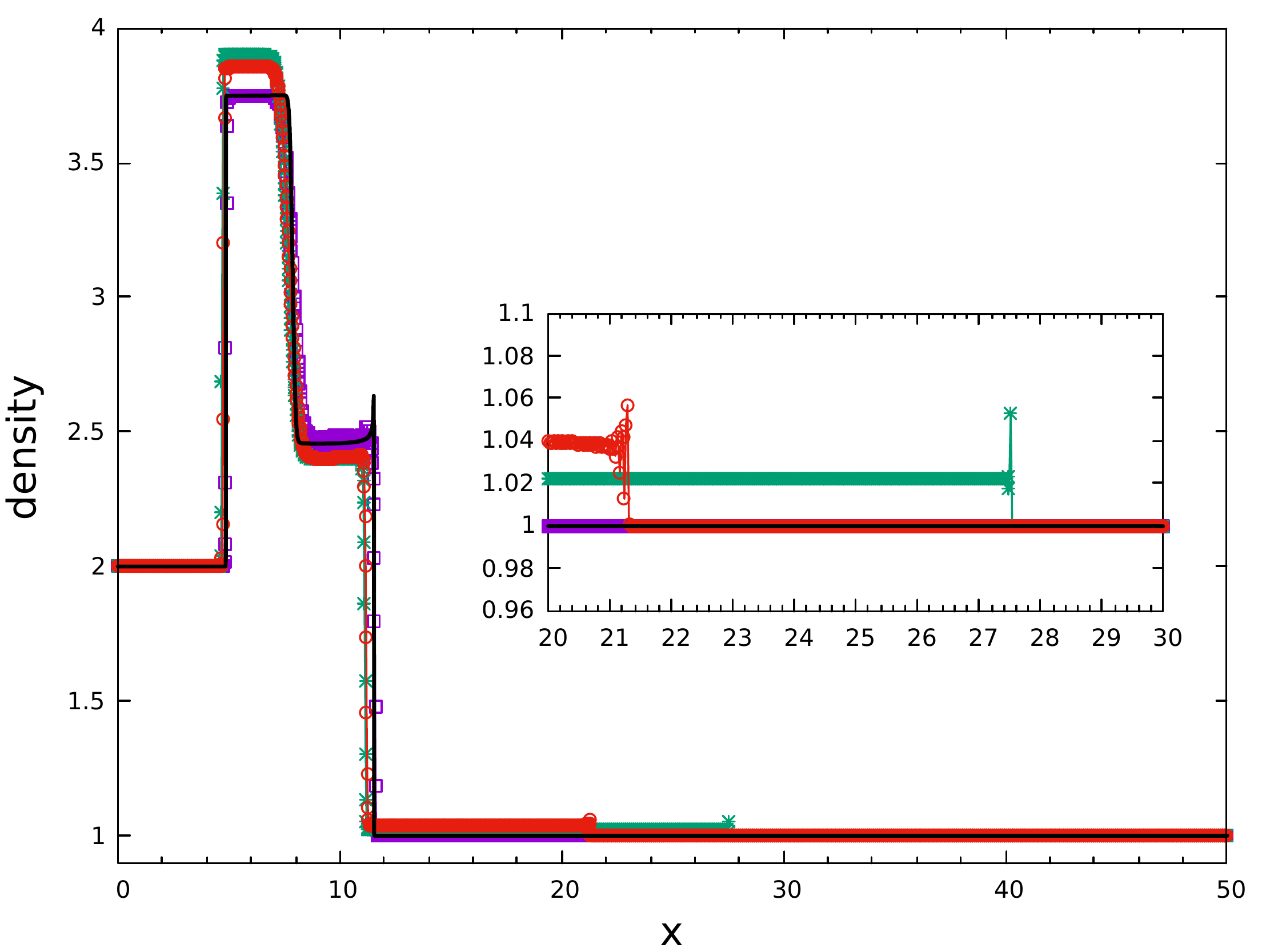} \\
  \includegraphics[scale=0.3]{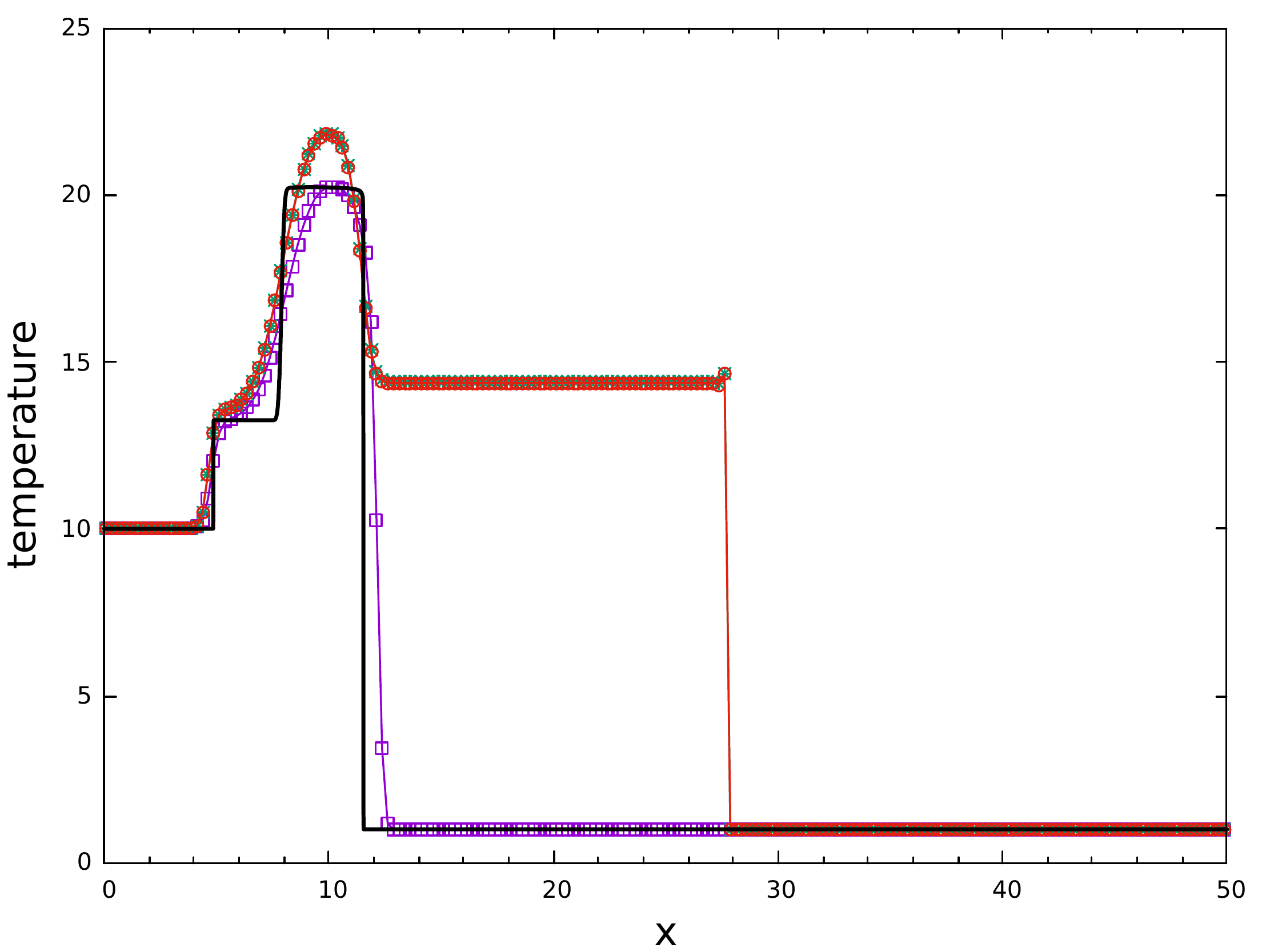} 
  \includegraphics[scale=0.3]{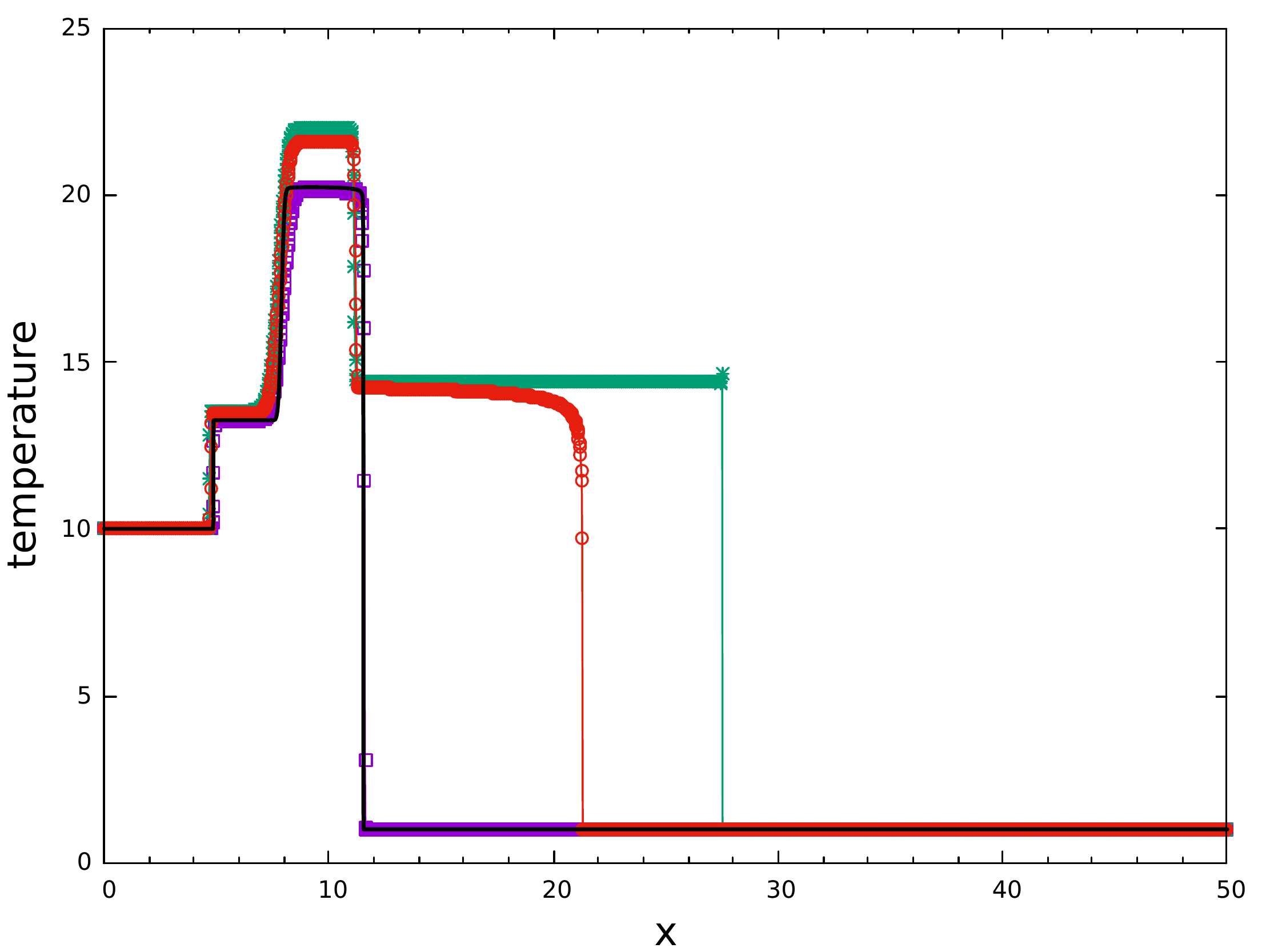} \\
  \includegraphics[scale=0.3]{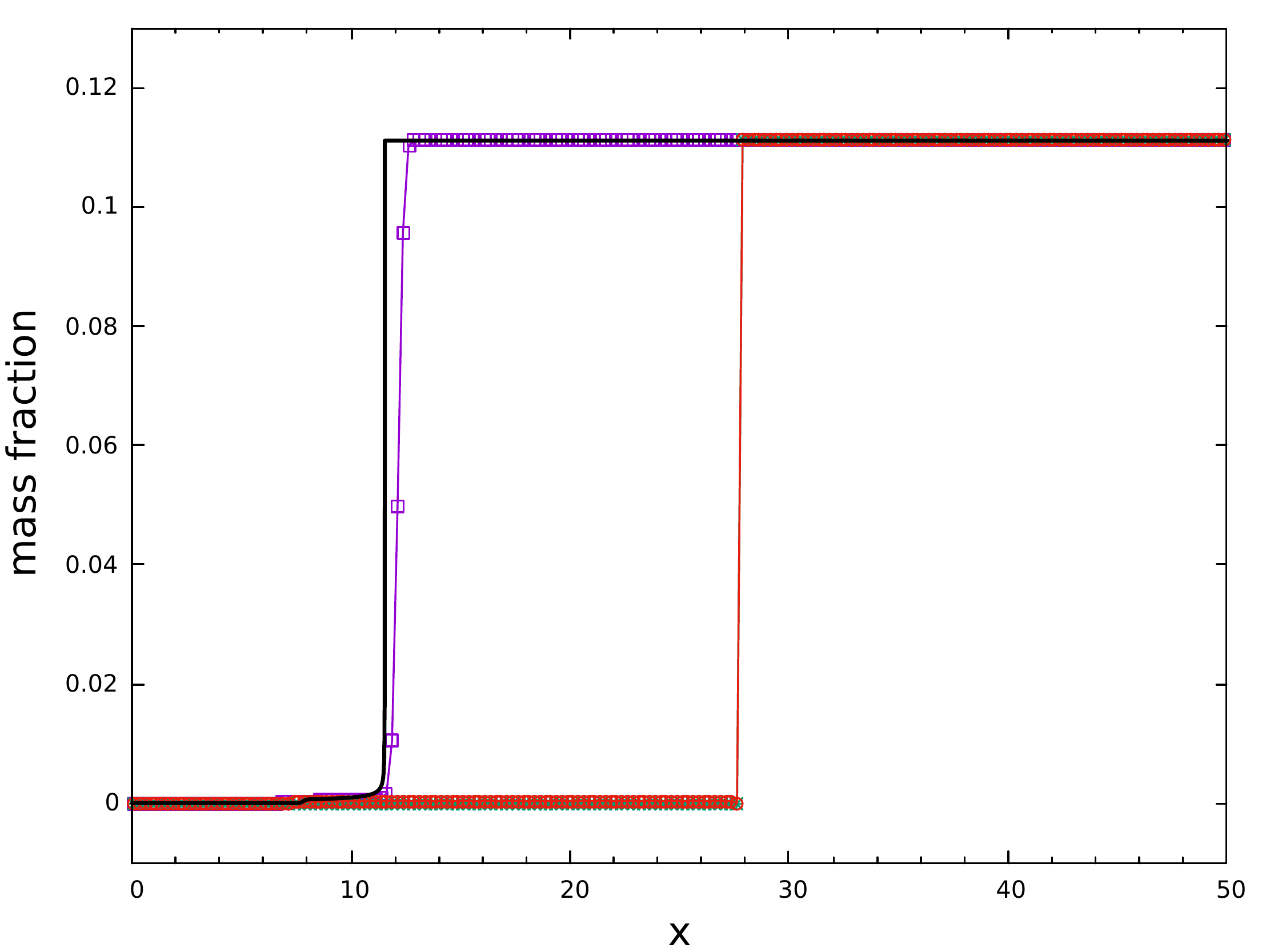} 
  \includegraphics[scale=0.3]{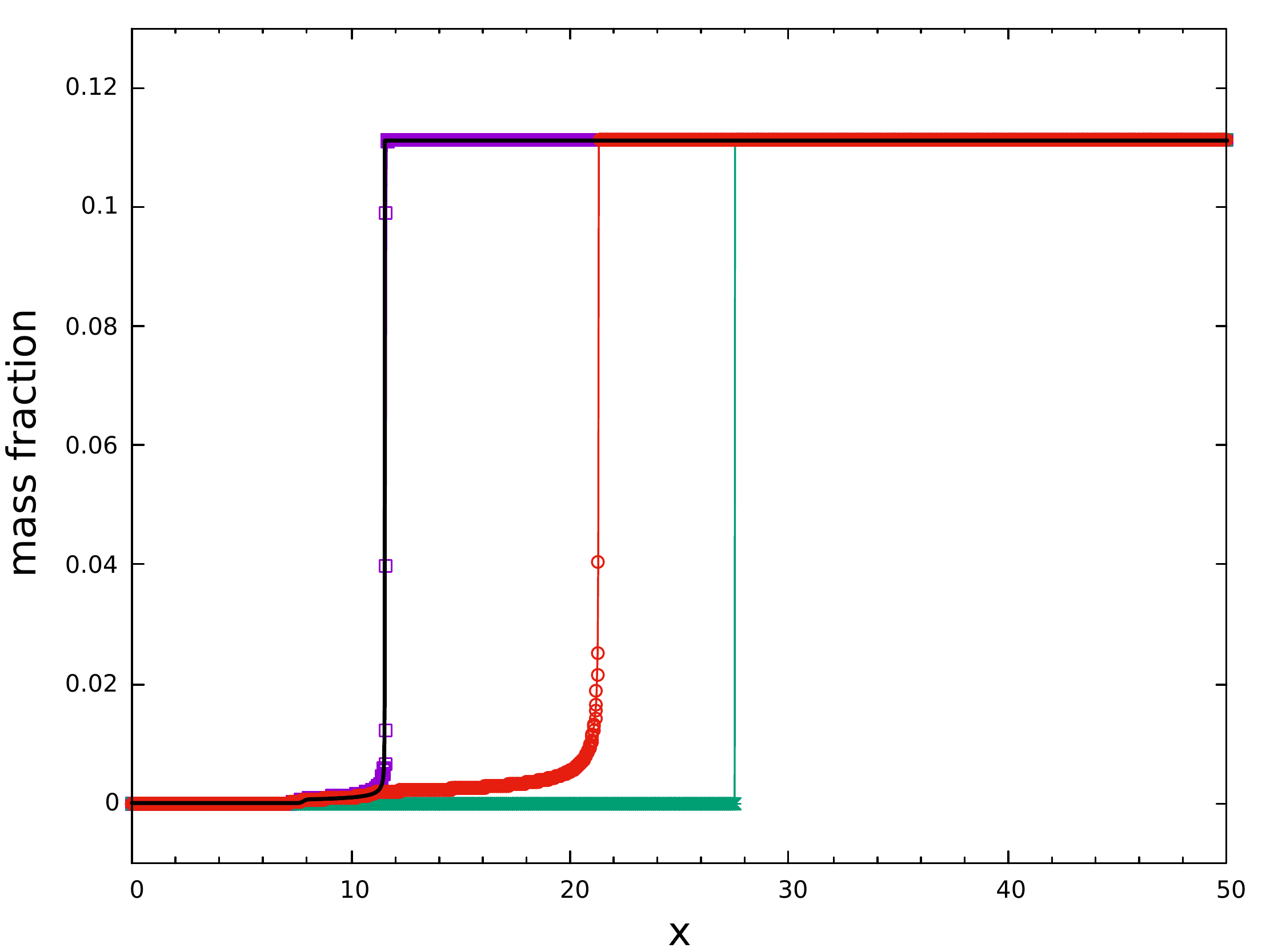} \\
  \caption{Example 2 one reaction, strong detonation at $t=1$: purple square line $\sim$ SRR solution; red circle line $\sim$ deterministic solution with Arrhenius kinetics; green cross line $\sim$ deterministic solution with Heviside kinetics; black solid line $\sim$ reference solution; left column $\sim$ $\Delta x=0.25$, $\Delta t=0.01$; right column $\sim$ $\Delta x=0.025$, $\Delta t=0.001$.}
  \label{example2}
\end{figure}

The exact solution consists of a detonation wave, followed by a contact discontinuity and a shock, all moving to the right. Similarly, we obtain the reference solution by the deterministic method using a resolved grid and a tiny timestep, and then compare the results by SRR and the deterministic method using a very coarse grid and another finer grid with proper timesteps, as explained in Fig. \ref{example2}. Note that in the deterministic method, we adopt both the Arrhenius model and Heaviside model for the chemical kinetics. It is readily to see the proposed SRR method can capture all discontinuities effectively, while the deterministic method produces the spurious solutions in the same under-resolved conditions. In particular, using the Heaviside model, the deterministic method produces more severely incorrect solution due to its greater stiffness compared to the Arrhenius model (see the right column of Fig. \ref{example2}). 

EXAMPLE 3 (A Strong Detonation). This case considers a multi-step reaction mechanism with two one-way reactions and five species
\begin{equation*}\label{case3_model}
\begin{aligned}
&1) \qquad \text{H}_2 + \text{O}_2 \longrightarrow 2 \text{OH}, \\
&2) \qquad 2 \text{OH} + \text{H}_2 \longrightarrow 2 \text{H}_2\text{O}, 
\end{aligned}
\end{equation*} with $\text{N}_2$ as a dilute catalyst. Similar examples have been studied in \cite{bao2002random}.  

The parameters for the reaction model and species properties are
\begin{equation*}\label{case3_para}
\begin{aligned}
\left( \gamma ,q_{\text{H}_2}, q_{\text{O}_2}, q_{\text{OH}},  q_{\text{H}_2\text{O}}, q_{\text{N}_2}\right) &= \left( 1.4, 0, 0, -20, -100, 0  \right), \\
\left( W_{\text{H}_2}, W_{\text{O}_2}, W_{\text{OH}},  W_{\text{H}_2\text{O}}, W_{\text{N}_2}\right) &= \left( 2, 32, 17, 18, 28  \right), \\
\left( A^1, B^1, T_{ign}^1 \right) &= \left( 10^5, 0, 2 \right), \\
\left( A^2, B^2, T_{ign}^2 \right) &= \left( 2\times10^4, 0, 10 \right). \\
\end{aligned}
\end{equation*}
The initial condition of piecewise constants is given by
\begin{equation*}\label{case3_initial}
\begin{aligned}
\left( p,T,u,y_{\text{H}_2}, y_{\text{O}_2}, y_{\text{OH}},  y_{\text{H}_2\text{O}}, y_{\text{N}_2}\right) = 
\begin{cases}
\left( 40,20,10, 0, 0, 0.17, 0.63, 0.2 \right), & x<2.5, \\
\left( 1,1,0,0.08,0.72,0,0,0.2 \right), & x \geq 2.5. \\
\end{cases}
\end{aligned}
\end{equation*}
The left part gas is at the burnt equilibrium state and it is moving at a speed larger than $u_{CJ}$ relative to the stationary unburnt gas of the right part so that a strong detonation wave is to occur. This problem is solved on the interval $\left[0,50\right]$.
   
\begin{figure}
  \centering
  \includegraphics[scale=0.3]{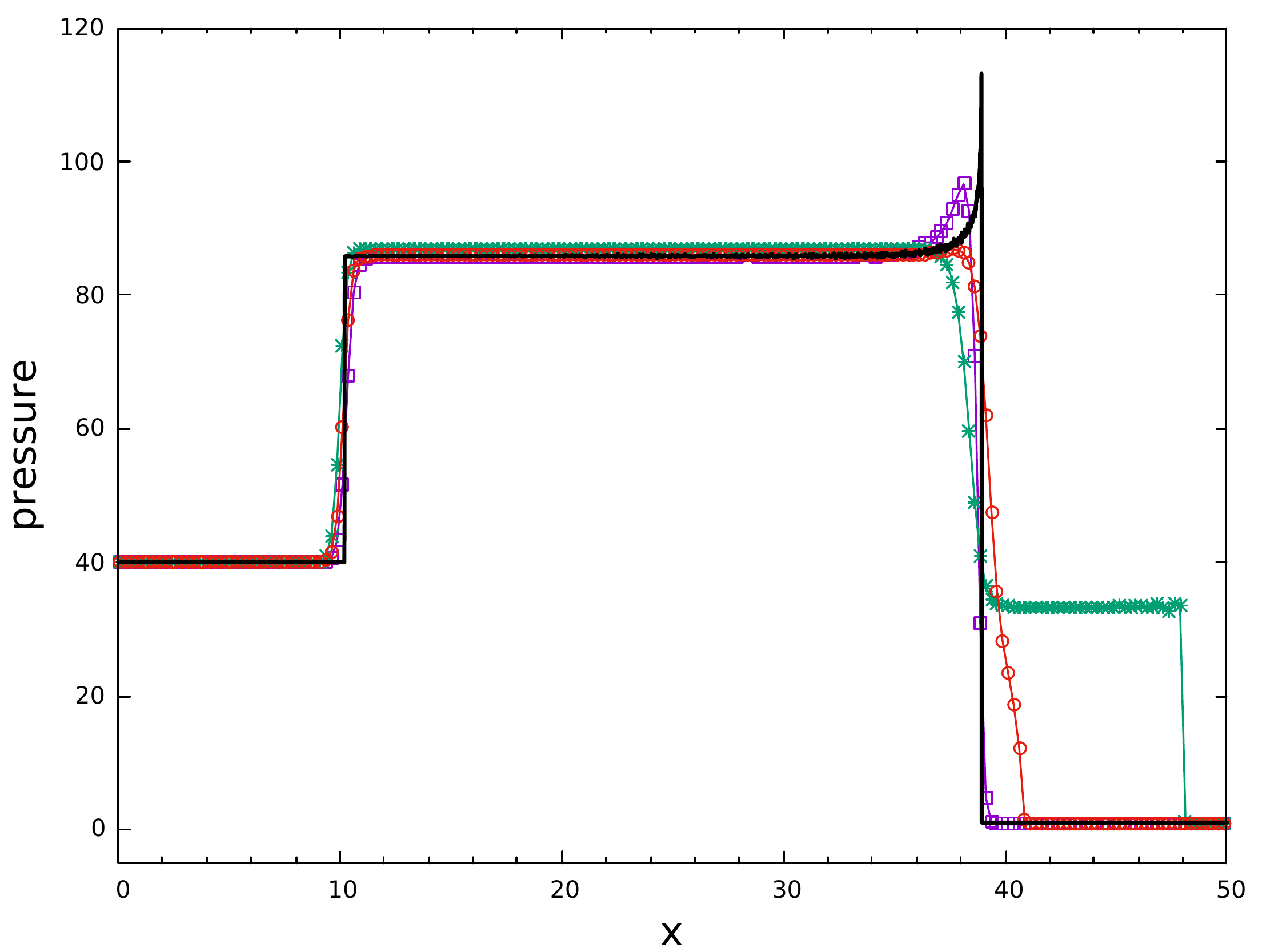} 
  \includegraphics[scale=0.3]{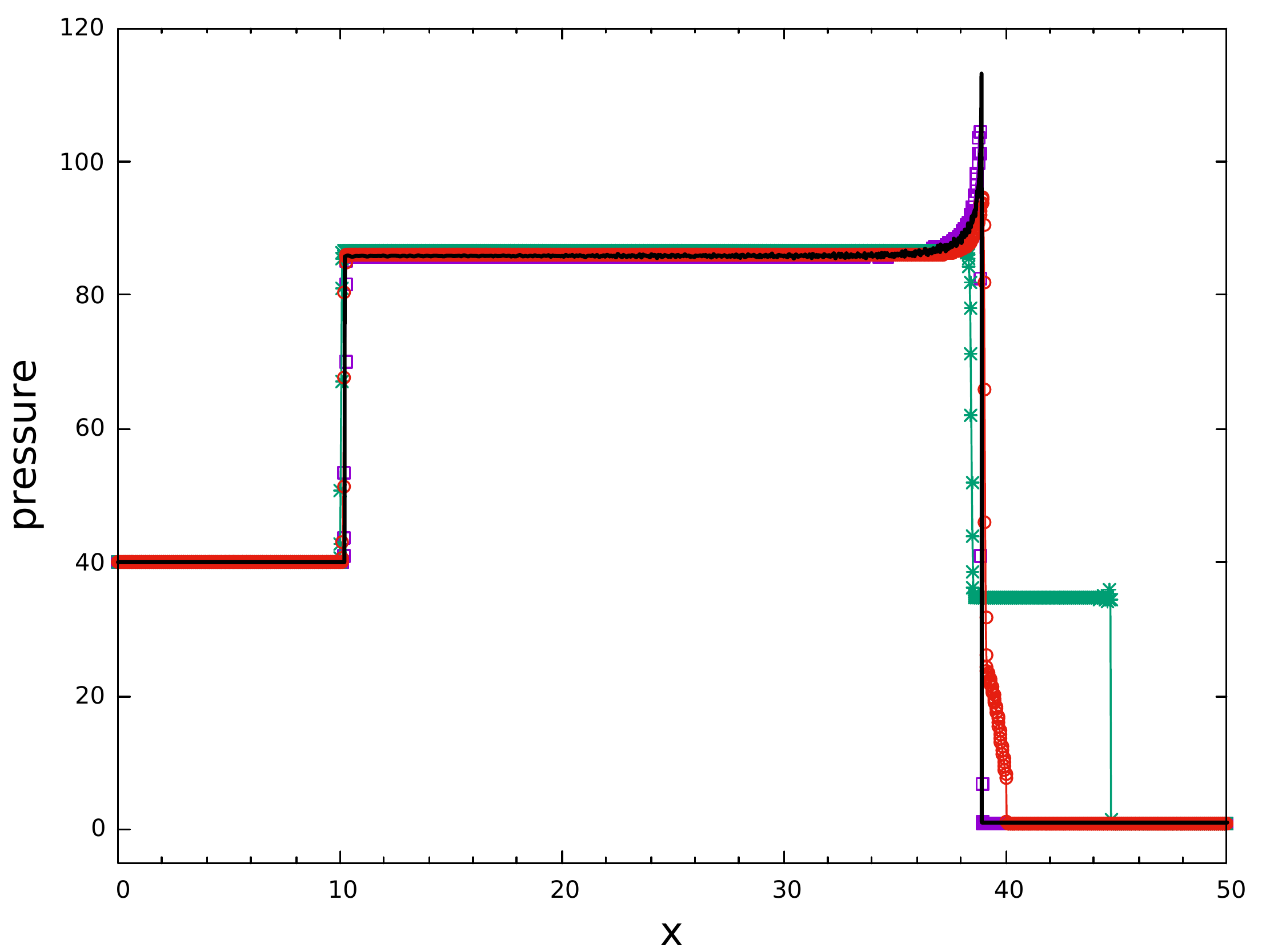} \\
  \includegraphics[scale=0.3]{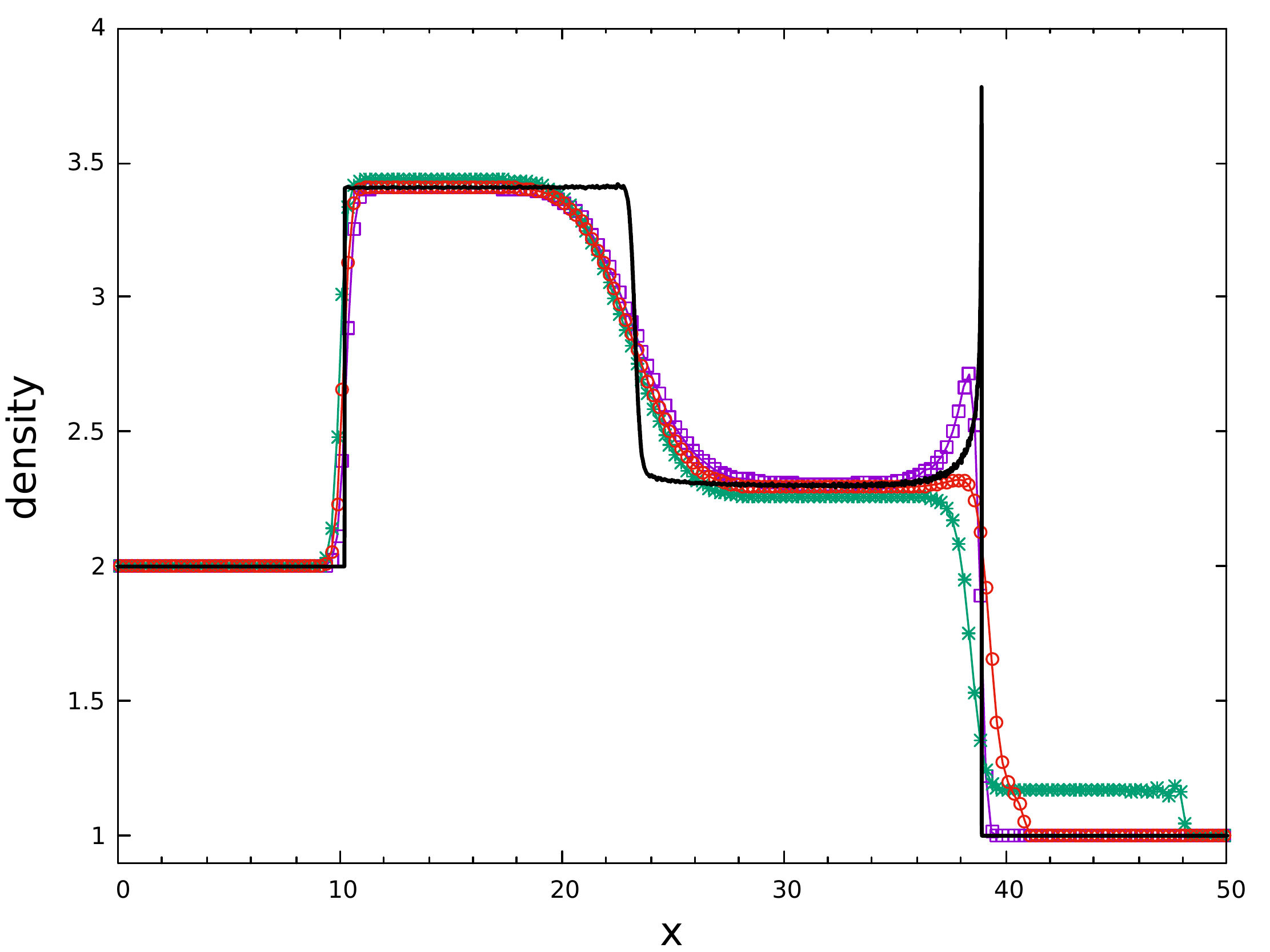} 
  \includegraphics[scale=0.3]{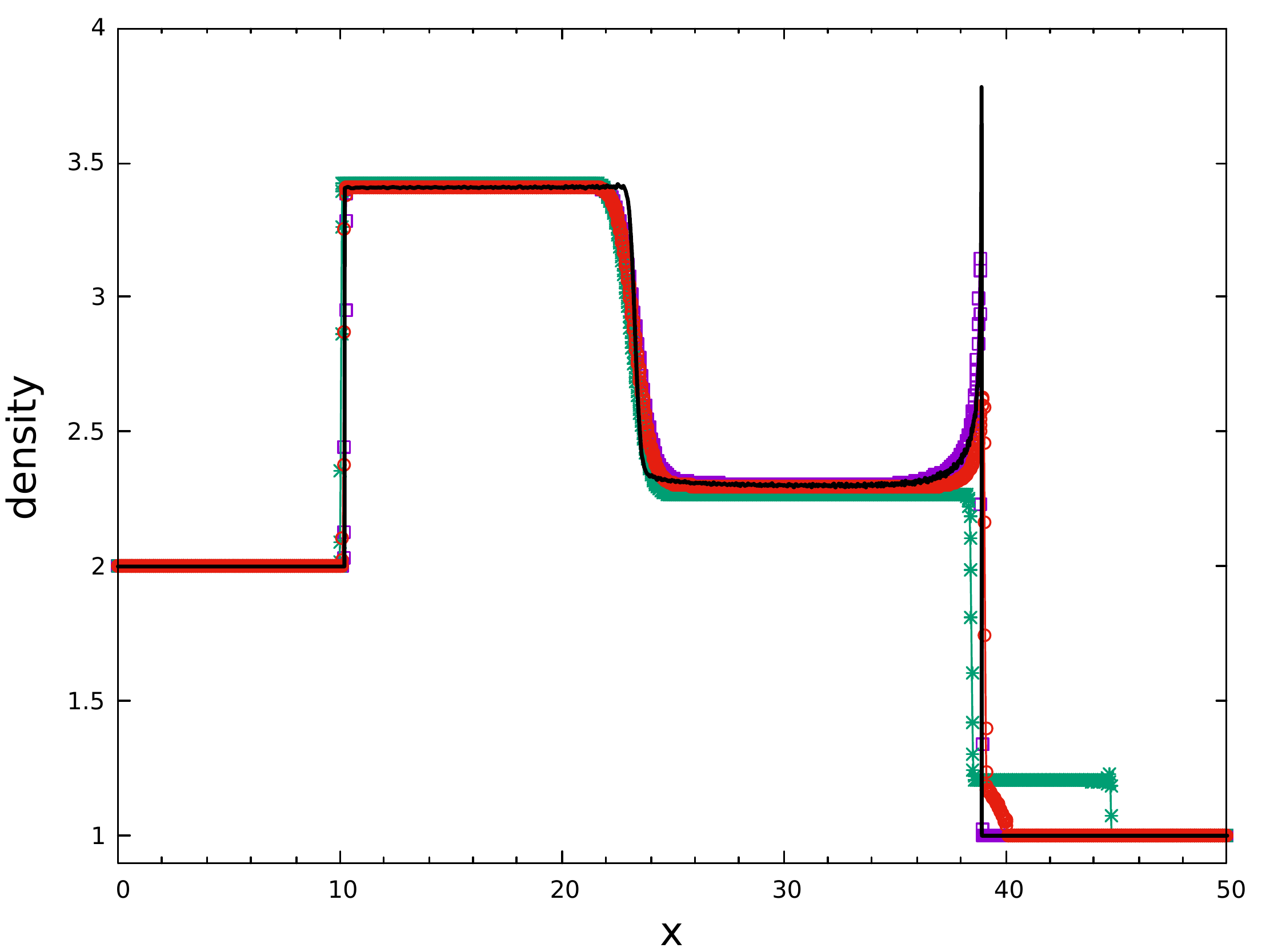} \\
  \includegraphics[scale=0.3]{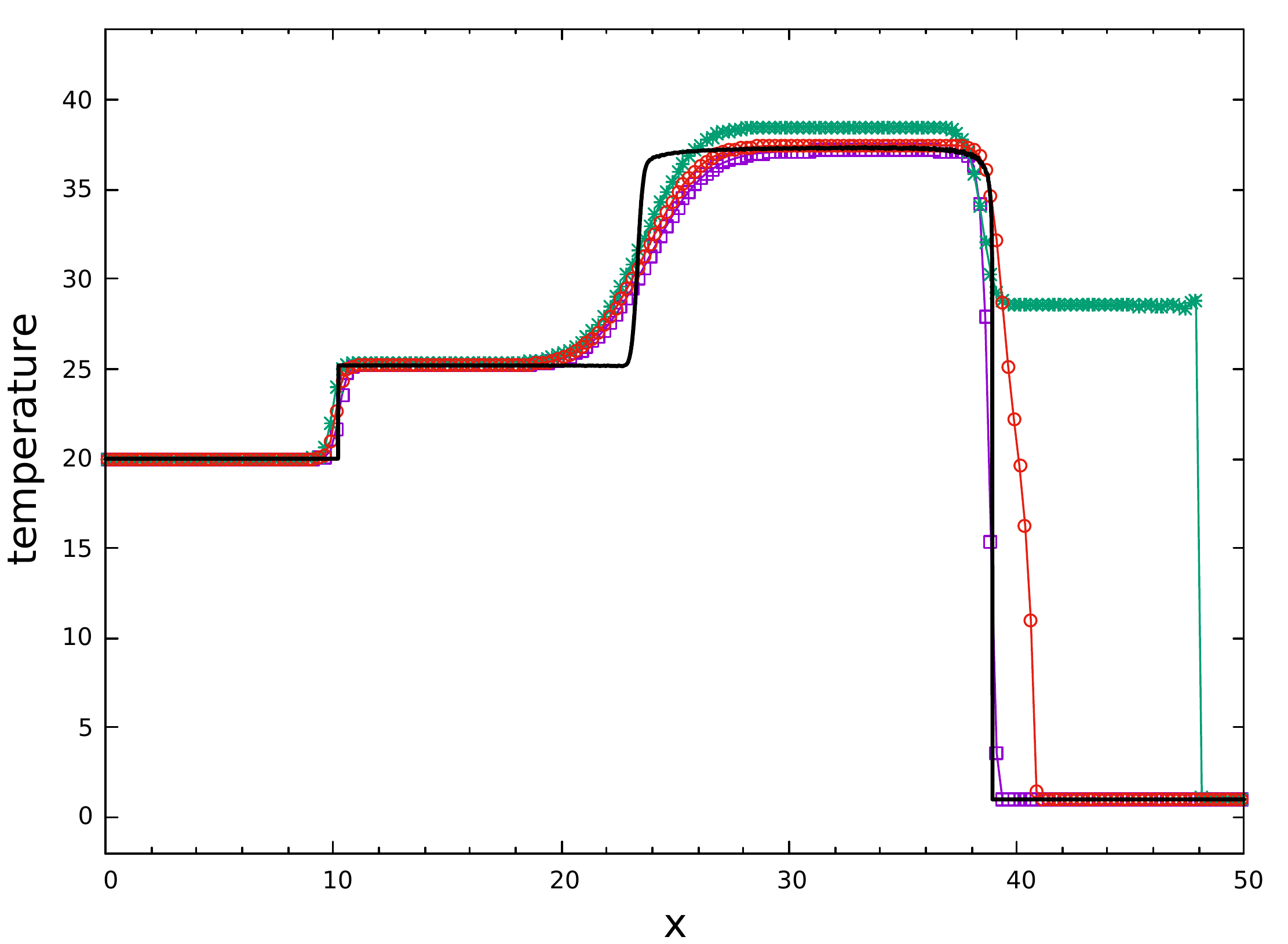} 
  \includegraphics[scale=0.3]{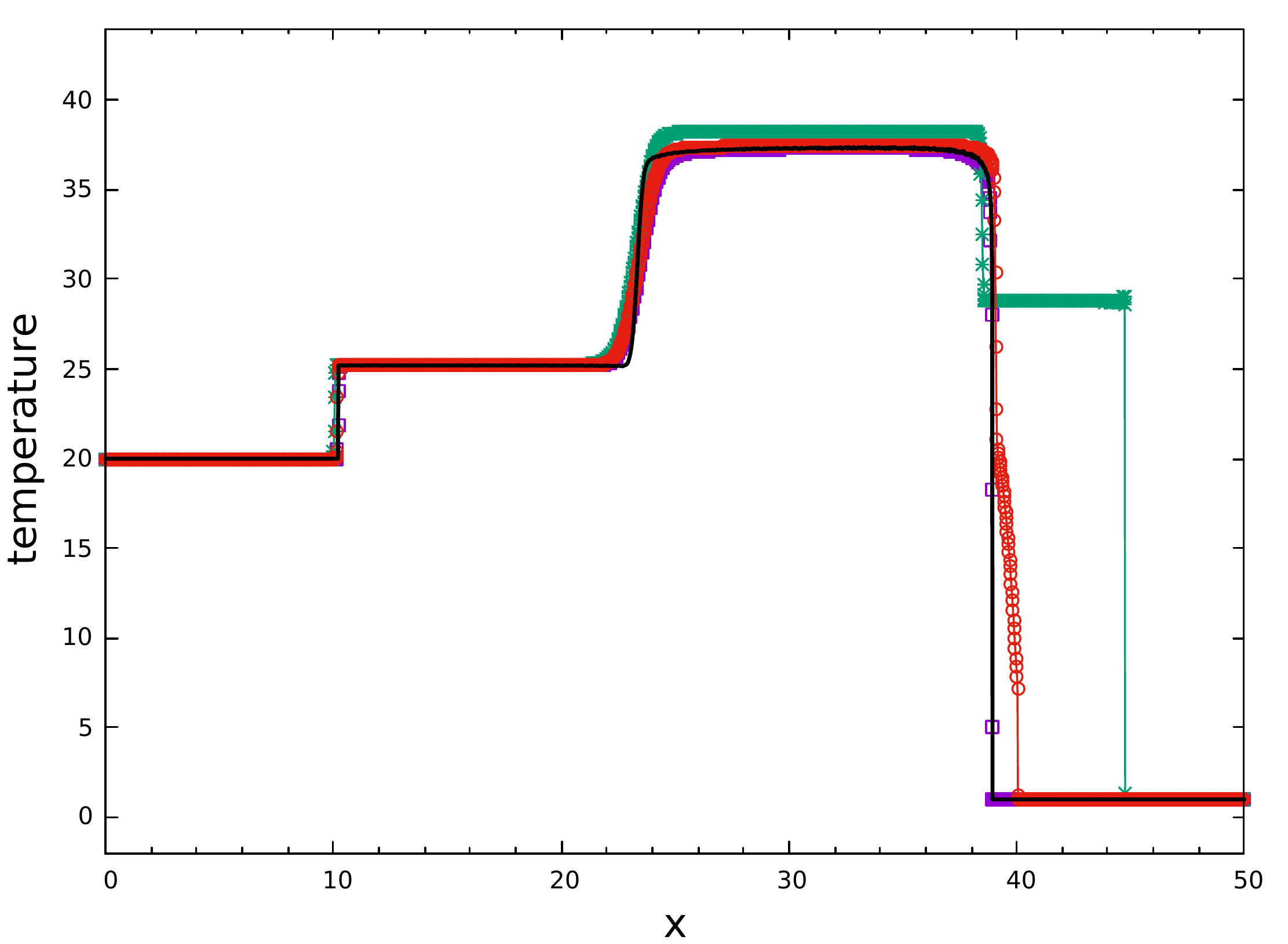} \\
  \includegraphics[scale=0.3]{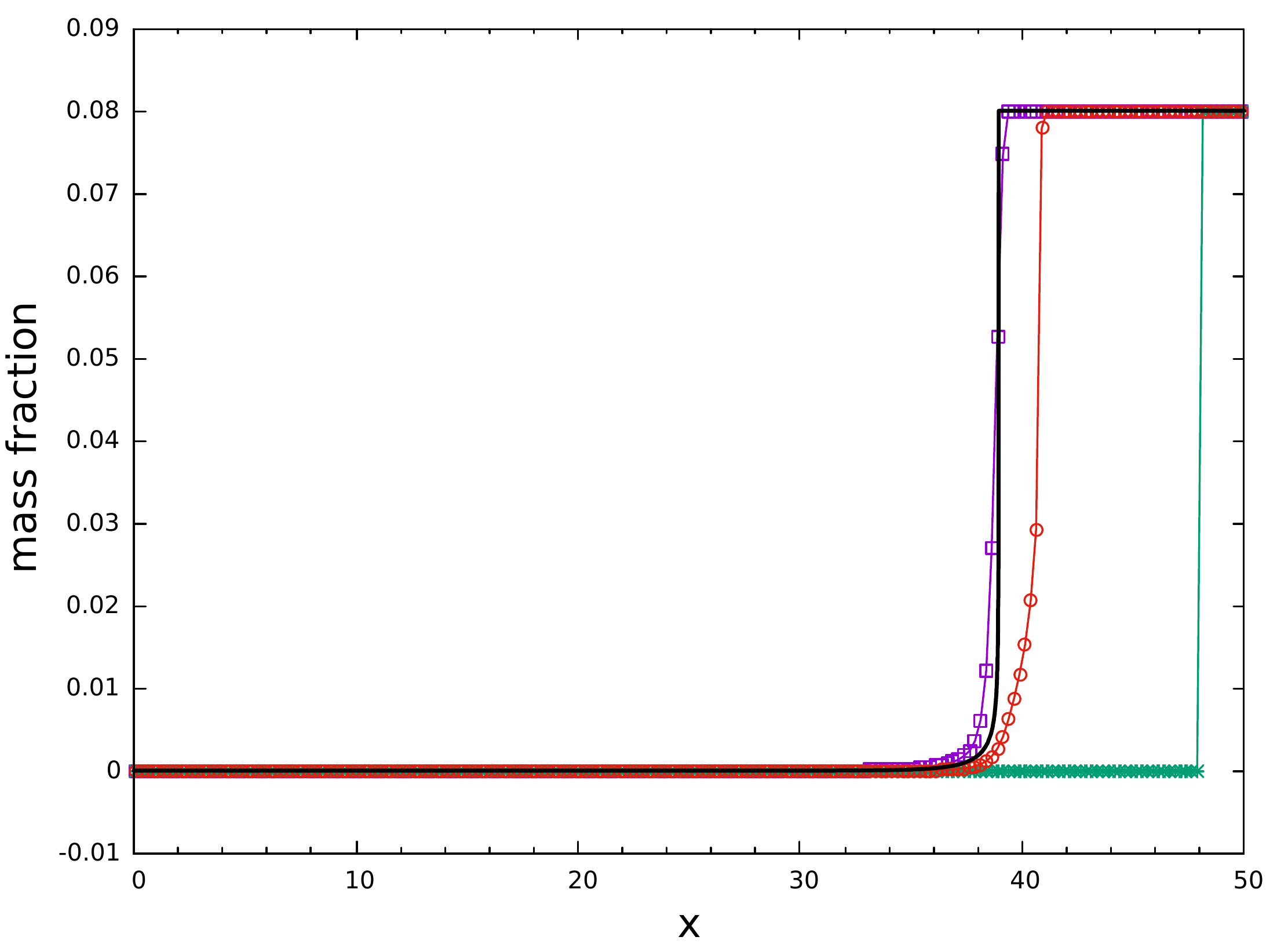} 
  \includegraphics[scale=0.3]{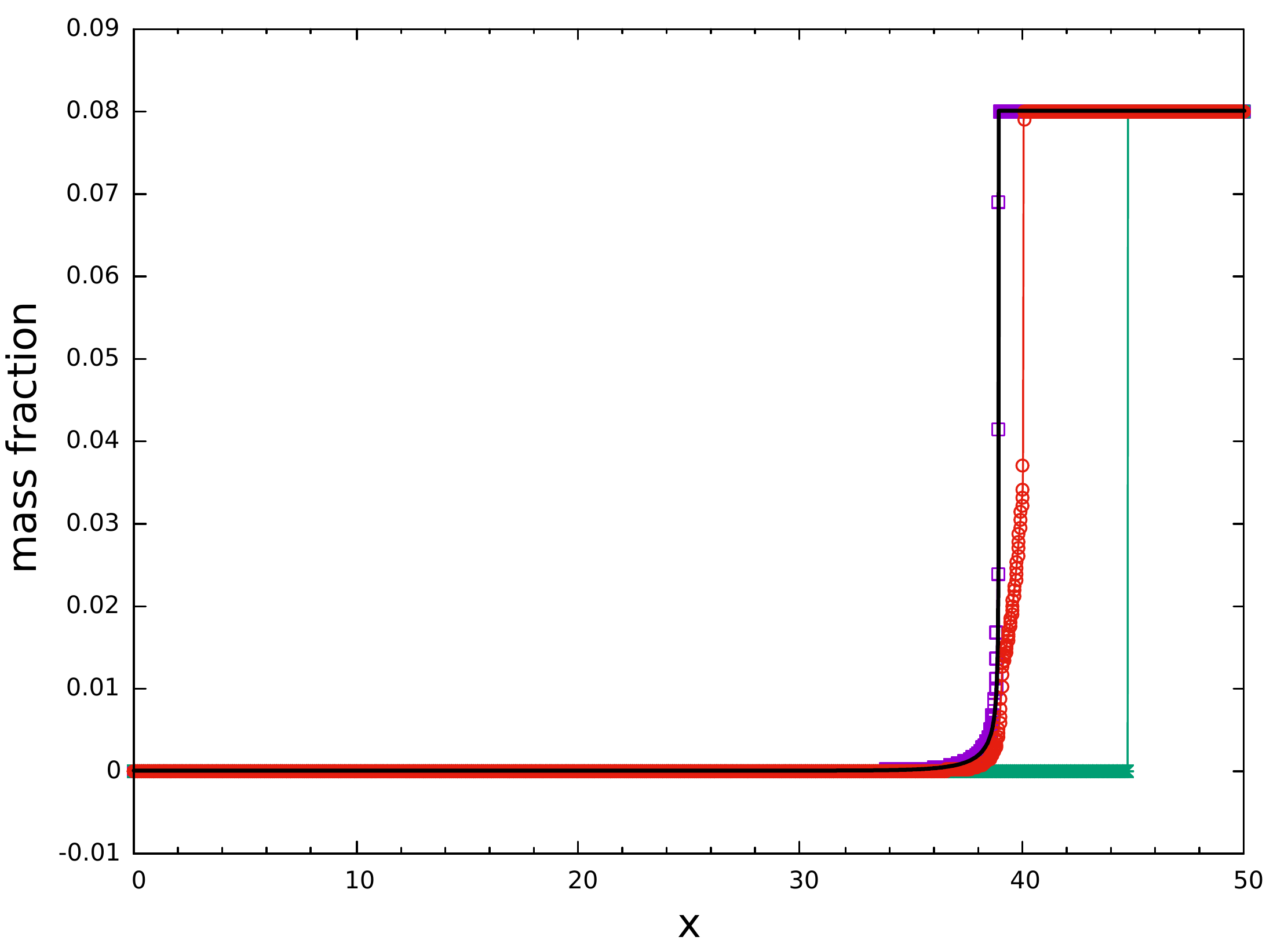} \\
  \caption{Example 3 two reactions, strong detonation at $t=3$: purple square line $\sim$ SRR solution; red circle line $\sim$ deterministic solution with Arrhenius kinetics; green cross line $\sim$ deterministic solution with Heviside kinetics;  black solid line $\sim$ reference solution; left column $\sim$ $\Delta x=0.25$, $\Delta t=0.01$; right column $\sim$ $\Delta x=0.025$, $\Delta t=0.001$.}
  \label{example3}
\end{figure}

The exact solution consists of a detonation wave, followed by a contact discontinuity and a shock, all moving to the right. Figure \ref{example3} presents the computational conditions and results obtained accordingly. All waves are captured with the correct speeds by the SRR method, in good agreement with the reference solution. However, the deterministic method obviously fails using the Heaviside model with the same under-resolved grids and timesteps. This is also because the stiffness of the Heaviside model as an infinite-rate reaction model is more severe and the deterministic method is poor to deal with stiffness unless both the space and time scales are resolved. Besides, the error of the spurious weak detonation by the deterministic method using the Arrhenius model grows with time, although its difference from the correct one is not very apparent at the present time point. 

EXAMPLE 4 (A Strong Detonation). This case considers a more complicated multi-step reaction model with three one-way reactions and five species involved
\begin{equation*}\label{case4_model}
\begin{aligned}
&1) \qquad \text{H}_2 \longrightarrow 2 \text{H}, \\
&2) \qquad 2\text{H} + \text{O}_2 \longrightarrow 2 \text{OH}, \\
&3) \qquad 2 \text{OH} + \text{H}_2 \longrightarrow 2 \text{H}_2\text{O}, 
\end{aligned}
\end{equation*} without $\text{N}_2$ here. The model is extended from the above two-reaction example, but with three distinct reaction rates (fast, medium and slow, respectively) to enlarge the stiffness due to multiple timescales.     

The parameters for the reaction model and species properties are
\begin{equation*}\label{case4_para}
\begin{aligned}
\left( \gamma ,q_{\text{H}_2}, q_{\text{O}_2}, q_{\text{OH}},  q_{\text{H}_2\text{O}}, q_{\text{H}}\right) &= \left( 1.4, 0, 0, -20, -100, 10  \right), \\
\left( W_{\text{H}_2}, W_{\text{O}_2}, W_{\text{OH}},  W_{\text{H}_2\text{O}}, W_{\text{H}}\right) &= \left( 2, 32, 17, 18, 1  \right), \\
\left( A^1, B^1, T_{ign}^1 \right) &= \left( 10^7, 0, 1.5 \right), \\
\left( A^2, B^2, T_{ign}^2 \right) &= \left( 10^5, 0, 2 \right), \\
\left( A^3, B^3, T_{ign}^3 \right) &= \left( 10^3, 0, 10 \right). \\
\end{aligned}
\end{equation*}
The initial condition of piecewise constants is given by
\begin{equation*}\label{case4_initial}
\begin{aligned}
\left( p,T,u,y_{\text{H}_2}, y_{\text{O}_2}, y_{\text{OH}},  y_{\text{H}_2\text{O}}, y_{\text{H}}\right) = 
\begin{cases}
\left( 40,20,10, 0, 0, 0.17, 0.72, 0.11 \right), & x<2.5, \\
\left( 1,1,0,0.2,0.8,0,0,0 \right), & x \geq 2.5. \\
\end{cases}
\end{aligned}
\end{equation*}
The left part gas is at the burnt equilibrium state and it is moving at a speed larger than $u_{CJ}$ relative to the stationary unburnt gas of the right part so that a strong detonation wave is to occur. This problem is solved on the interval $\left[0,50\right]$.

\begin{figure}
  \centering
  \includegraphics[scale=0.3]{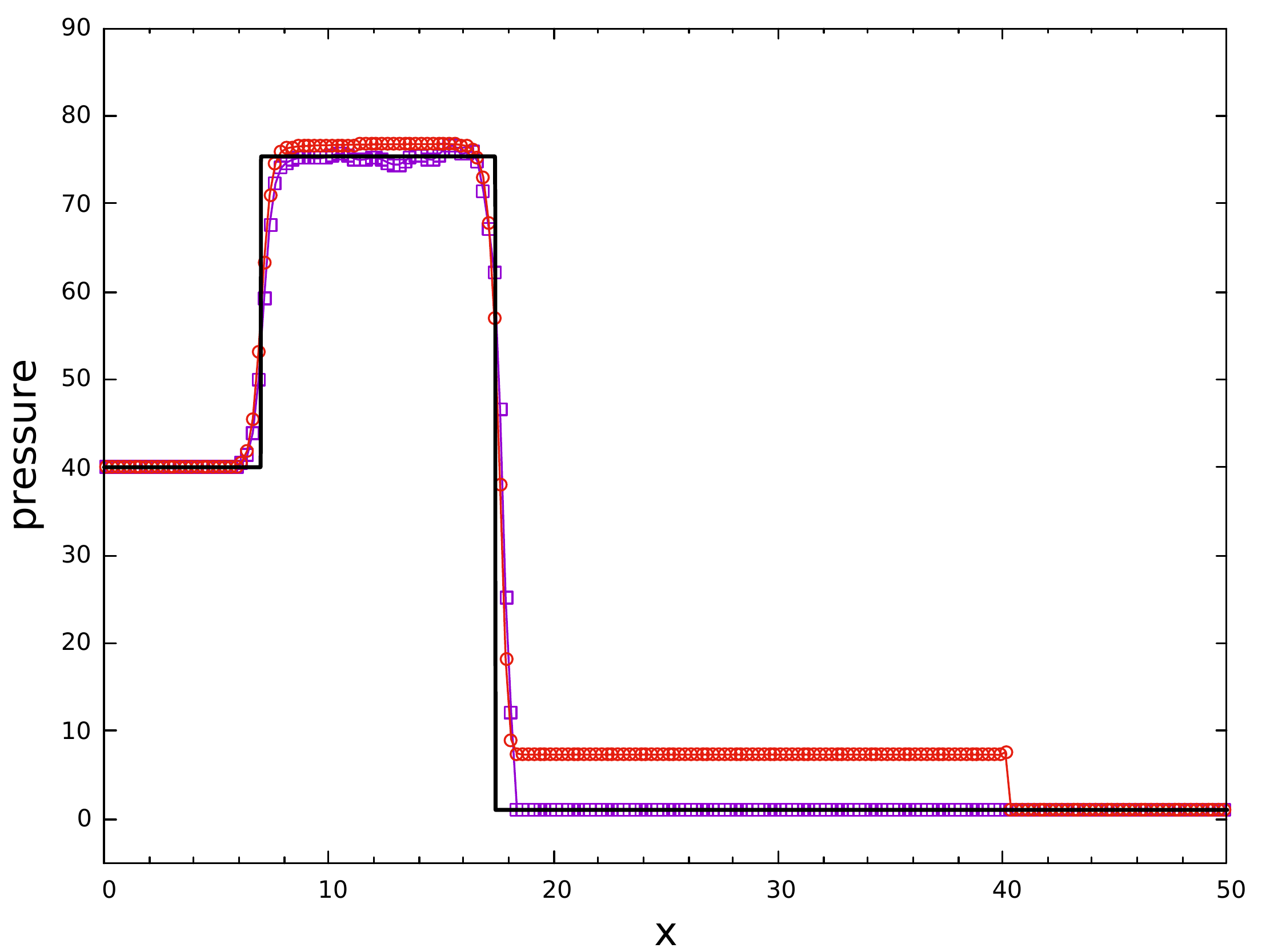} 
  \includegraphics[scale=0.3]{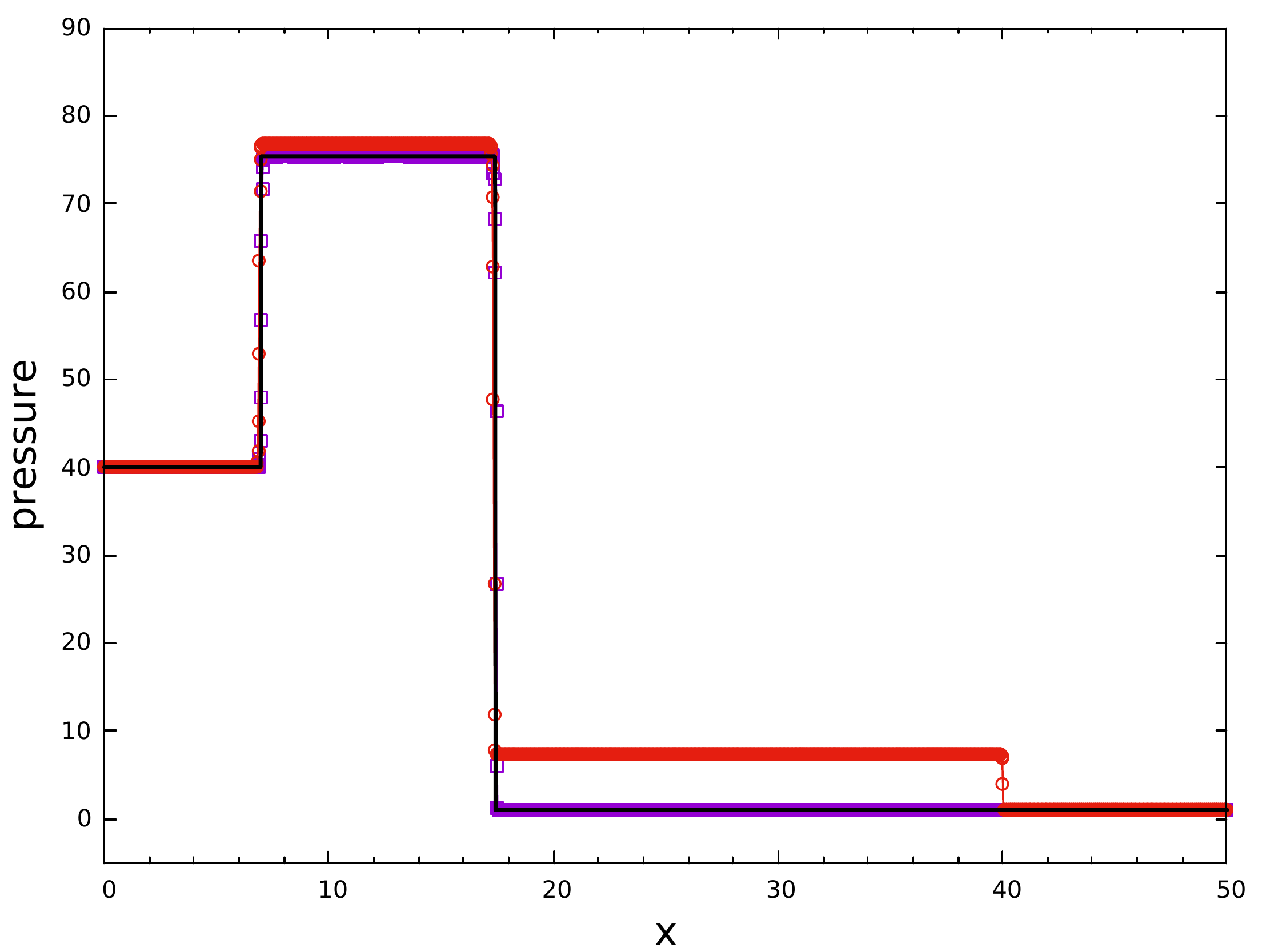} \\
  \includegraphics[scale=0.3]{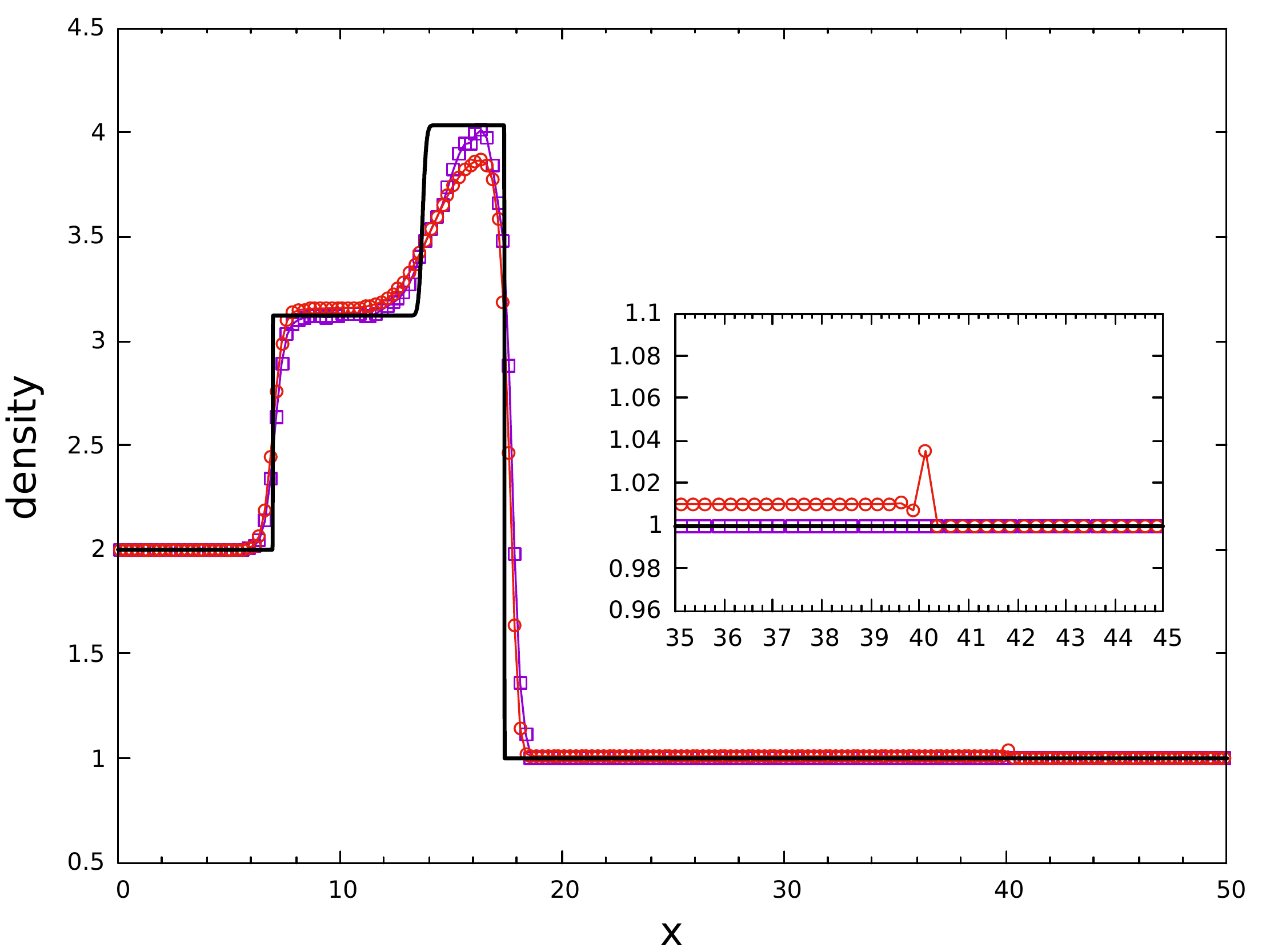} 
  \includegraphics[scale=0.3]{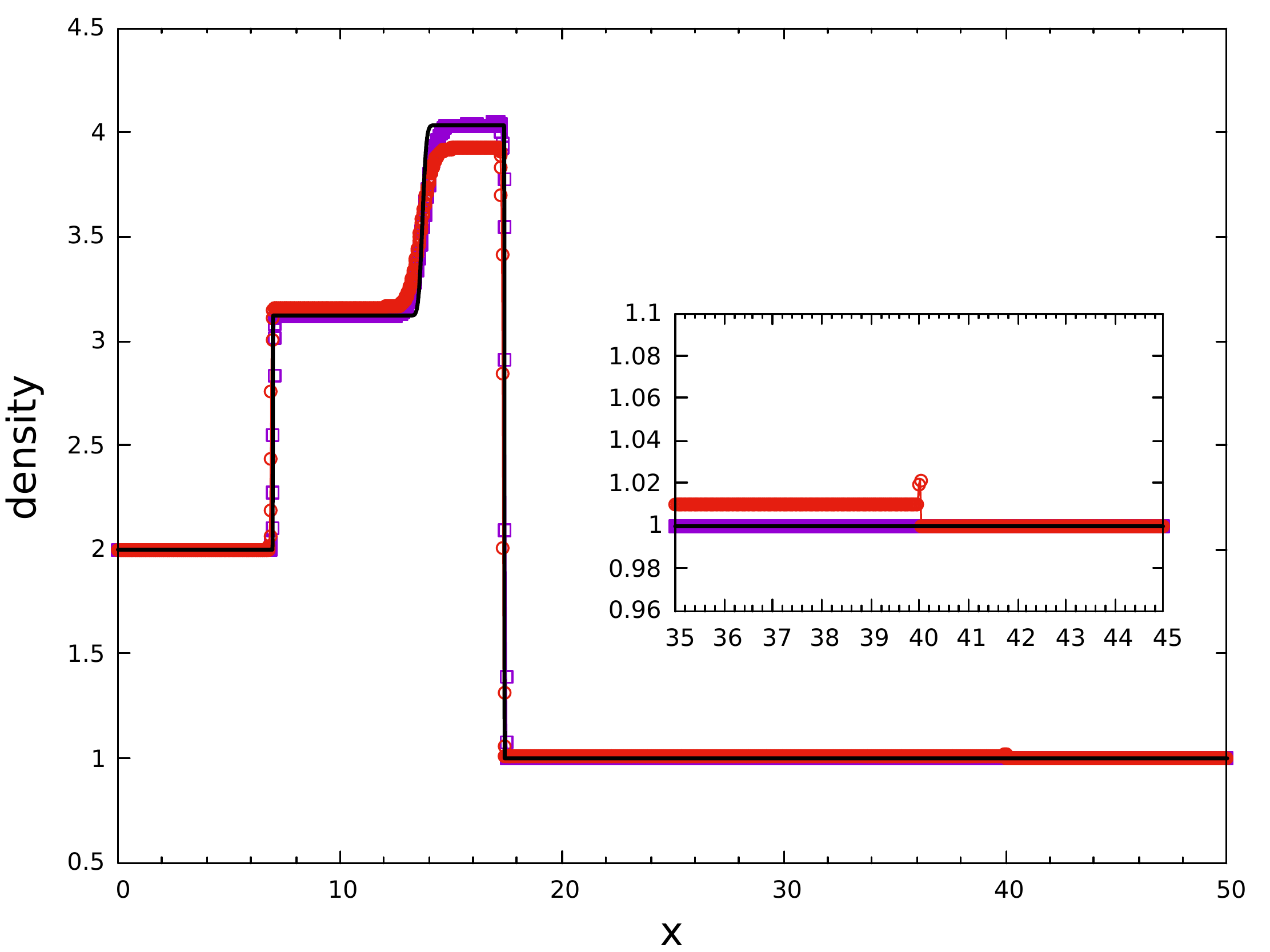} \\
  \includegraphics[scale=0.3]{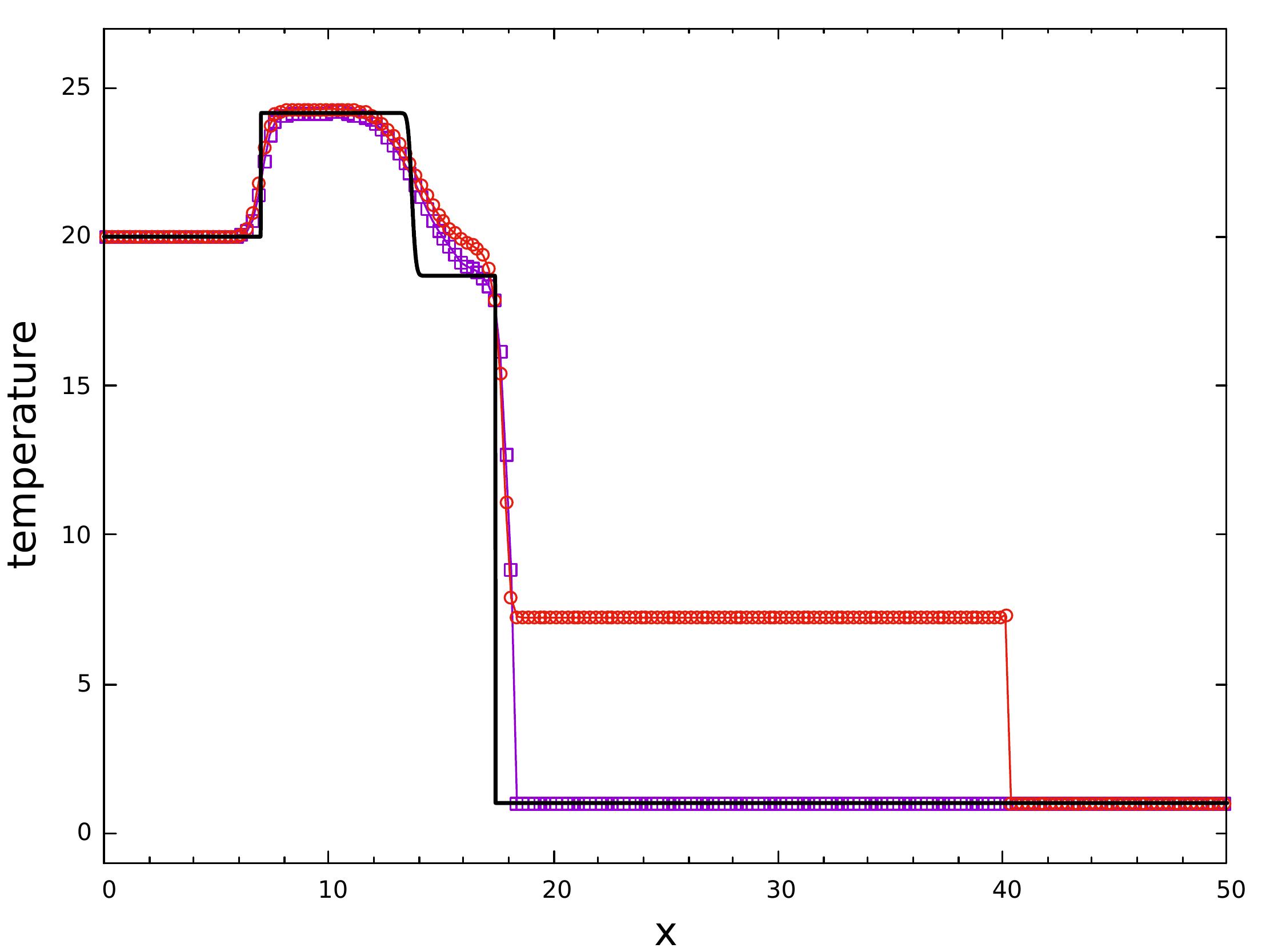} 
  \includegraphics[scale=0.3]{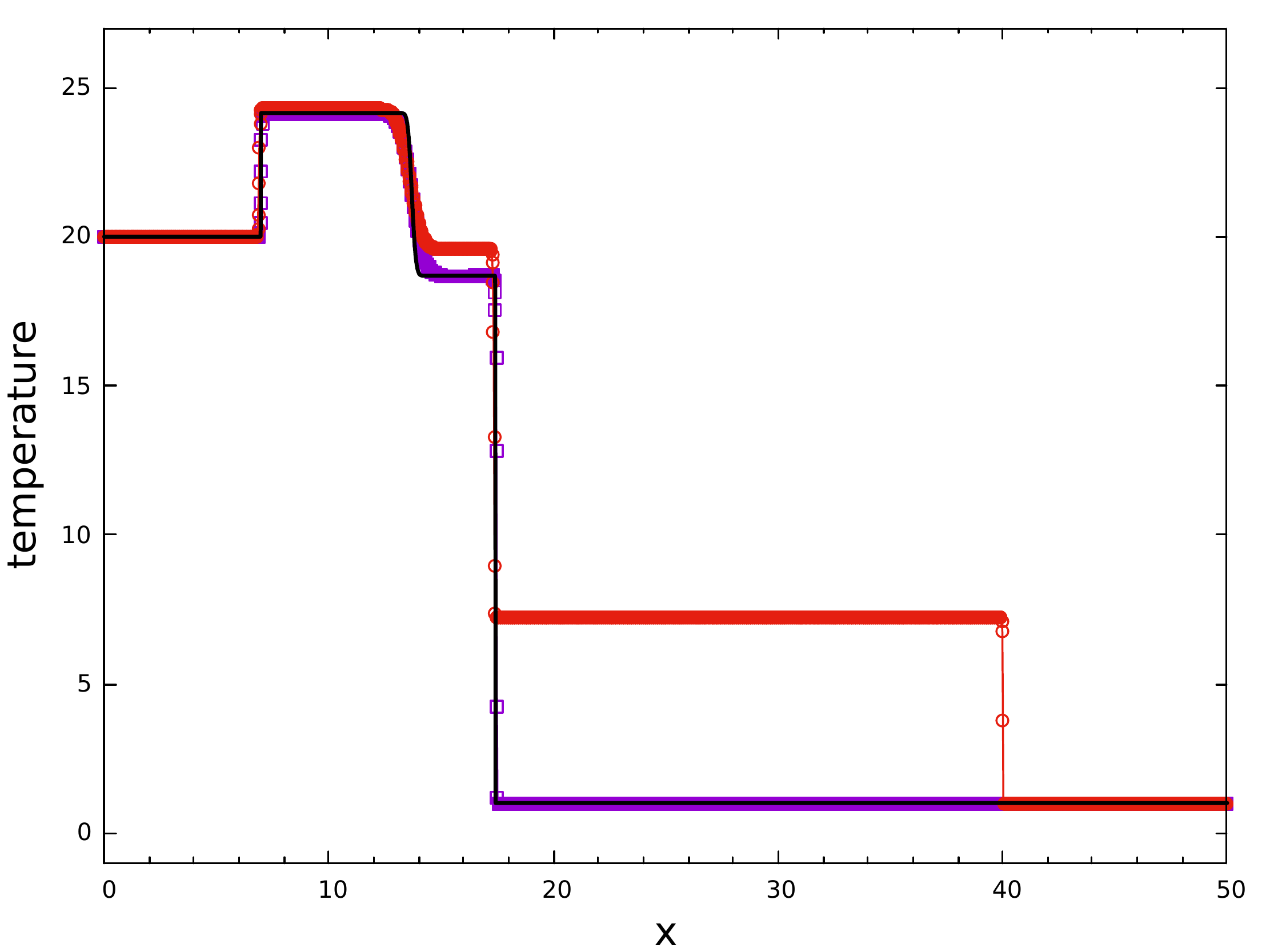} \\
  \includegraphics[scale=0.3]{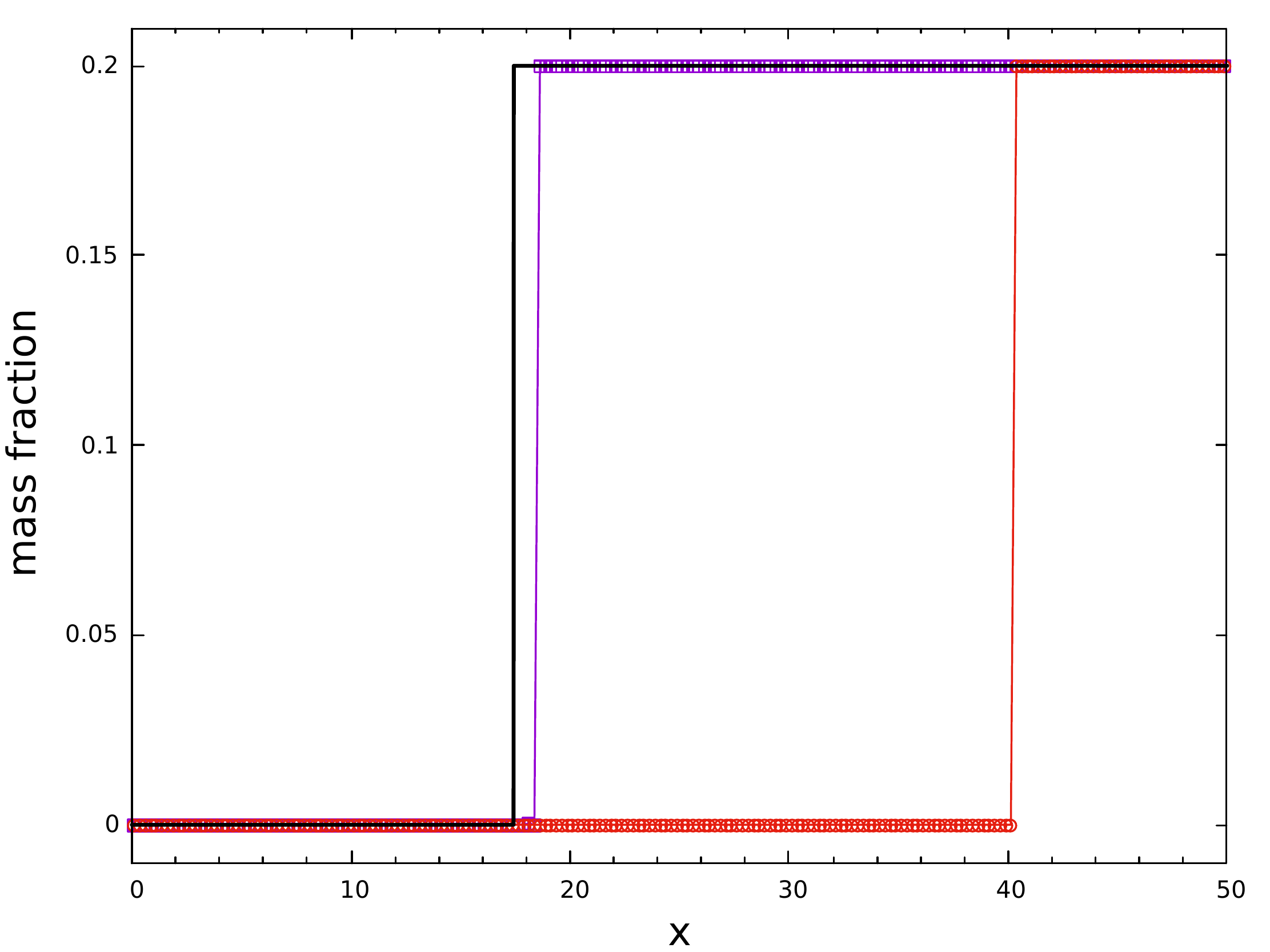} 
  \includegraphics[scale=0.3]{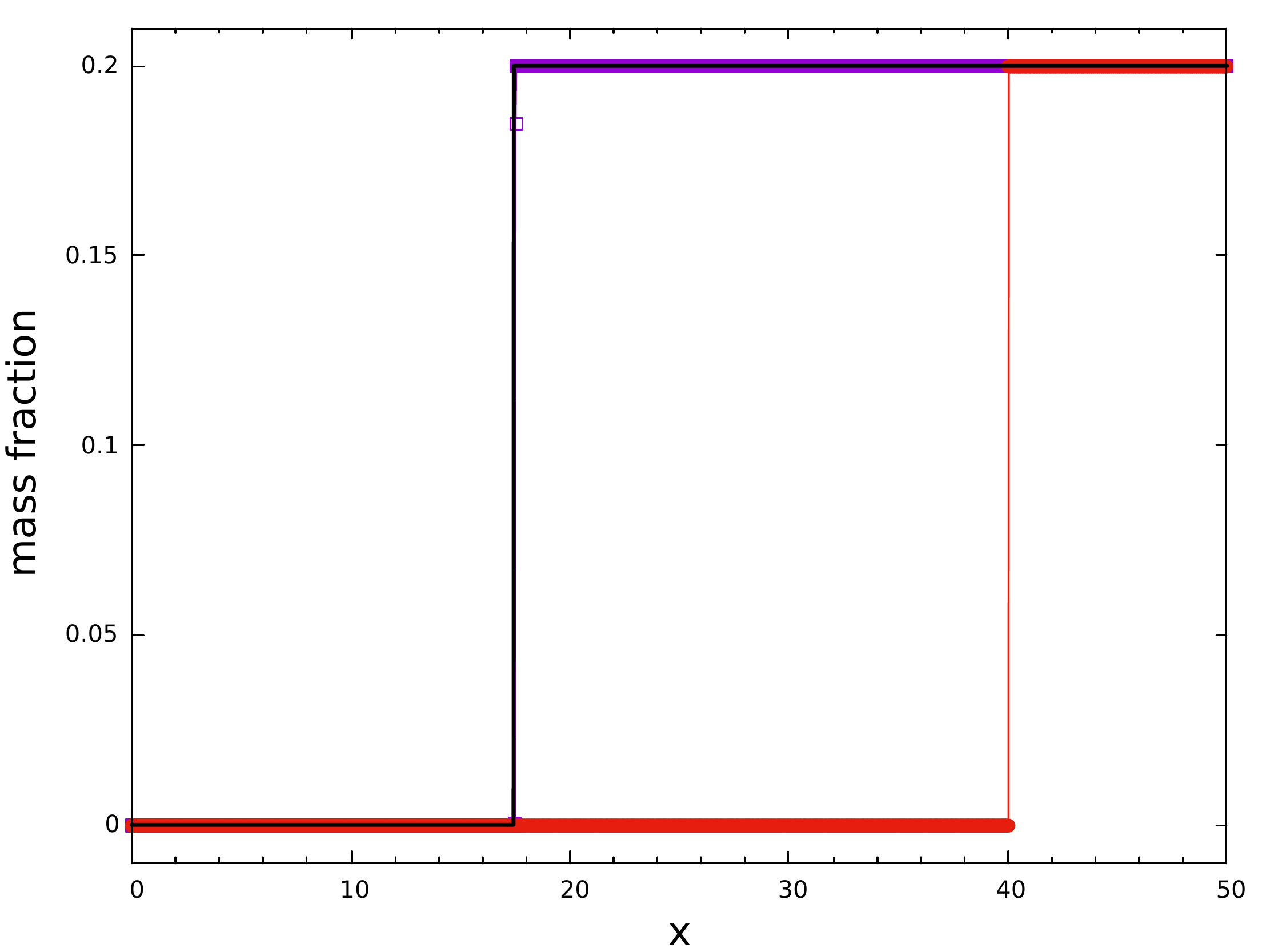} \\
  \caption{Example 4 three reactions, strong detonation at $t=1.5$: purple square line $\sim$ SRR solution; red circle line $\sim$ deterministic solution with Arrhenius kinetics; black solid line $\sim$ reference solution; left column $\sim$ $\Delta x=0.25$, $\Delta t=0.01$; right column $\sim$ $\Delta x=0.025$, $\Delta t=0.001$.}
  \label{example4}
\end{figure}

The exact solution shares the same wave pattern with the former example while the wave profiles differ greatly due to the change in the kinetics model. Figure \ref{example4} presents the computational conditions and computed results. All waves are captured with the correct speeds by the SRR method numerically, in good agreement with the reference solution with a location of the detonation wave at $x \approx 17$. However, the deterministic method obviously fails using the Arrhenius model with the same under-resolved grids and timesteps, by yielding a too fast weak detonation located at $x=40$. And the incorrect weak detonation wave by the deterministic method using the Heaviside model has already run out of the domain at $t=1.5$ (thus not shown in the plots).

EXAMPLE 5 (A CJ Detonation in 2D). This 2D case extends EXAMPLE 1 to model the radially symmetric point-source explosion, where $A$ in Eq. \eqref{case1_para} is amplified by $10000$ times to approximate the infinitely fast reaction with extreme stiffness. Similar tests have been studied in \cite{bao2002random, helzel2000modified}. 

With radial symmetry, 1/4 part of the explosion is convenient to take into use as in space, $\left[0,50\right] \times \left[0,50\right]$. The hot-spot area of the initial high-temperature high-pressure burnt gas is a circle with radius 10 and the reactive unburnt gas takes the outside. Initial condition is the same as in Example 1 except the initial velocity of the circle area is adjusted to along the radial direction, i.e.
\begin{equation*}\label{case5x_initial1}
\begin{aligned}
\left(u,v\right) = 
\begin{cases}
\left( 2.899x/r,2.899y/r \right), & r<10, \\
\left( 0,0 \right), & r \geq 10, \\
\end{cases}
\end{aligned}
\end{equation*}       
where $r=\sqrt{x^2+y^2}$.

\begin{figure}
	\centering
	\includegraphics[trim={0cm 0cm 0cm 0cm},scale=0.5]{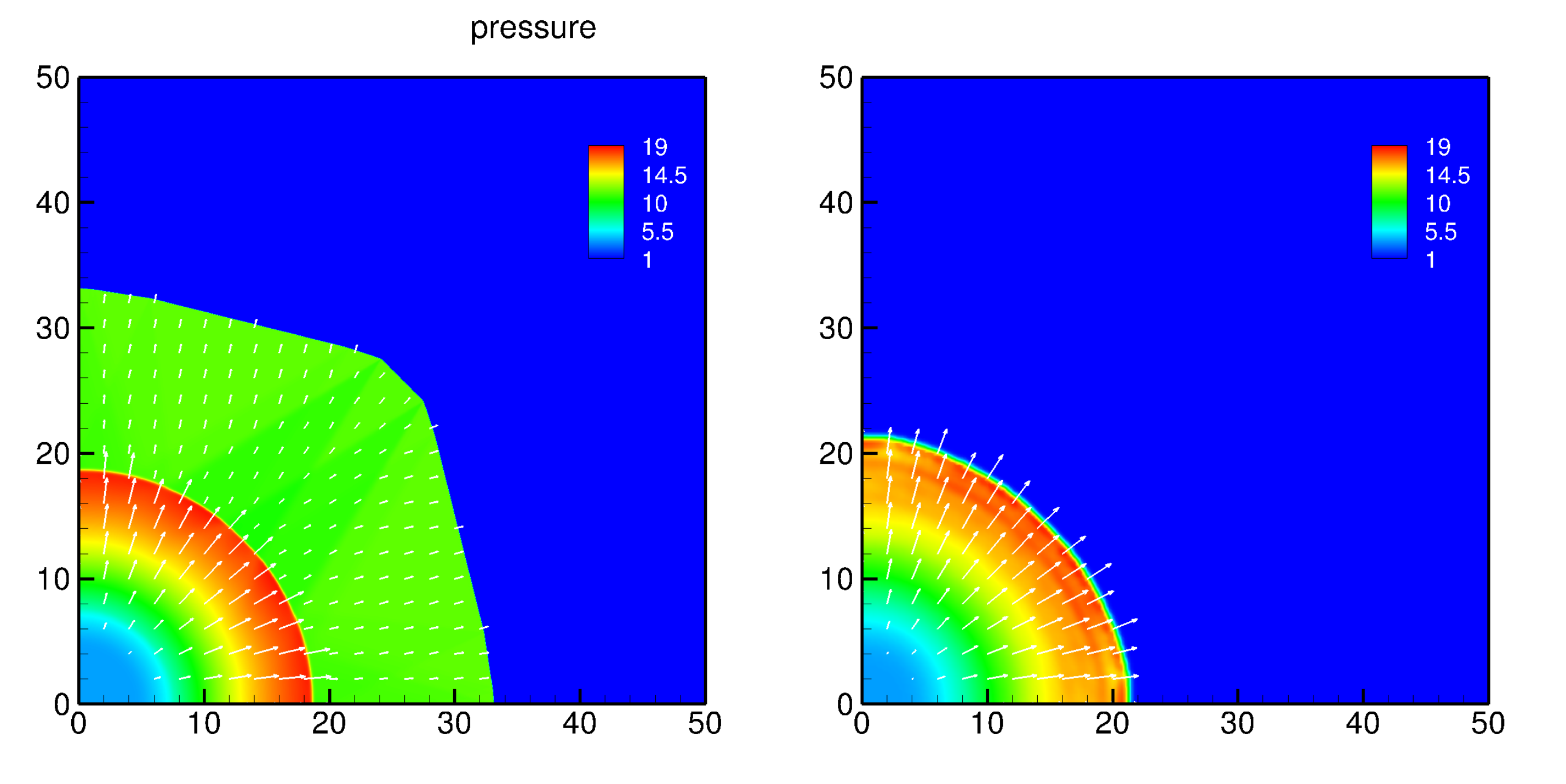} 
	\includegraphics[trim={0cm 0cm 0cm 0cm},scale=0.5]{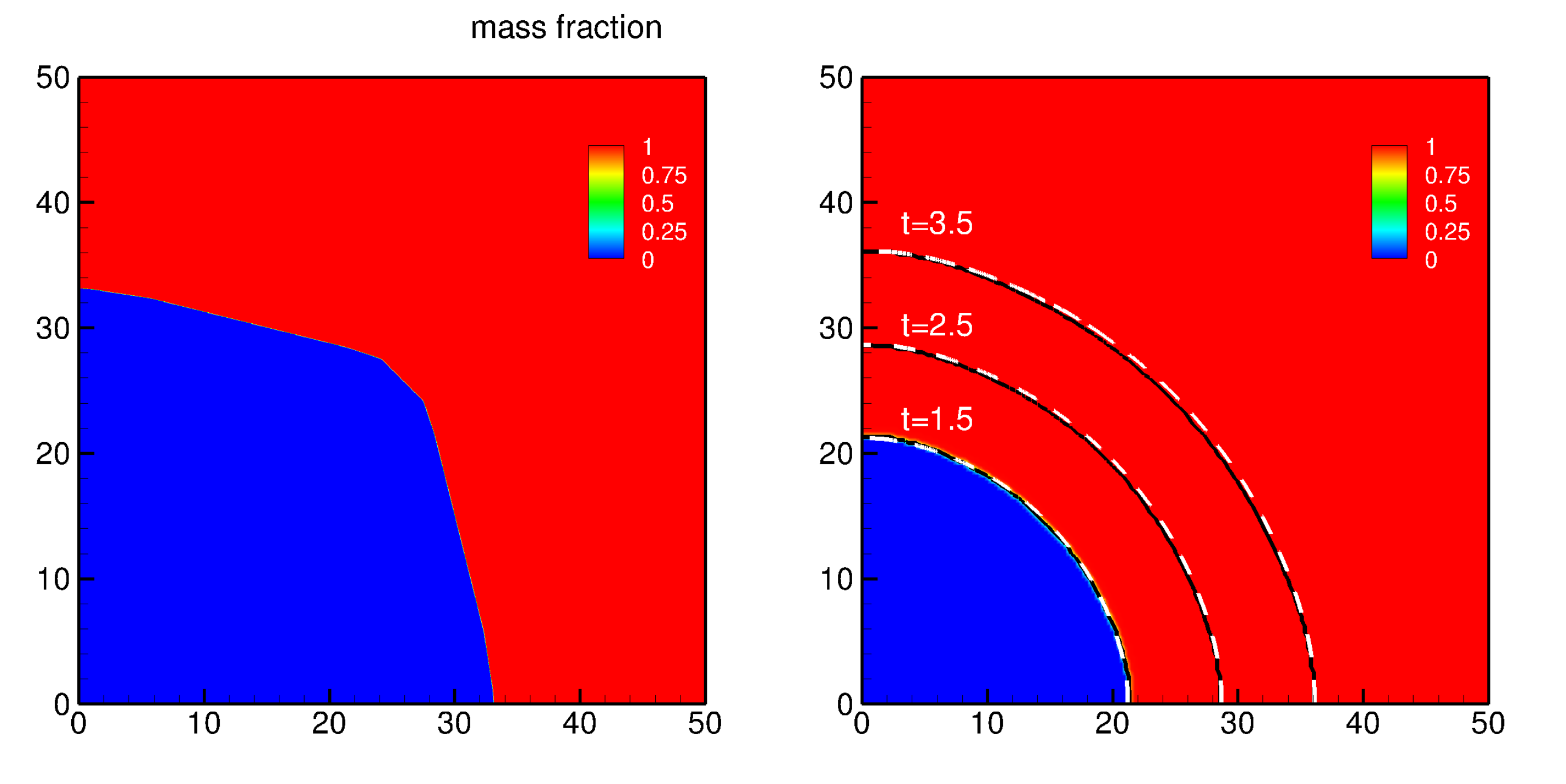} \\ 
	\caption{Example 5 2D case, one infinite-rate reaction, CJ detonation: left $\sim$ deterministic solution; right $\sim$ SRR solution. Locations of the CJ detonation wave at three times are marked by $y_{A}=0.5$: black solid line $\sim$ low resolution; white dashed line $\sim$ high resolution.}
	\label{example5}
\end{figure}

In our computations, a coarse grid ($200 \times 200$) and a finer grid ($2000 \times 2000$) are employed referring to Example 1. Corresponding timesteps are $\Delta t=1 \times 10^{-2}$ and $1 \times 10^{-3}$, respectively. Unfortunately, we cannot obtain the reference solution by the deterministic method with a further refined grid for this 2D case.
With the finer grid, the deterministic method still gives the obviously spurious solution at $t=1.5$, see the left column of Fig. \ref{example5}, in that a nonphysical weak detonation wave is generated and the reacting front is no more circular. In contrast, our SRR method can capture the shape and location of the CJ detonation front accurately, see the right column of the figure, by observing the radial velocity vector in the pressure contour even in the low resolution and the self-similarly circular outwards-developing detonation fronts in black/white lines of two resolutions at different times. The line-marked locations calculated by the random method in two resolutions agree excellently with each other and thus a grid convergence to the exact solution is reasonable to expect for the proposed SRR method. Besides, with ignorable curvature effects \cite{aslam1999detonation,short2016steady} as the detonation radius is large and the under-resolved reaction zone is infinitesimal, the calculated speed of the detonation front approaches the 1D theoretical speed of $D_{CJ}=7.1247$ as in Example 1.

EXAMPLE 6 (A Strong Detonation in 2D). The present case considers the same multi-step reaction mechanism as in EXAMPLE 3 except that $q_{\text{OH}}$ in Eq. \eqref{case3_para} changes into $-50$. This is also a multi-dimensional case used to prove the dimension-independent nature of the proposed method, unlike the original random projection method which requires a dimension-by-dimension scanning for local projection. The test is also studied in \cite{zhang2014equilibrium}.

\begin{figure}
  \centering
  \includegraphics[trim={5cm 9cm 5cm 9cm},scale=0.4]{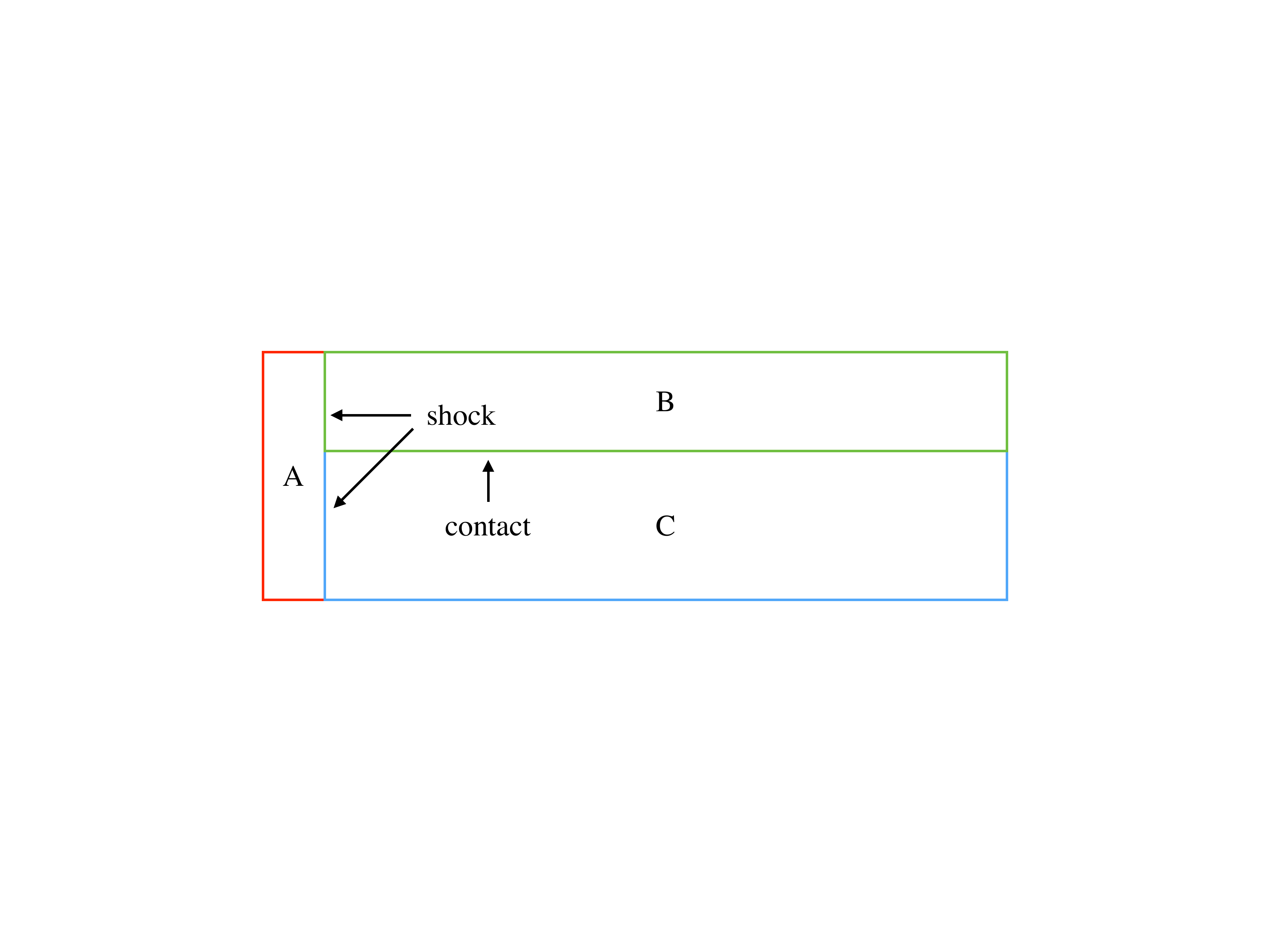}
  \caption{Schematic for the 2D domain in Example 6.}
  \label{schematic}
\end{figure}

The initial condition of piecewise constants in the $\left[0,6\right] \times \left[0,2\right]$ 2D domain consists of 
\begin{equation*}\label{case6_initial1}
\begin{aligned}
\left( p,T,u,v\right) = 
\begin{cases}
\left( 40,20,10, 0 \right), & x<0.5, \\
\left( 1,1,0,0 \right), & x \geq 0.5, \\
\end{cases}
\end{aligned}
\end{equation*}
\begin{equation*}\label{case6_initial2}
\begin{aligned}
\left( y_{\text{H}_2}, y_{\text{O}_2}, y_{\text{OH}},  y_{\text{H}_2\text{O}}, y_{\text{N}_2} \right) = 
\begin{cases}
\left( 0,0,0.17,0.63,0.2 \right), & x<0.5, \\
\left( 0,0,0.17,0.63,0.2 \right), & x \geq 0.5, y \geq 1.2, \\
\left( 0.08,0.72,0,0,0.2 \right), & x \geq 0.5, y < 1.2, \\
\end{cases}
\end{aligned}
\end{equation*}
as shown in Fig. \ref{schematic}. We can see that the computational domain is composed of three parts (zone A, B and C) with shock and contact surface. Both zone A and B are filled with burnt gas and zone C is filled with the reactive unburnt gas.

\begin{figure}
  \centering
  \includegraphics[scale=0.45]{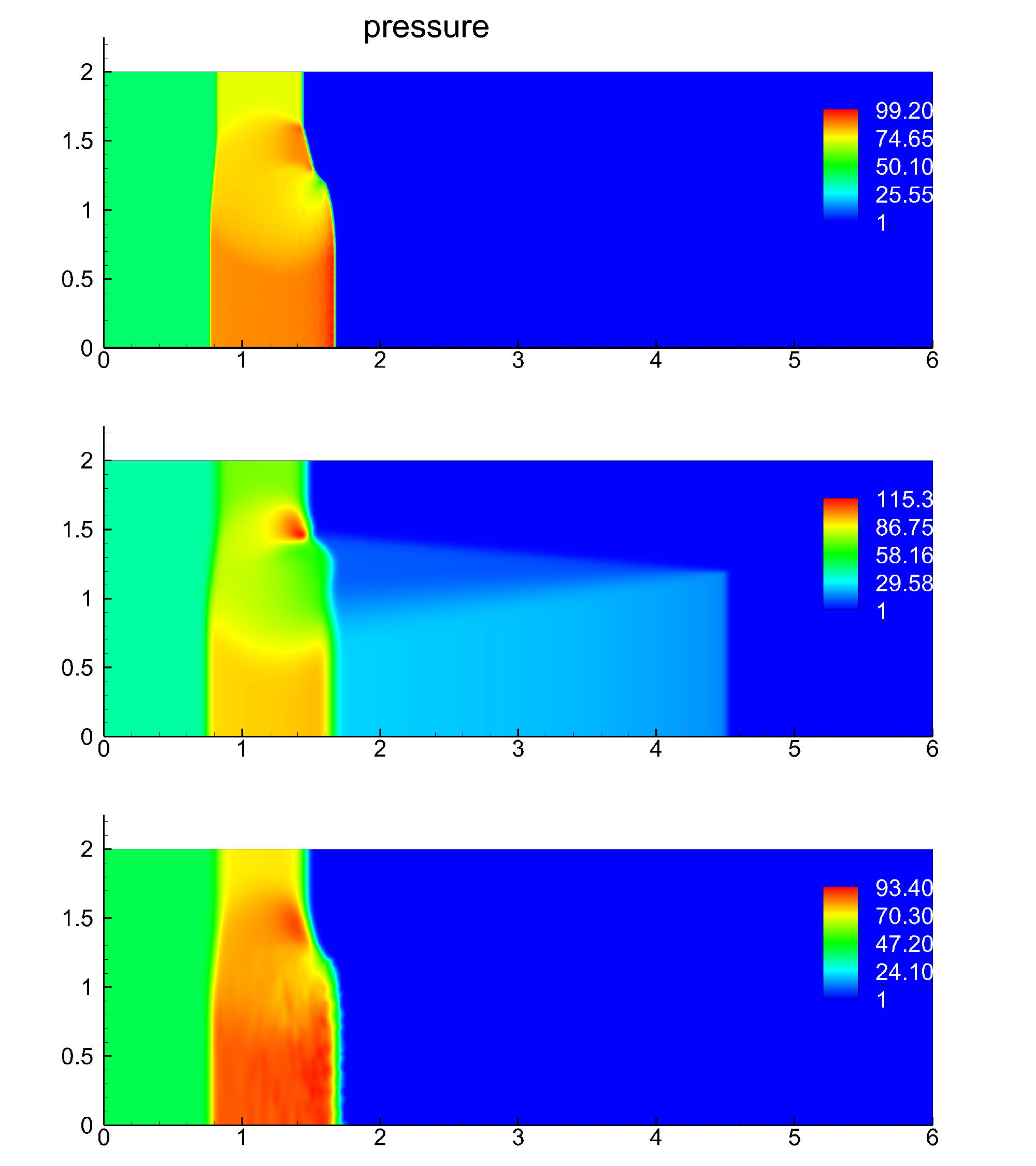} 
  \includegraphics[scale=0.45]{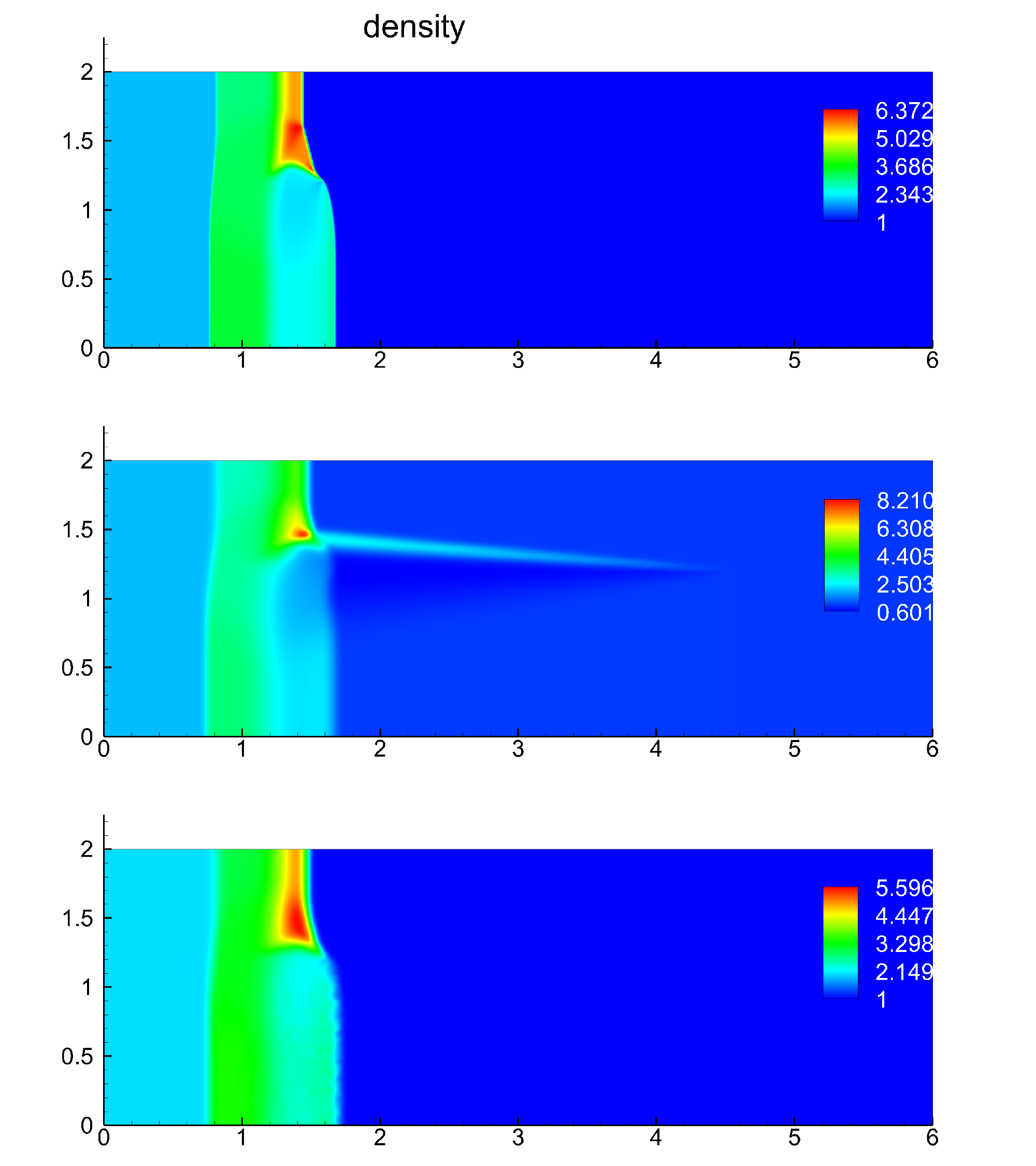} \\
  \includegraphics[scale=0.45]{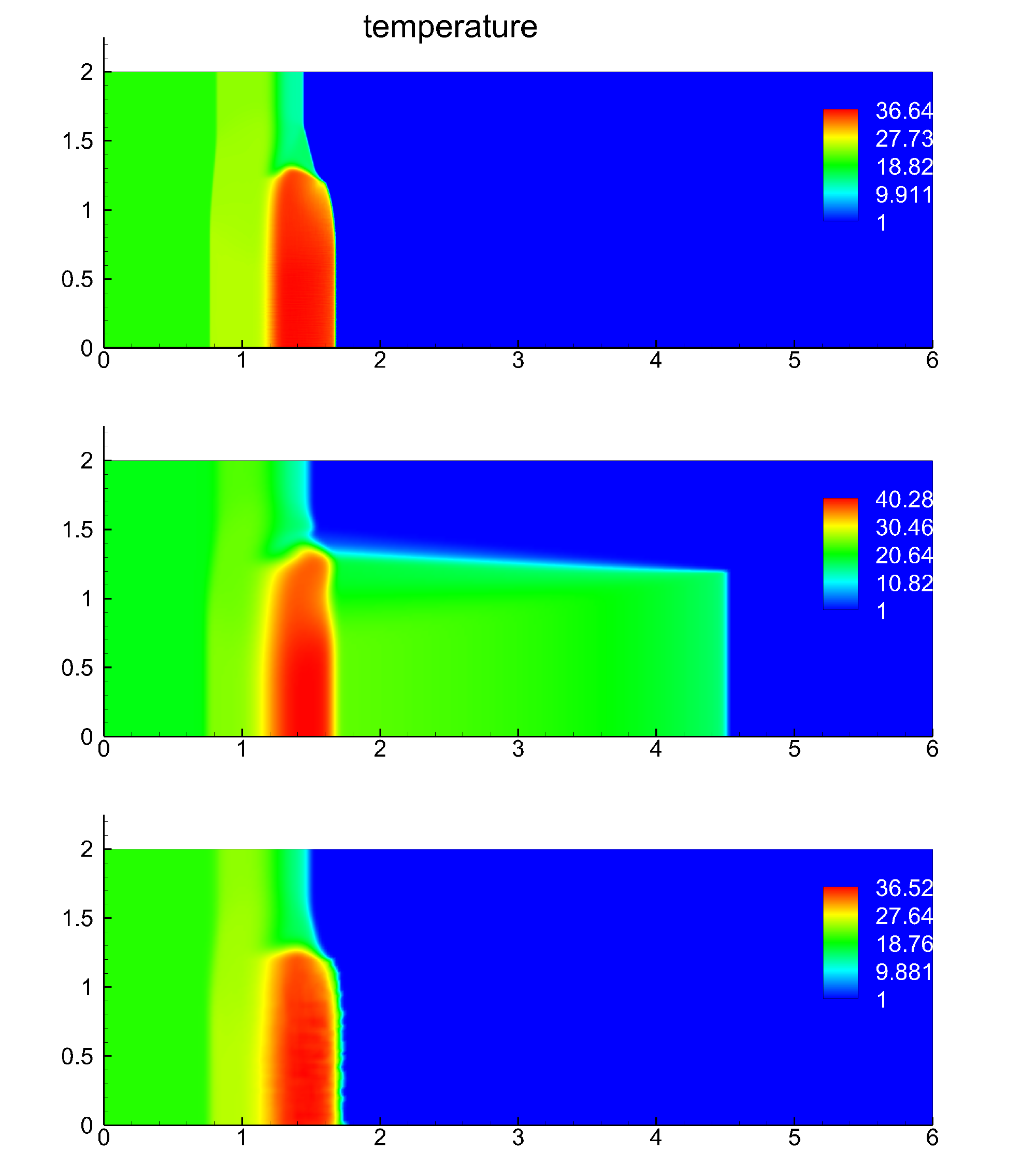} 
  \includegraphics[scale=0.45]{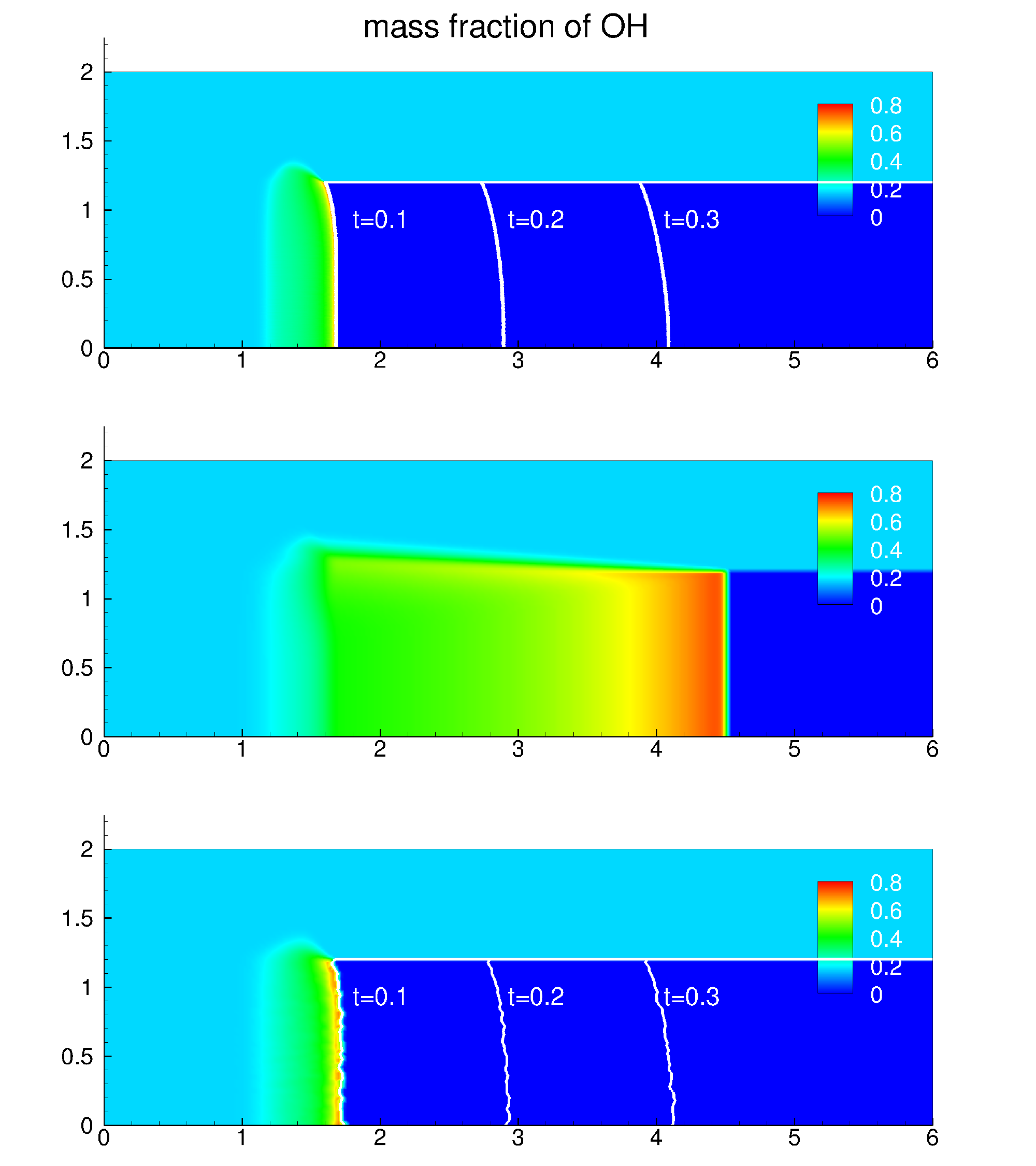} \\  
  \caption{Example 6 2D case, two reactions, strong detonation at $t=0.1$: top $\sim$ reference solution; middle $\sim$ deterministic solution with Arrhenius kinetics; bottom $\sim$ SRR solution; in the mass fraction contour, locations of the detonation front at $t=0.1, 0.2, 0.3$ are additionally marked by setting $y_{\text{O}_2}=0.5$ in white solid lines.}
  \label{example6}
\end{figure}

In our computations, a uniformly distributed coarse grid ($300 \times 100$) and a refined grid ($3000 \times 1000$) are employed. Corresponding timesteps are $\Delta t=5 \times 10^{-4}$ and $5 \times 10^{-5}$, respectively. The reference solution is obtained by the deterministic method using the fine grid and tiny timestep. The comparison of the SRR method and deterministic method on capturing stiff detonation waves is based on the under-resolved grid and timestep. In Fig. \ref{example6}, it is readily to see at $t=0.1$ the spurious solution given by the deterministic method on the coarse grid contains a too fast weak detonation wave, which has passed half of the domain. However, the correct detonation waves from the SRR method on the same resolution and the deterministic method on a fine grid agree with each other excellently and fall far behind the spurious weak detonation wave. Good agreement of the self-similar propagation of the detonation wave from $t=0.1$ to $0.3$ is also can be seen in the mass fraction contour given by the reference solution and the under-resolved SRR solution, respectively. The slight difference between the two correct solutions lies in some small around-shock statistical fluctuations due to the random nature of the method \cite{bao2000random}.     

\newgeometry{left=1.5cm,bottom=2cm,right=1.5cm,top=2cm}
\restoregeometry

\subsection{Reactive Euler equations with real-world nonequilibrium kinetics}

In this subsection, we try to validate the SRR method in capturing stiff detonation waves governed by the reactive Euler equations coupled with real-world chemical nonequilibrium kinetics, in which the much more complicated reaction mechanism will introduce multiple temperature-dependent reactions with distinct timescales. To our knowledge, both the two test cases below are reported for the first time, taking into account the detailed hydrogen-air combustion mechanism as in Subsection \ref{h2combustion}. Two different scenarios with the CJ detonation and strong detonation wave, respectively, are simulated in 1D or 2D domain, regardless of the dimensional independence property of the proposed method.  

The convection operator adopts an ordinary shock capturing scheme as in the former subsection, and the reaction step is solved by the proposed SRR method and the popular CHEMEQ2 integrator as the deterministic method to make a comparison. In particular, reaction splitting in the SRR method is based on the 2nd-order Strang's scheme to reduce splitting errors.         

EXAMPLE 7 (A Realistic CJ Detonation). The setup of this case consists of two parts divided by a shock moving to the right in a 1D domain of length $L=4$m: the left part is post-shock and filled with high-temperature high-pressure burnt gas while the right part is pre-shock and filled with reactive unburnt gas in one atmosphere pressure and room temperature, see details in Table \ref{example7_IC}. The theoretical CJ detonation states for the unburnt gas can be generated using the NASA Chemical Equilibrium Analysis (CEA) program \cite{gordon1994computer} and according to the CJ condition \cite{bao2000random,yee2013spurious,zhang2014equilibrium}, i.e.
\begin{equation*}
D_{CJ} = u_{CJ} + (\gamma p_b/\rho_b)^{1/2},
\end{equation*}
we adopt $u_b=800\text{m/s} \approx u_{CJ}$ for the initial velocity of the burnt gas, to generate a CJ detonation wave sweeping the stationary unburnt gas. The shock is initially located at $x=0.5$m. Boundary condition for the left/right end is simply extrapolation from the mirror image points inside the domain. All simulations stop at $t=1.2 \times 10^{-3}$s.         

The exact solution is a steady self-similar CJ detonation wave travelling from left to right, in similar with the model problem of Example 1. We obtain the reference exact solution by the deterministic method using a very fine grid with 10000 points and a fixed tiny timestep of $\Delta t = 5 \times 10^{-9}$s. Two sets of under-resolved grid and timestep are considered, i.e. $\Delta x=0.08\text{m}, \Delta t=1 \times 10^{-6}\text{s}$ and $\Delta x=0.02\text{m}, \Delta t=2.5 \times 10^{-7}\text{s}$, respectively. 

\begin{figure}
  \centering
  \includegraphics[scale=0.3]{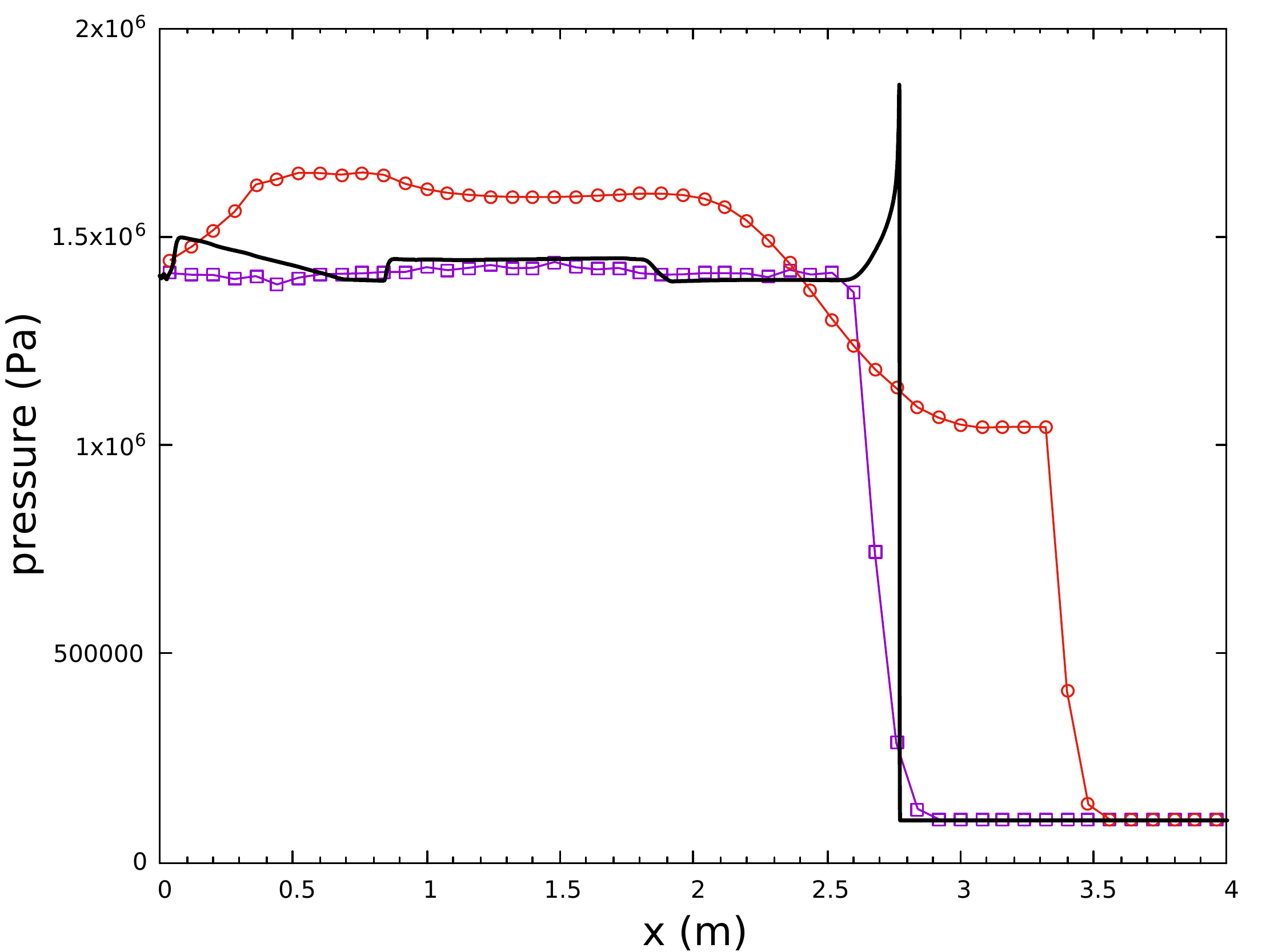} 
  \includegraphics[scale=0.3]{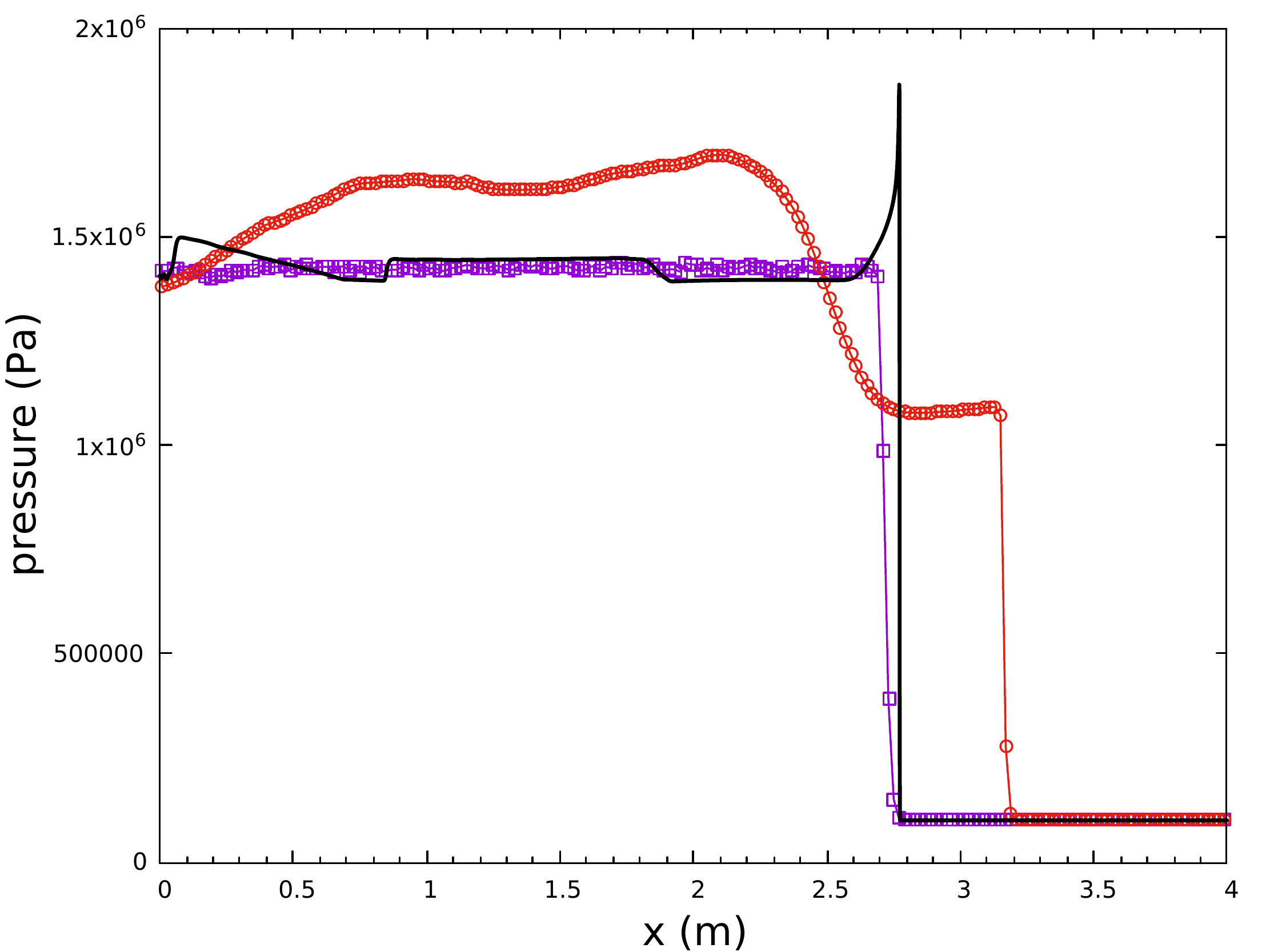} \\
  \includegraphics[scale=0.3]{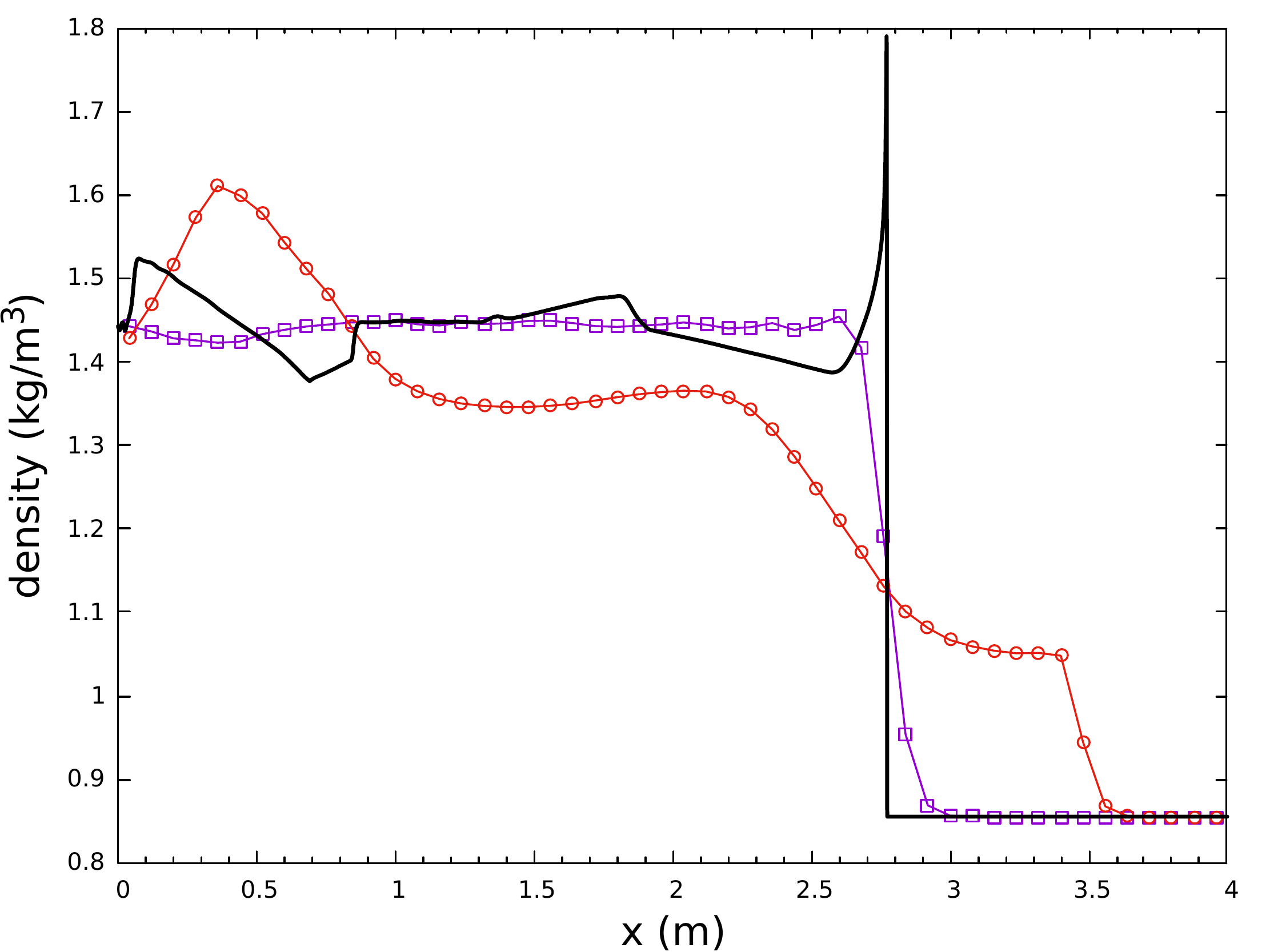} 
  \includegraphics[scale=0.3]{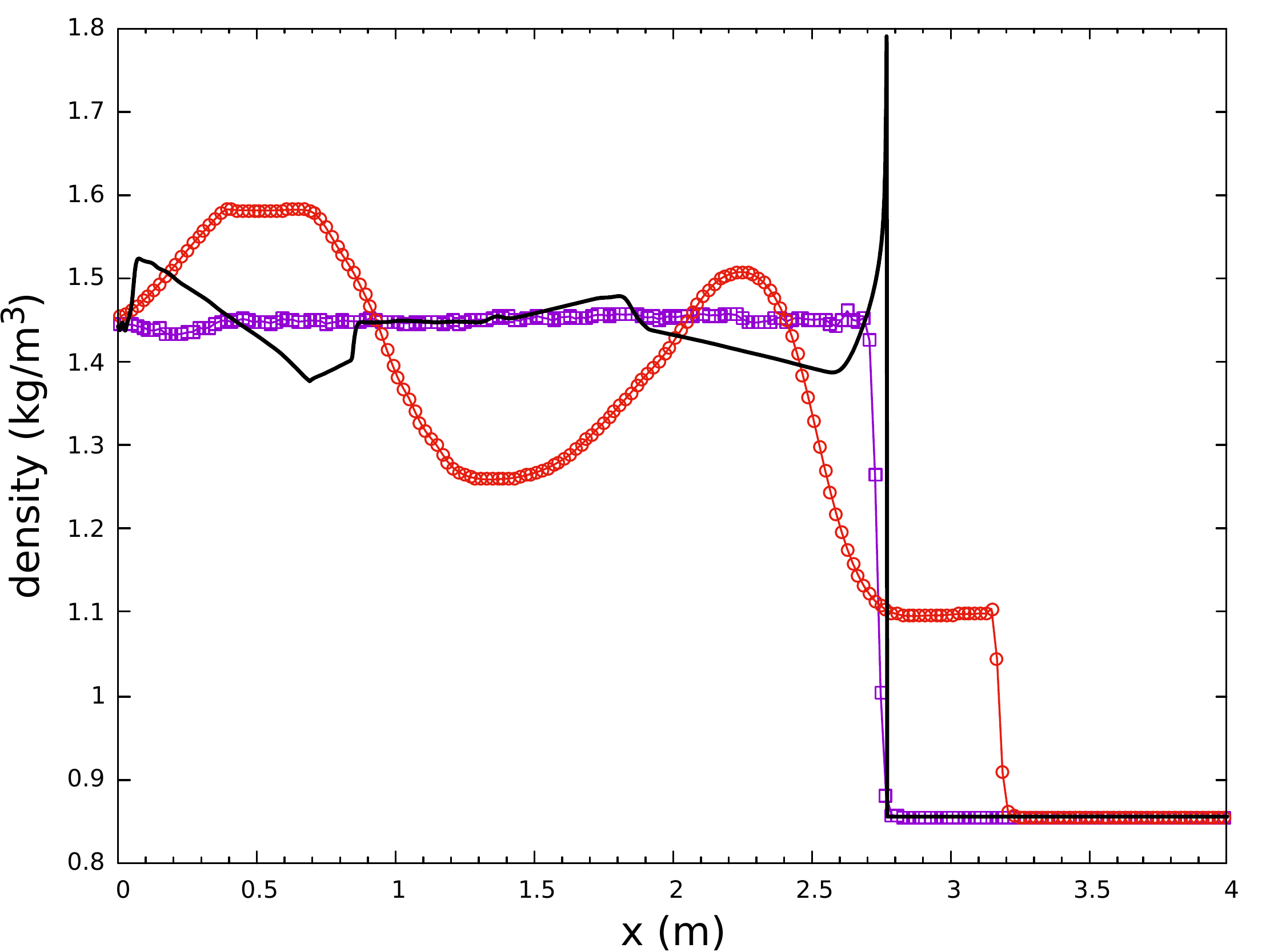} \\
  \includegraphics[scale=0.3]{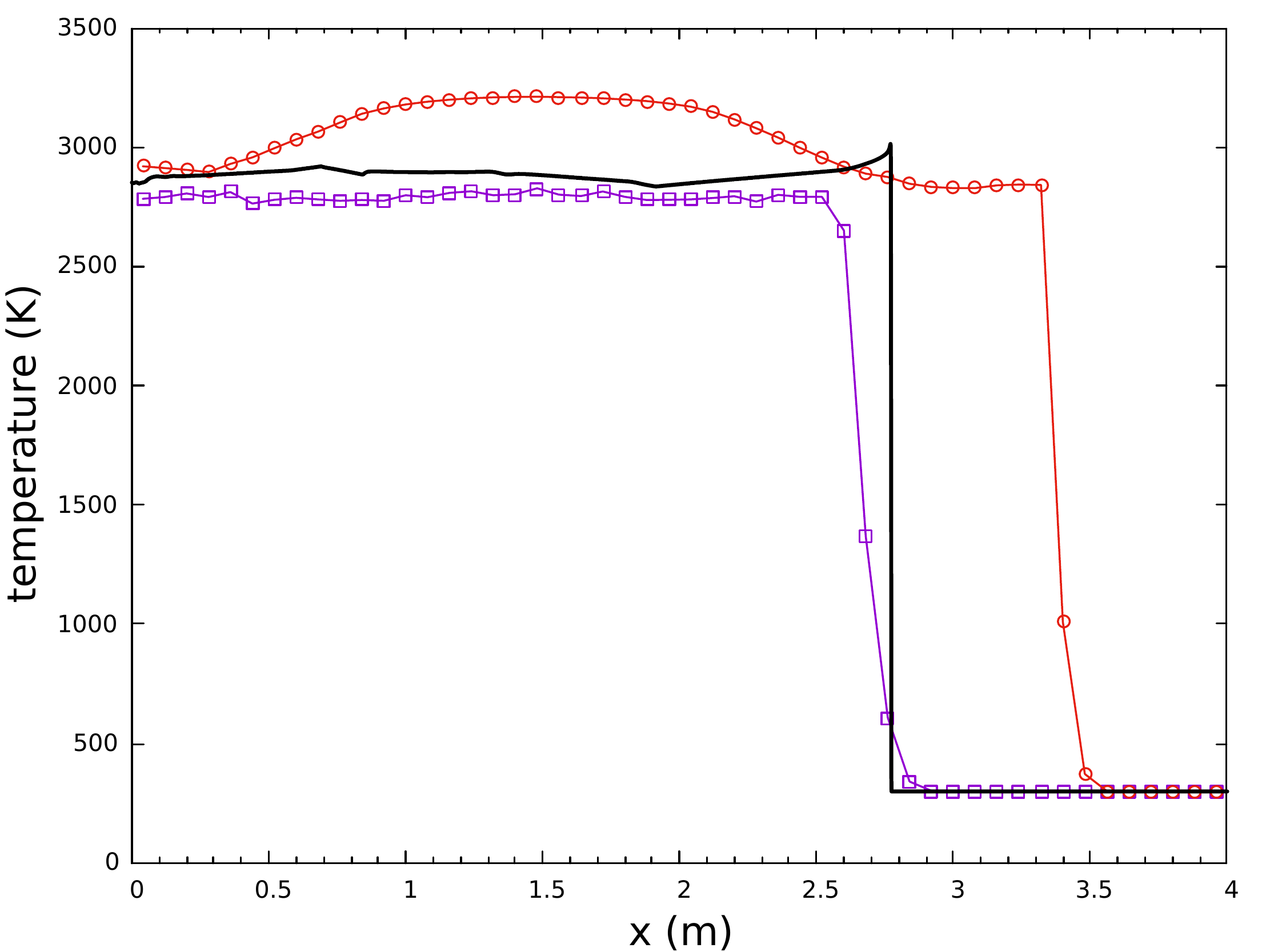} 
  \includegraphics[scale=0.3]{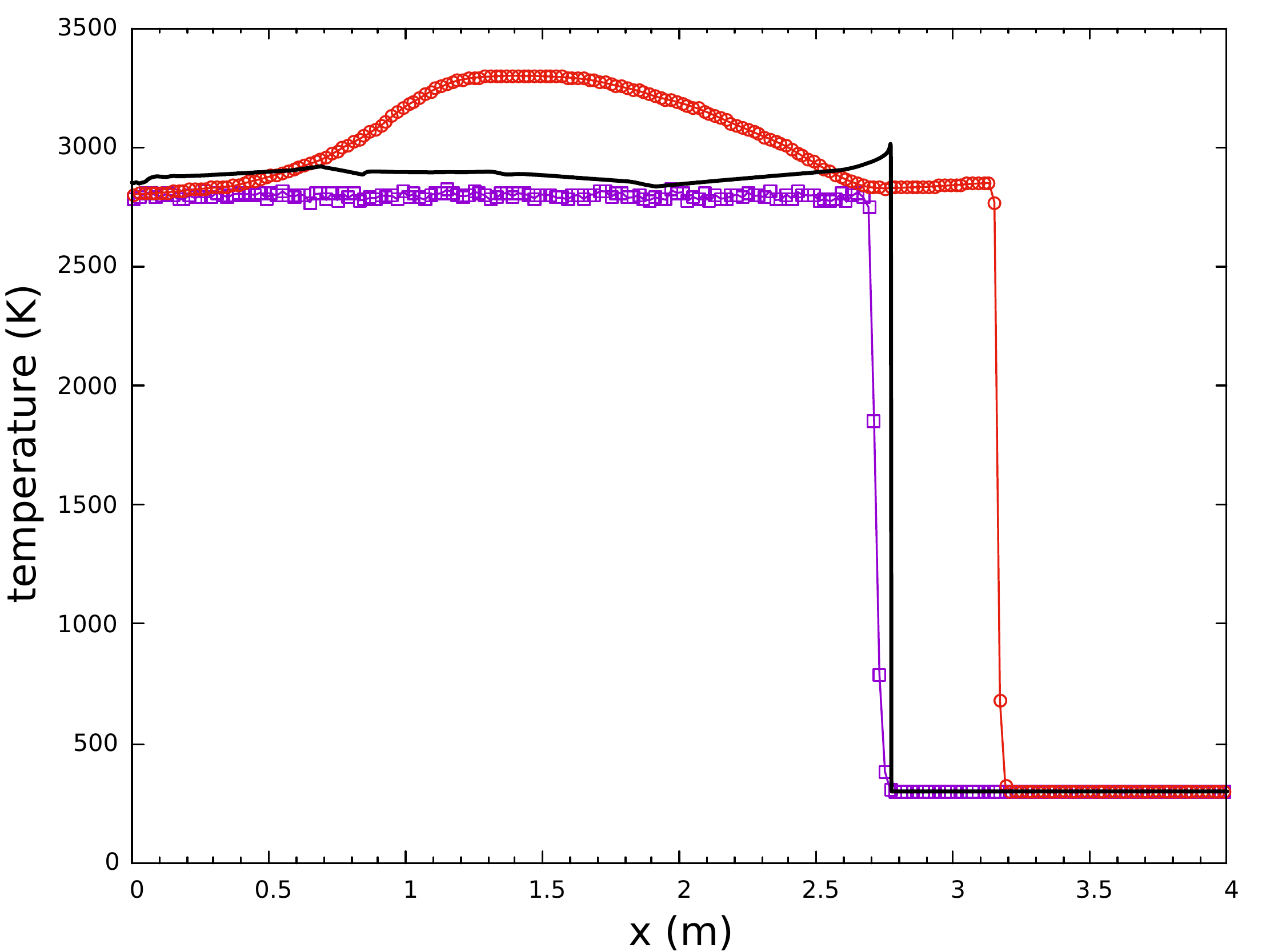} \\
  \includegraphics[scale=0.3]{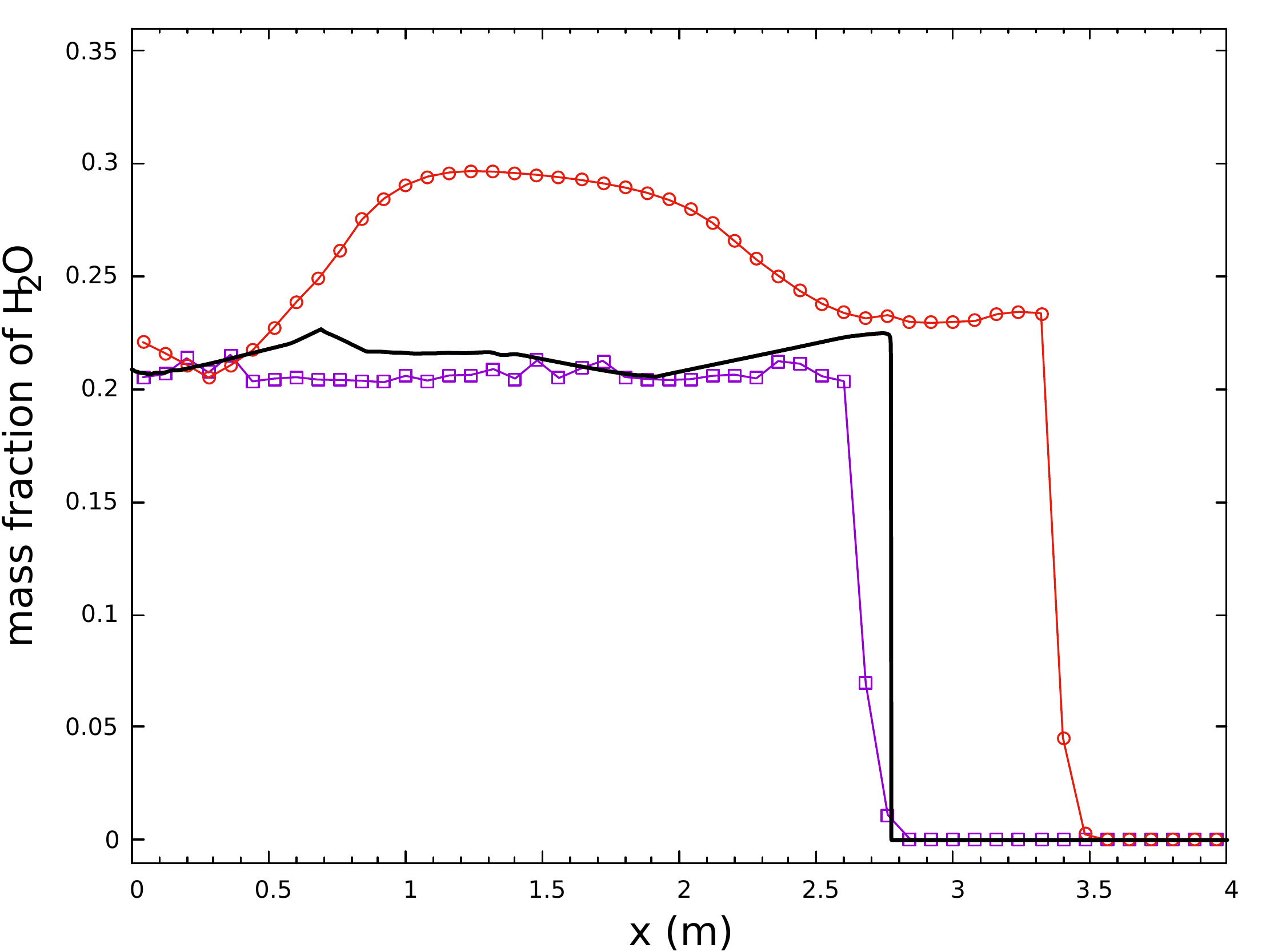} 
  \includegraphics[scale=0.3]{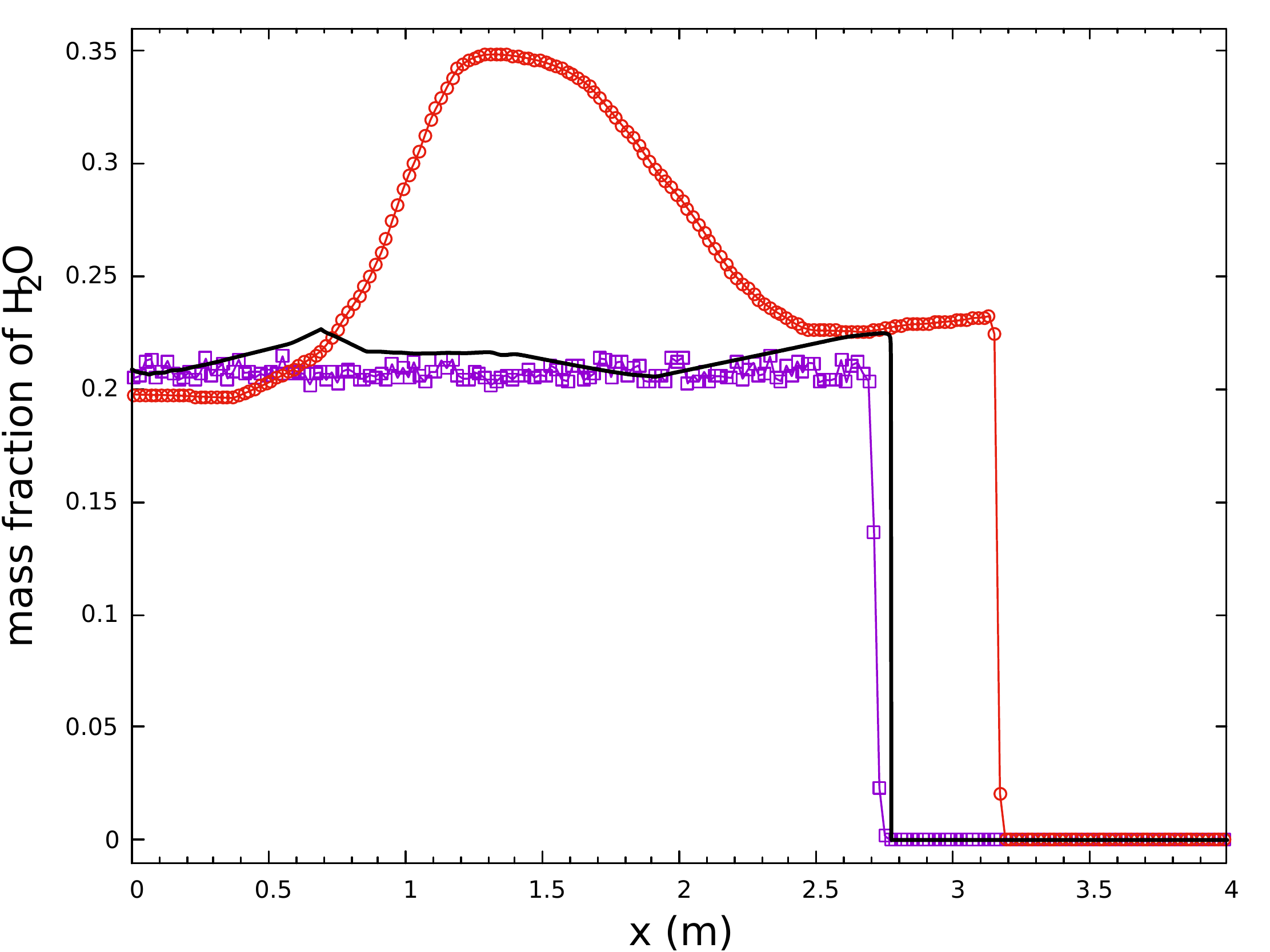} \\
  \caption{Example 7 9-species 23-reaction hydrogen-air CJ detonation at $t=1.2 \times 10^{-3}$s: purple square line $\sim$ SRR solution; red circle line $\sim$ deterministic solution by CHEMEQ2; black solid line $\sim$ reference solution; 
 left column $\sim$ $\Delta x=0.08\text{m}$, $\Delta t=1 \times 10^{-6}\text{s}$; 
 right column $\sim$ $\Delta x=0.02\text{m}$, $\Delta t=2.5 \times 10^{-7}\text{s}$.}    
  \label{example7_1}
\end{figure}

\begin{figure}
  \centering
  \includegraphics[scale=0.5]{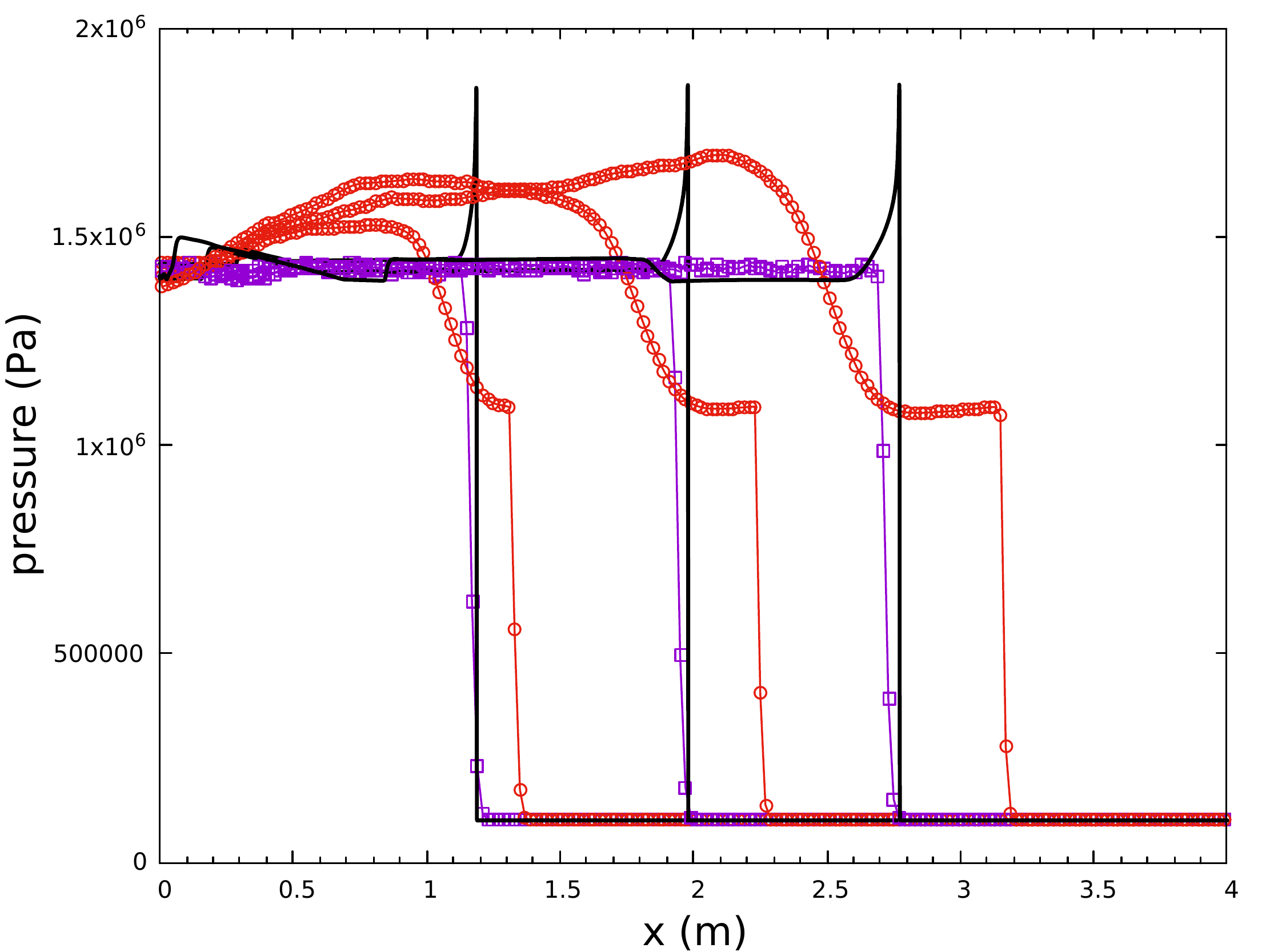} 
  \caption{Example 7 9-species 19-reaction hydrogen-air CJ detonation at $t=0.4, 0.8, 1.2 \times 10^{-3}$s: purple square line $\sim$ SRR solution; red circle line $\sim$ deterministic solution by CHEMEQ2; black solid line $\sim$ reference solution; both solutions $\sim$ $\Delta x=0.02\text{m}$, $\Delta t=2.5 \times 10^{-7}\text{s}$.}    
  \label{example7_2}
\end{figure}

We can see that in Fig. \ref{example7_1} at the given time: although the resolution of the grid and timestep is far lower than the resolved solution, the SRR method predicts the properties of the flowfield in quite good agreement with the reference solution, including the location of the detonation wave and the variable profiles. The obtained profiles tend to converge to the reference solution with the increase of the resolution (and the decrease of stiffness), which also indicates the proposed method can recover nonstiff problems by reducing to the deterministic reference solution under high resolutions, as stated previously. In contrast, using the same under-resolved grid and timestep, the deterministic method yields the spurious nonphysical weak detonation ahead of the shock and the flowfield profiles are totally changed in an incorrect way. In Fig. \ref{example7_2}, wave propagation at different times is presented by looking into the pressure distribution. Despite the deviation by few grid points, the SRR method can always capture the correct wave location while the error in the location of reaction front by the deterministic method is deteriorating in the form of a too fast weak detonation wave. Note that the von Neumann spike inside the reaction zone of the reference solution can be calculated only by very fine resolution both in space and time.

\begin{table}
\setlength{\belowcaptionskip}{10pt}
\caption{Initial condition for Example 7 9-species 23-reaction hydrogen-air CJ detonation}
\centering
\label{example7_IC}
\begin{tabular}{lcc}
\hline
							& post-shock gas 	& pre-shock gas		\\
\hline
pressure (Pa)   			& 1481999.362037  	&	101325			\\
temperature (K)  			& 2941.677242 		&	298				\\
velocity (m/s)	 			& \underline{800} ($\approx u_{CJ}$)   &	0				\\
mass fraction				&					&					\\
$y_{\text{H}}$				& 0.000247			&	0				\\
$y_{\text{O}}$				& 0.001617			&	0				\\
$y_{\text{H}_2\text{O}}$ 	& 0.225404 			&	0				\\
$y_{\text{OH}}$				& 0.014915 			&	0				\\
$y_{\text{O}_2}$			& 0.013336 			&	0.226362		\\
$y_{\text{H}_2}$			& 0.002429 			&	2.852103E-2		\\
$y_{\text{H}_2\text{O}_2}$	& 2.601600E-6 		&	0				\\
$y_{\text{HO}_2}$			& 1.857550E-5 		&	0				\\
$y_{\text{N}_2}$ 		   	& 0.742031 			&	0.745117		\\
\hline
\end{tabular}
\end{table}

\begin{figure}
  \centering
  \includegraphics[scale=0.6]{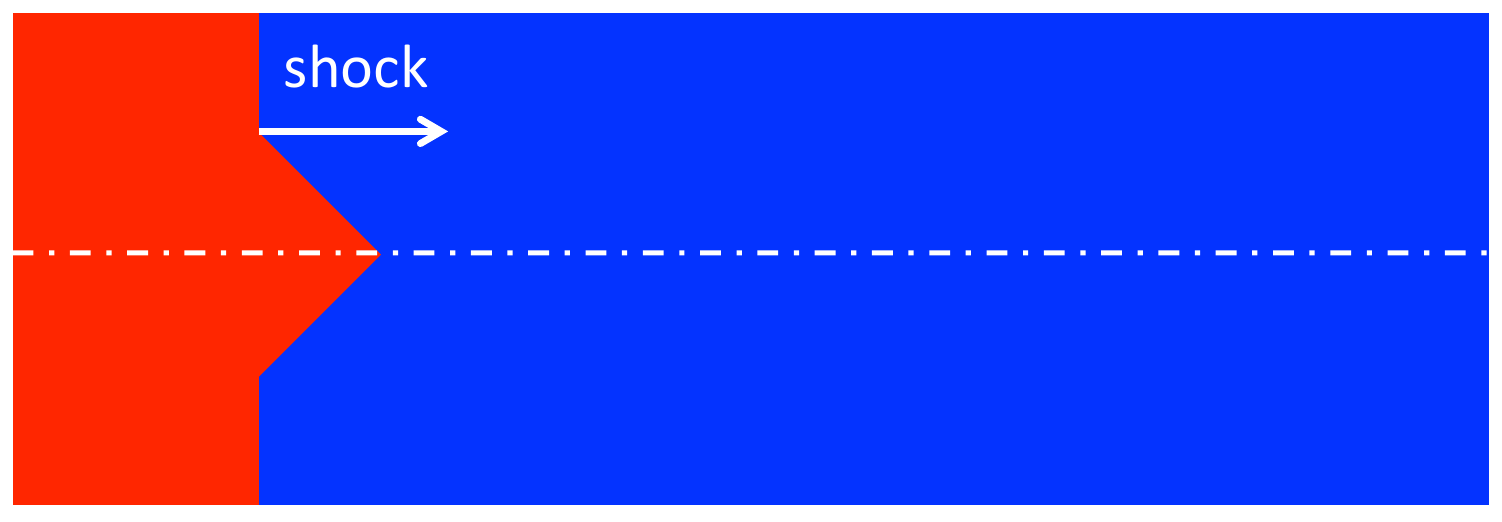} 
  \caption{Schematic for the 2D domain in Example 8.}
  \label{schematic2}
\end{figure}

EXAMPLE 8 (A Realistic Strong Detonation in 2D). The setup of this case consists of two parts divided by a shock travelling to the right in a 2D domain of $[0,3]\text{m} \times [0,1]\text{m}$, as in Fig. \ref{schematic2}: the left red part is post-shock and filled with high-temperature high-pressure burnt gas while the right blue part is pre-shock and filled with reactive unburnt gas in one atmosphere pressure and room temperature. Geometry of the post-shock burnt gas part follows
\begin{equation*}
\{|y-0.5|>0.25, x<0.5 \} \cup \{|y-0.5| \leq  0.25, x-0.25<y<1.25-x \},
\end{equation*}
and the unburnt gas occupies the rest of domain before the initial shock.
Initial states are identical with those in Example 6 except the x-velocity of the post-shock part is increased to $u_b=2000\text{m/s}>u_{CJ}$, to create a strong detonation wave. The boundary condition for the left/right end is simply extrapolation from the mirror image points inside the domain and the top/bottom boundary is considered as a slip wall. All simulations stop at $t=1 \times 10^{-3}$s.         

\begin{figure}
  \centering
  \includegraphics[scale=0.5]{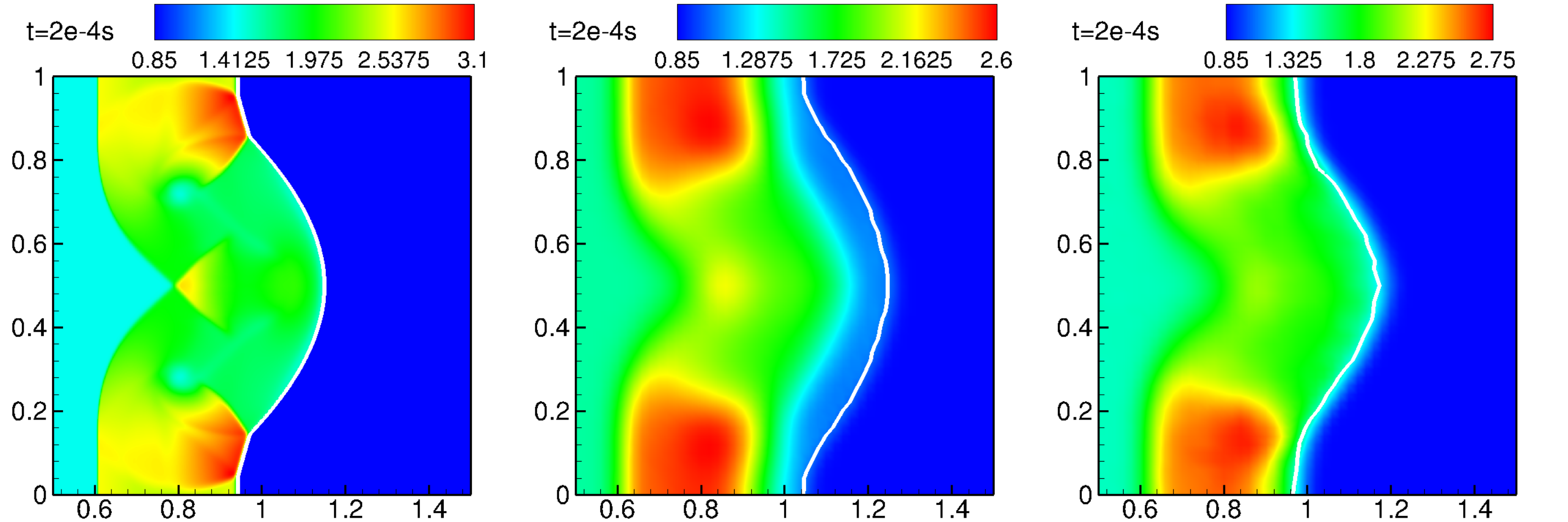} \\
  \includegraphics[scale=0.5]{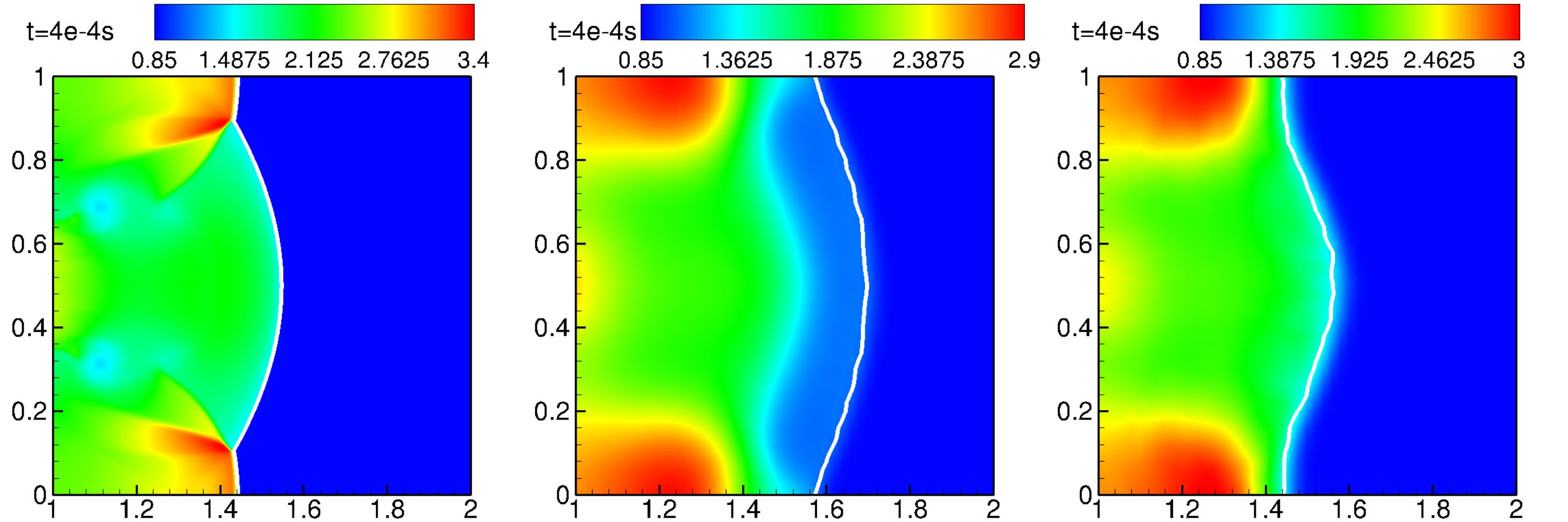} \\
  \includegraphics[scale=0.5]{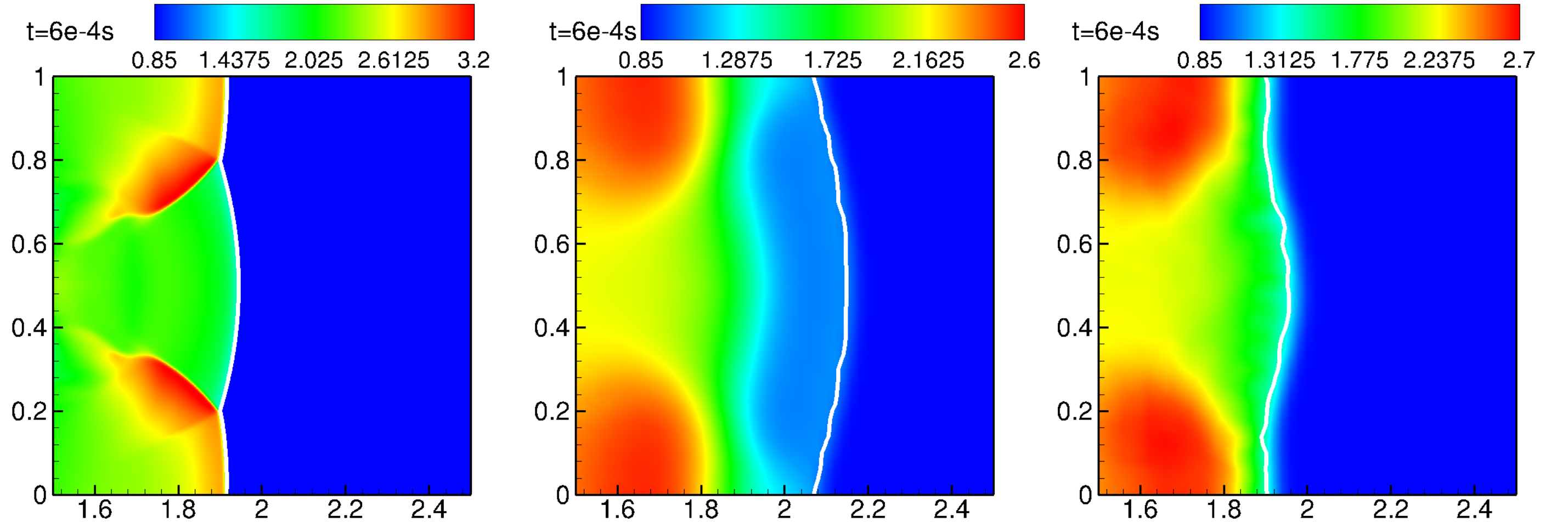} \\
  \includegraphics[scale=0.5]{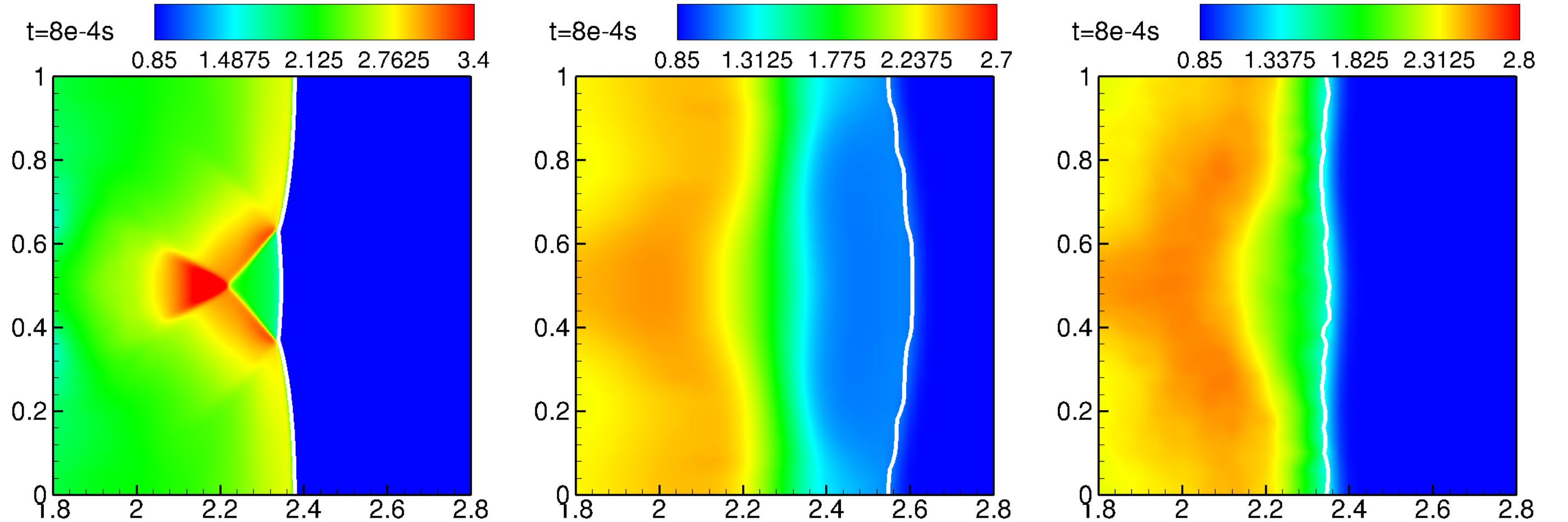} \\
  \includegraphics[scale=0.5]{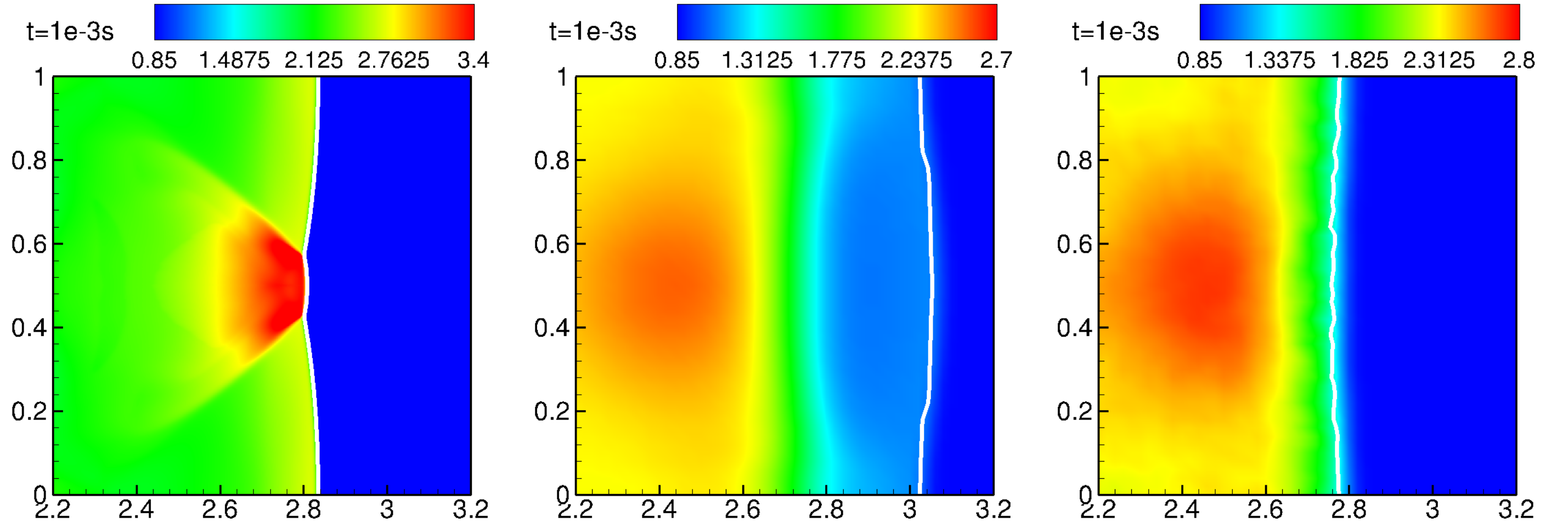} \\
  \caption{Example 8 the density distribution and the detonation front location at different times: left $\sim$ reference solution; middle $\sim$ deterministic solution by CHEMEQ2; right $\sim$ SRR solution; the location of the reacting front is marked by the white solid line with $y_{\text{H}_2\text{O}}=0.1$. }    
  \label{example8_1}
\end{figure}

We obtain the reference exact solution by the deterministic method using a very fine grid with $3000 \times 1000$ points and a fixed tiny timestep of $\Delta t = 2.5 \times 10^{-8}$s. In comparison, a set of under-resolved uniform grid and timestep is considered, i.e. $150 \times 50, \Delta t=2.5 \times 10^{-7}$s (we found using the linearly scaled $\Delta t=5 \times 10^{-7}$s corresponding to the $150 \times 50$ grid appears too large to integrate the ODEs system by CHEMEQ2 stably without any parameter tuning). From Fig. \ref{example8_1}, it is clearly to see the density distributions along with locations of the detonation wave at different times in three solutions. In comparison with the reference solution, the SRR method computes the reasonable locations of the reacting front at all times. Due to the considerably low resolution used in the SRR method, detailed characteristics presented in the reference solution such as the triple points, slip lines, small vortices and peak values of density are diffused while the overall flowfield including the profile of reacting front has been correctly captured. In stark contrast, for the deterministic method with the same resolution, a developing spurious weak detonation wave can be easily detected with a maximum error of nearly 10\% of the domain length in only 1 millisecond. It not only validates the wider effectiveness of the proposed method but also implies even tiny numerical dissipation is potential to be dangerous in a long-term development of reacting flows for ordinary shock-capturing schemes in under-resolved conditions.                          
 
\newgeometry{bottom=0.5cm,top=0.5cm}
\restoregeometry

\section{Conclusions}

A new fractional step method for simulating chemically reacting flows, especially for capturing stiff detonation waves in under-resolved conditions has been developed. 
Two procedures based on operator splitting are included: for the convection part of the reactive Euler equations, any standard shock-capturing scheme is free to utilize; for the reaction step the multi-species multi-reaction ODEs system in the source terms is further split to solve in a reaction-by-reaction manner, from which exact mass conservation, strict positivity preserving and almost unconditional stability are guaranteed. Unlike deterministic methods that integrate the ODEs directly or the random projection method that requires two presumed equilibrium states, each reaction in the reaction system either proceeds a timestep forward or stops according to a local random temperature in the proposed method. Chemical reaction, e.g. ignition, in the smeared discontinuities due to numerical viscosity in the shock-capturing method is therefore a random process 
, rather than a deterministic one with growing error accumulation. A wide range of numerical experiments including not only simple model kinetics but also real-world nonequilibrium chemistry such as the temperature-dependent finite-rate hydrogen-air combustion are considered in 1D and 2D flows, demonstrating the proposed method can effectively predict the correct propagation of discontinuities as well as the overall flowfield information in under-resolved conditions. 
Besides, 
the diminishing randomness by adding a shift term to generate a random temperature below its local smeared value enables the regression of the proposed random method into a deterministic method in terms of nonstiff cases with fine resolutions in space and time. Also, its dimensional independence makes further 3D extension of the proposed method straightforward.

\section*{Acknowledgements}
The financial support from the EU Marie Sk{\l}odowska-Curie Innovative Training Networks (ITN-ETN) (Project ID: 675528-IPPAD-H2020-MSCA-ITN-2015) for the first author is gratefully acknowledged.

\appendix

\newpage
\section{Analytical solutions to some simple forms of a one-way reaction equation} \label{appendix1}

For the simplest form of a reaction in Eq. \eqref{one-way-simple}, 
\begin{equation}
\begin{aligned}
A \longrightarrow \text{products}, \\
\end{aligned}
\end{equation}
we simply have an ODE for the molar concentration $\left[ A \right]$, as
\begin{equation}\label{A1}
\frac{d \left[ A \right]}{dt} = - k \left[ A \right],
\end{equation}
with $k$ being the rate constant and initial value of $\left[ A \right]_0$ at $t = t_0$. The above ODE written in the expression of molar concentration is equivilent to Eq. \eqref{Rj} using density and mass fraction since
\begin{equation}
 \left[ A \right] = \frac{\rho_A}{W_A} = \frac{\rho y_A}{W_A}.
\end{equation}
The solution to Eq. \eqref{A1} by seperation of varibles is
\begin{equation}
 \left[ A \right] =  \left[ A \right]_0 e^{-k (t-t_0)}.
\end{equation}

For the reaction form
\begin{equation} \label{A+B}
\begin{aligned}
A + B \longrightarrow \text{products}, \\
\end{aligned}
\end{equation}  
we have the ODEs system as  
\begin{equation} \label{A,B}
\begin{aligned}
\frac{d \left[ A \right]}{dt} = - k \left[ A \right]\left[ B \right], \\
\frac{d \left[ B \right]}{dt} = - k \left[ A \right]\left[ B \right]. \\
\end{aligned}
\end{equation}
This also means that 
\begin{equation}
\begin{aligned}
d \left[ A \right] = d \left[ B \right]
\end{aligned}
\end{equation}  
holds for any time interval $dt$ and thus
\begin{equation} \label{A=B}
\begin{aligned}
\left[ A \right] - \left[ A \right]_0 = \left[ B \right] - \left[ B \right]_0. 
\end{aligned}
\end{equation}  
Substituting relation \eqref{A=B} into Eq. \eqref{A,B}, we have 
\begin{equation}
\begin{aligned}
\frac{d \left[ A \right]}{dt} = - k \left[ A \right] ( \left[ A \right] + \Delta_{AB} ),
\end{aligned}
\end{equation}
where $ \Delta_{AB} = \left[ B \right]_0 - \left[ A \right]_0 $, leading to the solution of $\left[ A \right]$ as
\begin{equation}
\begin{aligned}
\left[ A \right] =
\begin{cases}
\frac{\Delta_{AB}}{ \frac{\left[B\right]_0}{\left[A\right]_0} e^{\Delta_{AB} k(t-t_0)} - 1  }, \quad & \text{if} \, \Delta_{AB} \neq 0, \\
\frac{1}{ k(t-t_0) + \frac{1}{ \left[ A \right]_0 }  }, \quad & \text{otherwise}.  \\
\end{cases}
\end{aligned}
\end{equation}

For reaction 
\begin{equation}
\begin{aligned}
2A \longrightarrow \text{products}, \\ 
\end{aligned}
\end{equation}
it is a special case for reaction \eqref{A+B} and the solution is 
\begin{equation}
\begin{aligned}
 \left[ A \right] &= \frac{1}{ k(t-t_0) + \frac{1}{ \left[ A \right]_0 }  }.  \\
\end{aligned}
\end{equation}

For a more complicated third-order reaction 
\begin{equation}
\begin{aligned}
 A + B + C \longrightarrow \text{products}, \\ 
\end{aligned}
\end{equation}
we also ultilize the relations 
\begin{equation}
\begin{aligned}
\left[ A \right] - \left[ A \right]_0 = \left[ B \right] - \left[ B \right]_0 = \left[ C \right] - \left[ C \right]_0 
\end{aligned}
\end{equation}
and perform seperation of varibles to get 
\begin{equation}
\begin{aligned}
\frac{d \left[ A \right]}{\left[ A \right] ( \left[ A \right] + \Delta_{AB} ) ( \left[ A \right] + \Delta_{AC} )} = - k dt .
\end{aligned}
\end{equation}
Finally, we can only have the implicit solution for $\left[ A \right]_0 \neq \left[ B \right]_0 \neq \left[ C \right]_0$ in general, obeying
\begin{equation}
\begin{aligned}
\left( \frac{\left[A\right]}{\left[A\right]+\Delta_{AC}}  \frac{\left[C\right]_0}{\left[A\right]_0} \right)^{\frac{1}{\Delta_{CB}\Delta_{AC}} }
-
\left( \frac{\left[A\right]}{\left[A\right]+\Delta_{AB}}  \frac{\left[B\right]_0}{\left[A\right]_0} \right)^{\frac{1}{\Delta_{CB}\Delta_{AB}} }
= e^{-k(t-t_0)}.
\end{aligned}
\end{equation}

Only when $\left[ A \right]_0 = \left[ B \right]_0 = \left[ C \right]_0$ or the special reaction
\begin{equation}
\begin{aligned}
3A \longrightarrow \text{products}, \\
\end{aligned}
\end{equation}
the explicit analytical solution exists, i.e.
\begin{equation}
\begin{aligned}
 \left[ A \right] = \sqrt{\frac{1}{\frac{1}{\left[A\right]_0^2} + 2k(t-t_0)}}.
\end{aligned}
\end{equation}

After the determination of the new state of the reactant species $\left[ A \right]$, states of the remaining species including all the products and other reactants can be updated by the law of mass conservation in Eq. \eqref{delta}.

\newpage
\section{Reaction mechanism for hydrogen-air combustion} \label{appendix2}

\begin{table*}[!htbp]
\centering
{
\begin{threeparttable}[b]
\begin{tabular}{lllll}
\hline
ID		& Elementary reaction & $A$ & $B$ & $E_a$  \\
\hline
1,2		& $\text{H} + \text{O}_2 \Longleftrightarrow \text{OH} + \text{O}$ & 1.91e+14 & 0.0 & 16.44 \\
3,4		& $\text{H}_2 + \text{O} \Longleftrightarrow \text{H} + \text{OH}_2 $ & 5.08e+04 & 2.67 & 6.292 \\ 
5,6		& $\text{H}_2 + \text{OH} \Longleftrightarrow \text{H} + \text{H}_2\text{O} $ & 2.16e+08 & 1.51 & 3.43 \\
7,8		& $\text{O} + \text{H}_2\text{O} \Longleftrightarrow \text{OH} + \text{OH} $ & 2.97e+06 & 2.02 & 13.4 \\ 
9,10\tnote{*}	& $\text{H}_2 + \text{M} \Longleftrightarrow \text{H} + \text{H} + \text{M} $ & 4.57e+19 & -1.4 & 105.1 \\ 
11,12\tnote{*}	& $\text{O} + \text{O} + \text{M} \Longleftrightarrow \text{O}_2 + \text{M} $ & 6.17e+15 & -0.5 & 0.0 \\
13,14\tnote{*}	& $\text{H} + \text{O} + \text{M} \Longleftrightarrow \text{OH} + \text{M} $ & 4.72e+18 & -1.0 & 0.0 \\
15,16\tnote{**}	& $\text{H} + \text{OH} + \text{M} \Longleftrightarrow \text{H}_2\text{O} + \text{M} $ & 4.50e+22 & -2.0 & 0.0\\
17,18\tnote{***} \quad & $\text{H} + \text{O}_2 + \text{M} \Longleftrightarrow \text{H}\text{O}_2 + \text{M} $ & 3.48e+16 & -0.41 & -1.12\\
19,20 	& $\text{H} + \text{O}_2 \Longleftrightarrow \text{H}\text{O}_2 $ & 1.48e+12 & 0.60 & 0.0 \\
21,22	& $\text{H} + \text{HO}_2 \Longleftrightarrow \text{H}_2 + \text{O}_2 $ & 1.66e+13 & 0.0 & 0.82 \\
23,24	& $\text{H} + \text{HO}_2 \Longleftrightarrow \text{OH} + \text{OH} $ & 7.08e+13 & 0.0 & 0.3 \\
25,26	& $\text{HO}_2 + \text{O} \Longleftrightarrow \text{OH} + \text{O}_2 $ & 3.25e+13 & 0.0 & 0.0 \\
27,28	& $\text{OH} + \text{HO}_2 \Longleftrightarrow \text{H}_2\text{O} + \text{O}_2 $ & 2.89e+13 & 0.0 & -0.5 \\
29,30	& $\text{H}\text{O}_2 + \text{H}\text{O}_2 \Longleftrightarrow \text{H}_2\text{O}_2 + \text{O}_2$ & 4.20e+14 & 0.0 & 11.98 \\
31,32	& $\text{H}\text{O}_2 + \text{H}\text{O}_2 \Longleftrightarrow \text{H}_2\text{O}_2 + \text{O}_2$ & 1.30e+11 & 0.0 & -1.629 \\
33,34\tnote{*}	& $\text{H}_2\text{O}_2 + \text{M} \Longleftrightarrow \text{OH} + \text{OH} + \text{M} $ & 1.27e+17 & 0.0 & 45.5 \\
35,36 	& $\text{H}_2\text{O}_2 \Longleftrightarrow \text{OH} + \text{OH} $ & 2.95e+14 & 0.0 & 48.4 \\
37,38	& $\text{H}_2\text{O}_2 + \text{H} \Longleftrightarrow \text{H}_2\text{O} + \text{OH} $ & 2.41e+13 & 0.0 & 3.97 \\
39,40	& $\text{H}_2\text{O}_2 + \text{H} \Longleftrightarrow \text{H}_2 + \text{HO}_2 $ & 6.03e+13 & 0.0 & 7.95 \\
41,42	& $\text{H}_2\text{O}_2 + \text{O} \Longleftrightarrow \text{OH} + \text{HO}_2 $ & 9.55e+06 & 2.0 & 3.97 \\
43,44	& $\text{H}_2\text{O}_2 + \text{OH} \Longleftrightarrow \text{H}_2\text{O} + \text{HO}_2 $ & 1.00e+12 & 0.0 & 0.0 \\
45,46	& $\text{H}_2\text{O}_2 + \text{OH} \Longleftrightarrow \text{H}_2\text{O} + \text{HO}_2 $ & 5.80e+14 & 0.0 & 9.56 \\
\hline
\end{tabular}
      \begin{tablenotes}[flushleft,para]
      \footnotesize{
      Third-body collision coefficiencies (default value is 1.0) in reactions with M: \\
        \item[*] $\text{H}_2\text{O}=12.0$, $\text{H}_2=2.5$; \\
        \item[**] $\text{H}_2\text{O}=12.0$, $\text{H}_2=0.73$;\\
        \item[***] $\text{H}_2\text{O}=14.0$, $\text{H}_2=1.3$.\\
      Units: cm$^3$, mol, s, kcal, K.
      }
      \end{tablenotes}
\end{threeparttable}
}
\end{table*}

\newpage
\section*{References}
\bibliographystyle{abbrv}
\bibliography{split-random}

%
%
%
%

\end{document}